\documentclass[twocolumn,twocolappendix,showpacs,superscriptaddress,groupedaddress,
  nofootinbib,floatfix]{aastex63}

\pdfoutput=1 
\usepackage{amsmath,amsfonts,amstext,amssymb}
\usepackage[T1]{fontenc}
\usepackage[utf8]{inputenc}
\usepackage{graphicx}
\usepackage{rotating}
\usepackage{hyperref}
\usepackage{xspace}
\usepackage{enumerate}

\usepackage{xcolor}

\newcommand{\be}{\begin{equation}}
\newcommand{\ee}{\end{equation}}

\newcommand{\tGRAthena}{\texttt{GR-Athena++}}
\newcommand{\GRAthena}{\tGRAthena\xspace}

\newcommand{\tAthena}{\texttt{Athena++}}
\newcommand{\Athena}{\tAthena\xspace}

\newcommand{\tGRoAthena}{\texttt{(GR-)Athena++}}
\newcommand{\GRoAthena}{\tGRoAthena}

\newcommand{\tBAM}{\texttt{BAM}}
\newcommand{\BAM}{\tBAM\xspace}
\newcommand{\tCactus}{\texttt{Cactus}}
\newcommand{\Cactus}{\tCactus\xspace}
\newcommand{\tGRChombo}{\texttt{GRChombo}}
\newcommand{\GRChombo}{\tGRChombo\xspace}

\newcommand{\tKokkos}{\texttt{Kokkos}}
\newcommand{\Kokkos}{\tKokkos\xspace}
\newcommand{\tKAthena}{\texttt{K-Athena}}
\newcommand{\KAthena}{\tKAthena\xspace}
\newcommand{\tDendrogr}{\texttt{Dendro-GR}}
\newcommand{\Dendrogr}{\tDendrogr\xspace}

\newcommand{\Cpp}{\texttt{C++}\xspace}
\newcommand{\AwA}{\texttt{AwA}\xspace}

\newcommand{\tTwoPunctures}{\texttt{TwoPunctures}}
\newcommand{\TwoPunctures}{\tTwoPunctures\xspace}

\def\GMc2{{\rm G M_{\odot} c^{-2}}}

\def\l{\ell}
\def\lm{{\ell m}}

\def\kt2{\kappa^\text{T}_2}

\def\kt2{\kappa^\text{T}_2}

\def\CFL{\mathrm{CFL}}
\def\KO{\sigma}

\def\D{\mathrm{D}}
\def\pd{\partial{}}

\def\Ng{\mathcal{N}_{\mathrm{g}}}
\def\Ncg{\mathcal{N}_{\mathrm{cg}}}

\def\2nd{2^\mathrm{nd}}
\def\4th{4^\mathrm{th}}
\def\6th{6^\mathrm{th}}
\def\8th{8^\mathrm{th}}
\def\sp{\delta x}
\def\defG{\widehat{\Gamma}}

\def\z4c{$\mathrm{Z}4\mathrm{c}$}
\def\z4oc{$\mathrm{Z}4(\mathrm{c})$}
\def\z4{$\mathrm{Z}4$}
\def\ccz4{$\mathrm{CCZ}4$}

\newcommand{\Mesh}{\texttt{Mesh}}
\newcommand{\MeshBlock}{\texttt{MeshBlock}}

\usepackage{color}
\definecolor{cyan}{rgb}{0,0.9,0.9}
\definecolor{orange}{rgb}{0.9,0.5,0}
\definecolor{magenta}{rgb}{1,0,1}
\definecolor{purple}{rgb}{0.8,0.4,0.8}
\definecolor{gray}{rgb}{0.5,0.5,0.5}

\begin{document}

\title{\GRAthena: puncture evolutions on vertex-centered oct-tree AMR}

\author{Boris \surname{Daszuta}}
\affiliation{
  Theoretisch-Physikalisches Institut, Friedrich-Schiller-Universit{\"a}t Jena, 07743, Jena, Germany}
\author{Francesco \surname{Zappa}}
\affiliation{
  Theoretisch-Physikalisches Institut, Friedrich-Schiller-Universit{\"a}t Jena, 07743, Jena, Germany}
\author{William \surname{Cook}}
\affiliation{
  Theoretisch-Physikalisches Institut, Friedrich-Schiller-Universit{\"a}t Jena, 07743, Jena, Germany}
\author{David \surname{Radice}}
\affiliation{Institute for Gravitation and the Cosmos, The Pennsylvania State University, University Park, PA 16802, USA}
\affiliation{Department of Physics, The Pennsylvania State University, University Park, PA 16802, USA}
\affiliation{Department of Astronomy and Astrophysics, The Pennsylvania State University, University Park, PA 16802, USA}
\author{Sebastiano \surname{Bernuzzi}}
\affiliation{
  Theoretisch-Physikalisches Institut, Friedrich-Schiller-Universit{\"a}t Jena, 07743, Jena, Germany}
\author{Viktoriya \surname{Morozova}}
\affiliation{Institute for Gravitation and the Cosmos, The Pennsylvania State University, University Park, PA 16802, USA}
\affiliation{
  Department of Physics, The Pennsylvania State University, University Park, PA 16802, USA}

\date{\today}

\begin{abstract}
  Numerical relativity is central to the investigation of
  astrophysical sources in the dynamical and strong-field gravity regime,
  such as binary black hole and neutron star coalescences.
  Current challenges set by gravitational-wave and multi-messenger
  astronomy call for highly performant and scalable codes on modern
  massively-parallel architectures.
  We present \GRAthena{}, a general-relativistic,
  high-order, vertex-centered solver that extends the oct-tree, adaptive
  mesh refinement capabilities of the astrophysical (radiation) magnetohydrodynamics
  code \Athena{}.
  To simulate dynamical space-times
  \GRAthena{} uses the \z4c{}
  evolution scheme of
  numerical relativity
  coupled to the moving puncture gauge.
  We demonstrate stable and accurate binary black hole merger evolutions
  via extensive convergence testing, cross-code validation, and
  verification against state-of-the-art effective-one-body waveforms.
  \GRAthena{} leverages the task-based parallelism paradigm
  of \Athena{} to achieve excellent scalability.
  We measure strong scaling
  efficiencies above $95\%$ for up to $\sim 1.2\times10^4$ CPUs and
  excellent weak scaling is shown up to $\sim 10^5$ CPUs in a
  production binary black hole setup with adaptive mesh
  refinement.
  \GRAthena{} thus allows for the robust simulation of
  compact binary coalescences and offers a viable path towards
  numerical relativity at exascale.
\end{abstract}

\section{Introduction}\label{sec:intro}
Numerical relativity (NR) provides robust techniques for constructing numerical solutions
to the Einstein field equations (EFE). The phenomenology of astrophysical inspiral and merger
events, such as those between
binary constituents involving (variously) black holes (BH) or neutron
stars (NS) can be succinctly described with
gravitational waves (GW) computed
with NR \cite{Pretorius:2005gq,Baker:2006ha,Campanelli:2005dd,Shibata:1999wm}.
This has crucially assisted in the recent detection of such events by the LIGO and Virgo
collaborations \cite{Abbott:2016blz,TheLIGOScientific:2016wfe,TheLIGOScientific:2017qsa}.
As the operating sensitivity of these detectors is improved \cite{Abbott_2020}
and new (KAGRA \cite{akutsu2020overviewkagracalibration}),
approved (LISA \cite{amaroseoane2017laserinterferometerspace}),
or
proposed (Einstein telescope \cite{Punturo:2010zz},
Cosmic explorer \cite{Evans:2016mbw}) designs come online
a concomitant enlargement of the physical parameter space that may be
experimentally probed is
offered. These experimental efforts are complemented by publicly available catalogs of simulation
data provided by the
NR community (see e.g.~\cite{Dietrich:2018phi,Boyle:2019kee,Healy:2019jyf,Jani:2016wkt}).

On the NR side there is thus a pressing requirement to better resolve and characterize the
underlying physics during simulation of binary black holes (BBH) in ever more extreme
configurations, such as higher mass ratio \cite{Lousto:2010qx,Nakano:2011pb}
or to provide discrimination between
candidate models in description of
binary NS \cite{Radice:2020ddv,Bernuzzi:2020tgt}
or BH-NS \cite{Shibata:2011jka} events.
Such simulations can be extremely demanding from the point of view of computational resources
and viability typically hinges upon the availability of high performance computing (HPC)
infrastructure \cite{huerta2019supportinghighperformancehighthroughput}.
Thus accurate NR codes that remain performant as HPC resources are scaled up
and simultaneously allow for the scope of input physics to be simply extended are crucial.

An important concern for NR investigations of the binary merger problem is treatment of
features sensitive to widely-varying length and time-scales. An approach inspired by
Berger-Oliger \cite{Berger:1984zza} (see also \cite{Berger:1989a}) is based on
the introduction of a sequence of hierarchically, well-nested
patches (usually boxes) of increasing resolution and decreasing diameter centered at the
constituents of the binary. The relative spatial positions of such patch configurations
may be arranged to automatically track the time-evolution of the aforementioned
compact objects.
This has been a common approach adopted for NR codes that build upon the open
source \Cactus{} framework \cite{Goodale:2002a} and
utilize the {\tt Carpet} thorn \cite{Schnetter:2003rb}
(see also the {\tt Einstein} toolkit \cite{Loffler:2011ay} for an overview tailored to
astrophysical applications).
Some notable code implementations based on \Cactus{} are
{\tt Llama} \cite{Pollney:2009yz,Reisswig:2012nc},
{\tt McLachlan} \cite{brown2009turduckeningblackholes},
{\tt LEAN} \cite{sperhake2007binaryblackholeevolutions},
{\tt LazEv} \cite{zlochower2005accurateblackhole},
and {\tt Maya} \cite{herrmann2007unequalmassbinary}
furthermore within this framework magnetohydrodynamics (MHD) may be coupled through
use of {\tt GRHydro} \cite{Moesta:2013dna}
or {\tt WhiskyTHC} \cite{Radice:2013xpa}.
Other examples of non-\Cactus{} codes adopting the Berger-Oliger approach include
\BAM{} \cite{Brugmann:2008zz,galaviz2010numericalevolutionmultiple,Thierfelder:2011yi},
{\tt AMSS-NCKU} \cite{cao2008reinvestigationmovingpunctured}.
Elements of this approach are also shared by the recent \GRChombo{} \cite{Clough:2015sqa}.
Thus far, all code-bases
discussed here make (at least some)
use of Cartesian grid coordinatizations and involve use of finite-difference (FD) approximants to
derivative operators.

Unfortunately here a priori specification of patch hierarchies
is usually required which makes capturing emergent features at unexpected locations challenging.
More importantly for Berger-Oliger the overhead of synchronization of solution data between
patches on differing levels can incur heavy performance penalties which spoil scaling in modern
highly-parallel HPC architectures \cite{stout1997adaptiveblockshigh}.

Other approaches such as those based on pseudo-spectral methods are represented by
{\tt SpeC} \cite{Szilagyi:2009qz} where
multi-patch decomposition of the computational domain is made using a combination of
topological spheres and cylinders. The SXS collaboration \cite{Boyle:2019kee} has
used {\tt SpeC} to produce some of the longest and most-accurate binary GW to date,
albeit BBH mass ratio (defined $q:=m_1/m_2$ where $m_i$ are constituent masses
and $m_1\geq m_2$) for publicly available GR templates remains confined
to $q\leq10$. Closely related are efforts based
on the discontinuous Galerkin (DG) method such
as {\tt bamps} \cite{Hilditch:2015aba,Bugner:2015gqa}
and {\tt SpECTRE} \cite{Kidder:2016hev}.
Another recently pursued alternative has been an attempt to eliminate the need for refinement
altogether through generation of adapted, problem-specific curvilinear-grids as recently
demonstrated in {\tt SENR/NRPy+} \cite{Mewes:2020vic,mewes2018numericalrelativityspherical,%
ruchlin2018senrnrpynumericala}.

A blend of benefits that ameliorates some of the disadvantages of the above approaches is offered
in block-based adaptive mesh refinement (AMR) strategies \cite{stout1997adaptiveblockshigh}.
Crucially, in contrast to hierarchical, nested patches, for block-based AMR each physical
position on a computational domain is covered by one and only one level. This reduces the
problem of synchronizing the data of a solution over differing levels to communication
between block boundaries (as in DG) which when logically arranged into an oct-tree
(in $3$ spatial dimensions -- see e.g.~\cite{Burstedde:2011a})
can greatly improve computational efficiency through preservation of data locality in memory.
Furthermore, making use of task-based
parallelism as the computational model greatly facilitates the overlap of communication and
computation.
Additionally, great flexibility is maintained in how the computational domain can be refined.
The recent work of \Dendrogr{} \cite{Fernando:2018mov} utilizes such an approach in treatment
of the vacuum sector of the EFE
with the BSSNOK formulation \cite{Nakamura:1987zz,Shibata:1995we,Baumgarte:1998te}
where solution
representation is in terms of adaptive wavelets and choice of regions to refine
controlled by the wavelet expansion itself \cite{holmstrom1999solvinghyperbolicpdes}.
Scaling in terms of HPC performance appears to have been convincingly
demonstrated with \Dendrogr{} for a mock $q=10$ BBH event,
although numerical accuracy and full scale evolutions, together with
 GW characteristics that would potentially be
calculated during production runs are not presented for any $q$.

In this work we present our effort to build upon a public version
of \Athena{} \cite{white2016extensionathenacode,%
felker2018fourthorderaccuratefinite,%
stone2020athenamathplusmathplus} where changes to core functionality have been made,
together with introduction of new modules targeted towards
solution of the EFE. We refer to these new features as \GRAthena{}.
Originally \Athena{} was conceived as a
framework for purely
non- and special-relativistic MHD, as well as GRMHD for stationary space-times,
which adopts many of the mature and robust numerical
algorithms of \cite{stone2008athenanewcode} in a modern \Cpp design centered around block-based
AMR.
Key design elements include native support for Cartesian and curvilinear coordinates
and a particular focus on future-proofing through code modularity. The computational model
is task-based and embeds hybrid parallelism through dual use of
message passing interface (MPI) and threading via OpenMP (OMP).
In addition to excellent
scaling properties on HPC infrastructure, modularity and modern code practices have allowed
for extension of \Athena{} to heterogeneous architectures through \Kokkos{}
\cite{carteredwards2014kokkosenablingmanycore} resulting in performant MHD
calculation on graphics processing units
with \KAthena{} \cite{grete2019kathenaperformanceportable} (see
also {\tt parthenon} \cite{parthenon:web}).

These attractive properties served as a strong motivation in development
of \GRAthena{} where we have implemented
the \z4c{} formulation \cite{Bernuzzi:2009ex,Ruiz:2010qj,Weyhausen:2011cg,Hilditch:2012fp}
of NR utilizing the (moving) puncture gauge \cite{Brandt:1997tf,Baker:2006ha,Campanelli:2005dd}.
We provide accurate and efficient extensions
to derivative approximants through (templated) arbitrary-order FD based
on \cite{Alfieri:2018a}.
Our introduction of vertex-centered (VC) variable treatment
(extending core cell- and face-centered functionality)
is motivated by a desire to match any selected FD order in calculations
that involve AMR. Furthermore
our implementation of the level to level transfer operators that occur in AMR
takes advantage of the particular structure over sampled nodes at differing
grid levels to simultaneously improve computational efficiency and accuracy.
Within \GRoAthena{} time-evolution is achieved through the standard method of lines
approach where scheme order may be specified flexibly. In this work the formal order
of the spatial discretizations considered are $\4th$ and $\6th$ whereas the temporal
treatment is at $\4th$ order.

As calculating quantities such as GW typically involves integration
over spherical surfaces we have introduced a module implementing geodesic spheres
based on \cite{Wang:2011}. A primary demonstration of this
functionality is presented in direct, cross-code validation against \BAM{} where extracted
GW are computed from the Weyl scalar $\Psi_4$ and the gravitational strain is examined for
the prototype BH and BBH calibration problems of \cite{Brugmann:2008zz}.
Additionally, it is important that simulations can be carried out that provide
data of physical relevance to detection efforts.
To this end we consider an equal mass BBH inspiral on an initially quasi-circular
(i.e low eccentricity) co-orbit that results in a merger event. Initial data is based on the
configuration of \cite{Hannam:2010ec}.
Verification is made utilizing the state-of-the-art,
NR informed, effective one body model of {\tt TEOBResumS} \cite{Nagar:2018zoe}.

As we prioritized HPC efficiency it is thus important
that \GRAthena{} preserves the already impressive behavior of \Athena{}
where over $80\%$ parallel efficiency is shown in weak scaling tests with uniform
grids employing up to $\sim 1.3\times10^5$ CPUs
for MHD/HD problems~\cite{stone2020athenamathplusmathplus}.
\Dendrogr{} code, in which BBH are evolved using an oct-tree grid as we do in
\GRAthena{} but with different strategies, demonstrates very good scaling
properties of $q = 10$ BBH evolutions utilizing up to $\sim 1.3\times10^5$ CPUs.
In these tests, however, re-mesh and inter-grid transfer operations are disabled.
In this work we aim to reach such performance for BBH evolutions with full AMR.

The rest of this paper is organized as follows: In \S\ref{sec:method} we provide further
details on the computational approach taken within \GRoAthena{} together with the various
extensions we have made to core functionality.
Subsequently in \S\ref{sec:z4csystem} an
overview of the \z4c{} system we use in our calculations is provided together with description
of the numerical algorithms employed. Refinement strategy and details concerning grids are
provided in \S\ref{sec:amr_criterion}. In \S\ref{sec:puncture_tests} we discuss results of
extensive testing of \GRAthena{} on BH and BBH problems performing cross-code validation and
assessing convergence properties whereupon in \S\ref{sec:scaling} computational performance
is detailed through strong and weak scaling tests.
Finally \S\ref{sec:final_word} summarizes and concludes.

\section{Method}\label{sec:method}
\GRAthena builds upon \Athena thus in order to specify nomenclature, provide
a self-contained description, and explain our extensions, we first briefly recount some details
of the framework (see also \cite{white2016extensionathenacode,%
felker2018fourthorderaccuratefinite,%
stone2020athenamathplusmathplus}).

In \GRoAthena{} overall details about the domain $\Omega$ over which a problem is
formulated are abstracted from the salient physics and contained
within a class called the \Mesh{}.
Within the \Mesh{} an overall representation of
the domain as a logical $n$-rectangle is stored, together with details of
coordinatization type (Cartesian or more generally curvilinear), number of points along each
dimension for the coarsest sampling $N_M=(N_{M_1},\,\cdots,\,N_{M_d})$, and physical boundary
conditions on
$\partial \Omega$.
In order to partition the domain
we first fix a choice $N_B=(N_{B_1},\,\cdots,\,N_{B_d})$ where each element of $N_B$ must divide
each element of $N_M$ component-wise. Then $\Omega$ is domain-decomposed through
rectilinear sub-division into a family of $n$-rectangles satisfying
$\Omega = \sqcup_{i\in Z} \Omega_i$, where $Z$ is the set of \MeshBlock{} indices, corresponding to the ordering described in \S \ref{sec:tree_structure}.
Nearest-neighbor elements are constrained to only differ by a single sub-division at most.
The \MeshBlock{} class stores properties of an element $\Omega_i$ of the
sub-division. In particular the number of points in the sampling of
$\Omega_i$ is controlled through the choice of $N_B$. For purposes of communication of data between
nearest neighbor \MeshBlock{} objects the sampling
over $\Omega_i$ is extended by a thin layer of so-called ``ghost nodes'' in each direction.
Furthermore the local values (with respect to the chosen, extended sampling
on $\Omega_i$) of any discretized, dependent field variables of interest are stored
within the \MeshBlock{}.

In both uniform grid $(\forall i\in Z)$ $\mathrm{vol}(\Omega_i) = C$ and refined meshes
$(\exists i,\,j\in Z)$ $\mathrm{vol}(\Omega_i) \neq \mathrm{vol}(\Omega_j)$ it is
crucial to arrange
inter-\MeshBlock{} communication efficiently -- to this end the relationships
between differing \MeshBlock{} objects are arranged in a tree data structure,
to which we now turn.

\subsection{Tree Structure of \Mesh{}}\label{sec:tree_structure}
For the sake of exposition
here and convenience in later sections we now particularize to a
Cartesian coordinatization though we emphasize that the general
picture (and our implementation) of the discussions here and in \S\ref{sec:vertex_centered}
carry over to the curvilinear context with only minor modification.

\GRoAthena{} stores the logical relationship between the \MeshBlock{} objects (i.e. $\Omega_i$)
involved in description of a domain $\Omega$ within a tree data structure. A binary-tree,
quad-tree
or oct-tree is utilized when $d:=\dim(\Omega)=1,\,2,\,3$ respectively. The relevant
tree is then
constructed by first selecting the minimum $N$ such that $2^N$ exceeds the largest number of
$\Omega_i$ along any dimension. The root of the tree is assigned a logical level of zero and the
tree terminates at level $N$ with every \MeshBlock{} assigned to an appropriate leaf, though
some leaves and nodes of the tree may remain empty. Data locality is enhanced, as references to
\MeshBlock{} objects are stored such that a post-order,
depth-first traversal of the tree follows Morton order (also termed Z-order)
\cite{morton1966computer}.
This order can be used to encode multi-dimensional coordinates into a linear index
parametrizing a Z-shaped, space-filling curve where small changes in the parameter imply spatial
coordinates that are close in a suitable sense \cite{burstedde2019numberfaceconnectedcomponents}.

\begin{figure}[!ht]
	\centering
		\includegraphics[width=\columnwidth]{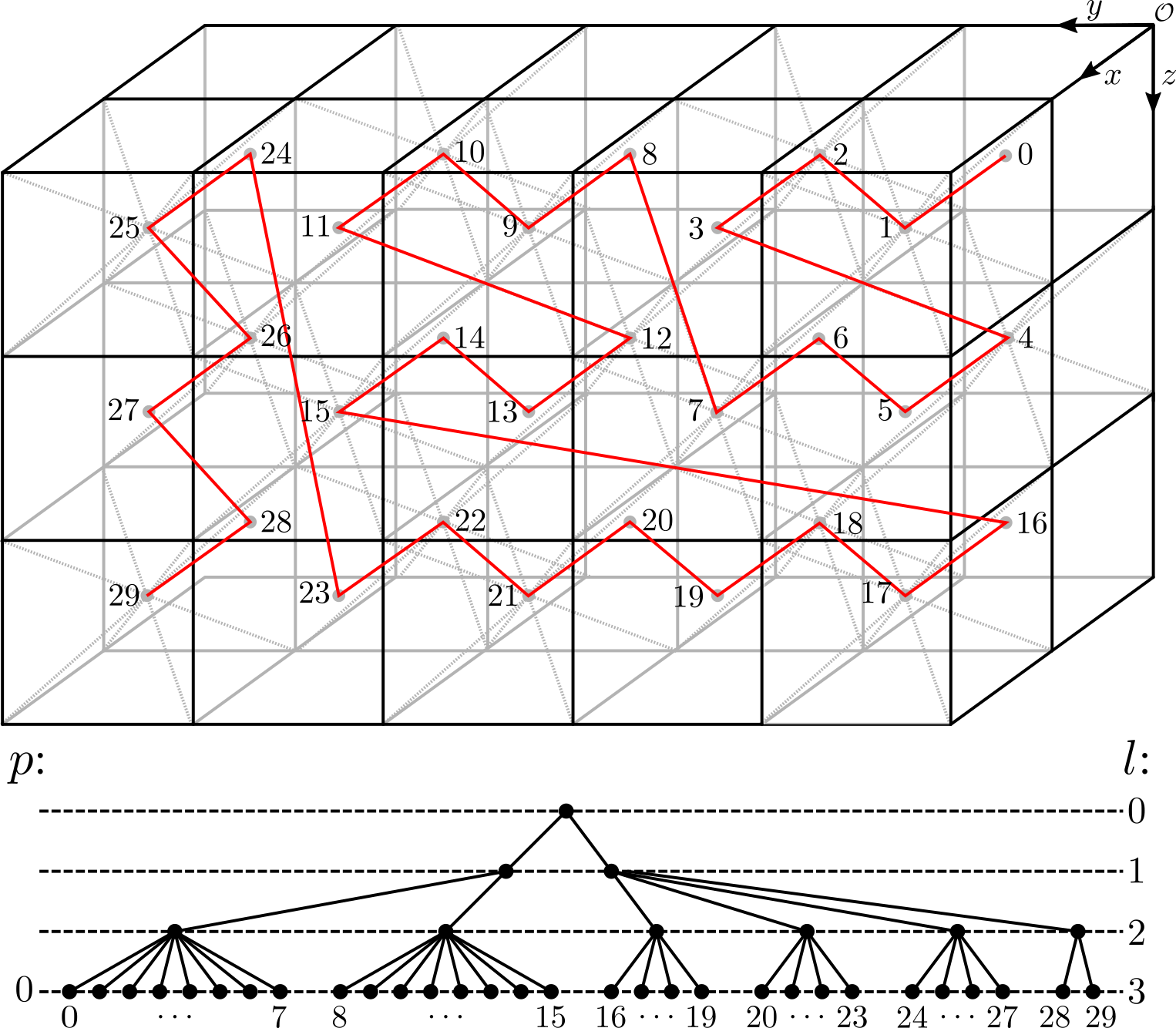}
    \caption{%
    Example of \Mesh{} partitioned uniformly by \MeshBlock{} objects indexed via Z-order and
    traced in red through constituent geometric centroids.
    The logical relationship
    between $\Omega_i$ is stored in an oct-tree. Empty leaves are suppressed though each populated
    node above logical level three has eight children.
    Notice that physical level $p$ and logical level $l$ are distinct.
    See text for further discussion.}
		\label{fig:oct-tree_partitioned}
\end{figure}

As an example we consider a three-dimensional \Mesh{} described by
$(N_x,\,N_y,\,N_z)=(2,\,5,\,3)$ \MeshBlock{}
objects in each direction at fixed physical level in Fig.\ref{fig:oct-tree_partitioned}.

Consider now a \Mesh{} with refinement. Function data at a fixed physical level is transferred
one level finer through use of a prolongation operator $\mathcal{P}$; dually, function data may
be coarsened by one physical level through restriction $\mathcal{R}$. The number of
physical refinement levels added to a uniform level, domain-decomposed $\Omega$ is
controlled by the parameter $N_L$.
By convention $N_L$ starts at zero. Subject to satisfaction of
problem-dependent refinement criteria, there may exist physical levels at $0,\,\cdots,\,N_L$.
When a given \MeshBlock{} is refined (coarsened) $2^d$ \MeshBlock{} objects are constructed
(destroyed). This is constrained to satisfy a $2:1$ refinement ratio where
nearest-neighbor \MeshBlock{} objects can differ by at most one physical level.

\begin{figure}[!ht]
	\centering
		\includegraphics[width=\columnwidth]{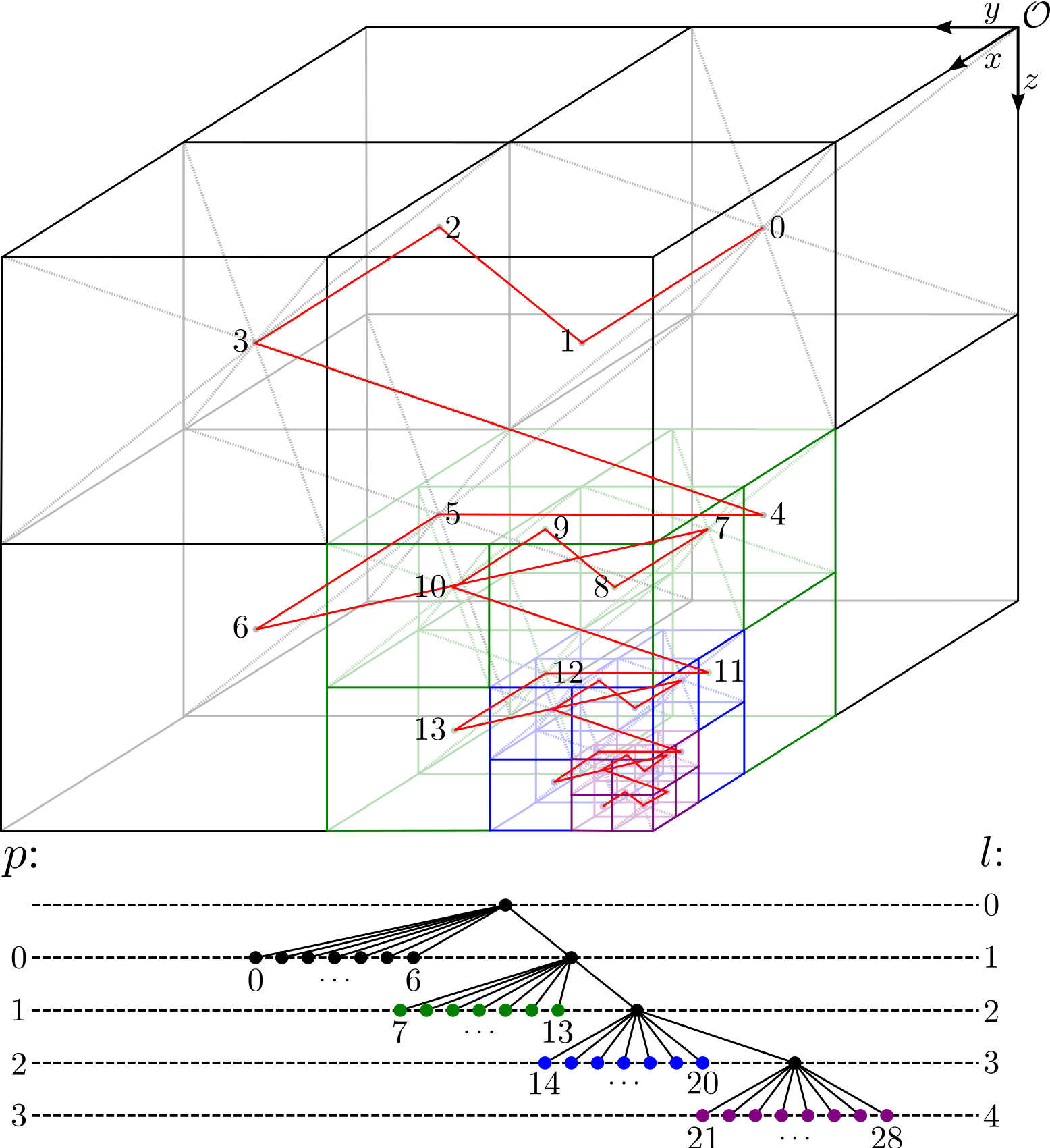}
    \caption{%
    Example of \Mesh{} partitioned and refined by \MeshBlock{} objects indexed (labels explicitly
    indicated up to logical level two) via Z-order and
    traced in red through constituent geometric centroids. The logical relationship
    between ${}^p\Omega_i$ and neighbors
    is stored in an oct-tree. There are no unpopulated leaves.
    Notice that physical level $p$ and logical level $l$ are distinct; coloring corresponds to
    physical level: $p=0$ in black, $p=1$ in dark green, $p=2$ in blue, and $p=3$ in purple.
    See text for further discussion.}
		\label{fig:oct-tree_refined}
\end{figure}

In Fig.\ref{fig:oct-tree_refined} we consider an example of a non-periodic $\Omega$ described by
$N_x=N_y=N_z=2$ \MeshBlock{} objects with $N_L=3$ selected with refinement introduced
at the corner $x_{\max}$, $z_{\max}$.
If periodicity conditions are imposed on $\partial \Omega$ then additional refinement may be
required for boundary intersecting \MeshBlock{} objects so as to maintain the aforementioned
inter-\MeshBlock{} $2:1$ refinement ratio.

\subsection{Vertex-centered Discretization}\label{sec:vertex_centered}
Natively \Athena supports cell-centered (CC) and face-centered (FC) description of variables,
together with calculation of line-averages on cell edges \cite{stone2020athenamathplusmathplus}.
\GRAthena extends support to allow for vertex-centering (VC). The modifications required to achieve
this are extensive as core code must be changed in such a way so as to complement existing
functionality. The modularity and good code practices of \Athena{} greatly facilitated matters.
Our motivation for introduction of VC is a desire to ensure each stage of our numerical
scheme maintains consistent (high) order while simultaneously maintaining efficiency of $\mathcal{R}$
and $\mathcal{P}$ operator choice and implementation. In the remainder of this section we briefly
describe this newly introduced functionality.

\subsubsection{VC and Communication: Fixed Physical Level}\label{sssec:vc_com_fixed}
Unless otherwise stated in all remaining sections we fix $N_M$ and $N_B$ to be uniform in each
dimension and represent each of these tuples with a single scalar. As a preliminary, $x\in[a,b]$
is said to be vertex-centered when discretized as $x_I=a+I\sp$
where $\sp=(b-a)/N_B$ and $I=0,\,\dots,\,N_B$ yielding $N_B+1$ total samples. In practice,
to this an additional $\Ng$ so-called ghost nodes are appended which extend
the interval by $\Ng\sp$ on both sides. When $d=\dim(\Omega_j)=2,3$ an
appropriate tensor product of such extended one-dimensional discretizations is utilized. When
a field component $\mathcal{V}$
is sampled on such grids it
is said to be VC. The additional ghost nodes form a layer that enables imposition of physical
boundary conditions and inter-\MeshBlock{} communication.

Consider a domain decomposed into multiple \MeshBlock{} objects. Discretized variable data must
be communicated. An additional intricacy
however arises due to the sharing of vertices at neighboring \MeshBlock{} interfaces that are not
part of the ghost-layer. The number of \MeshBlock{} objects a node is shared between is referred to
as the node-multiplicity.

\begin{figure}[t]
	\centering
		\includegraphics[width=0.6\columnwidth]{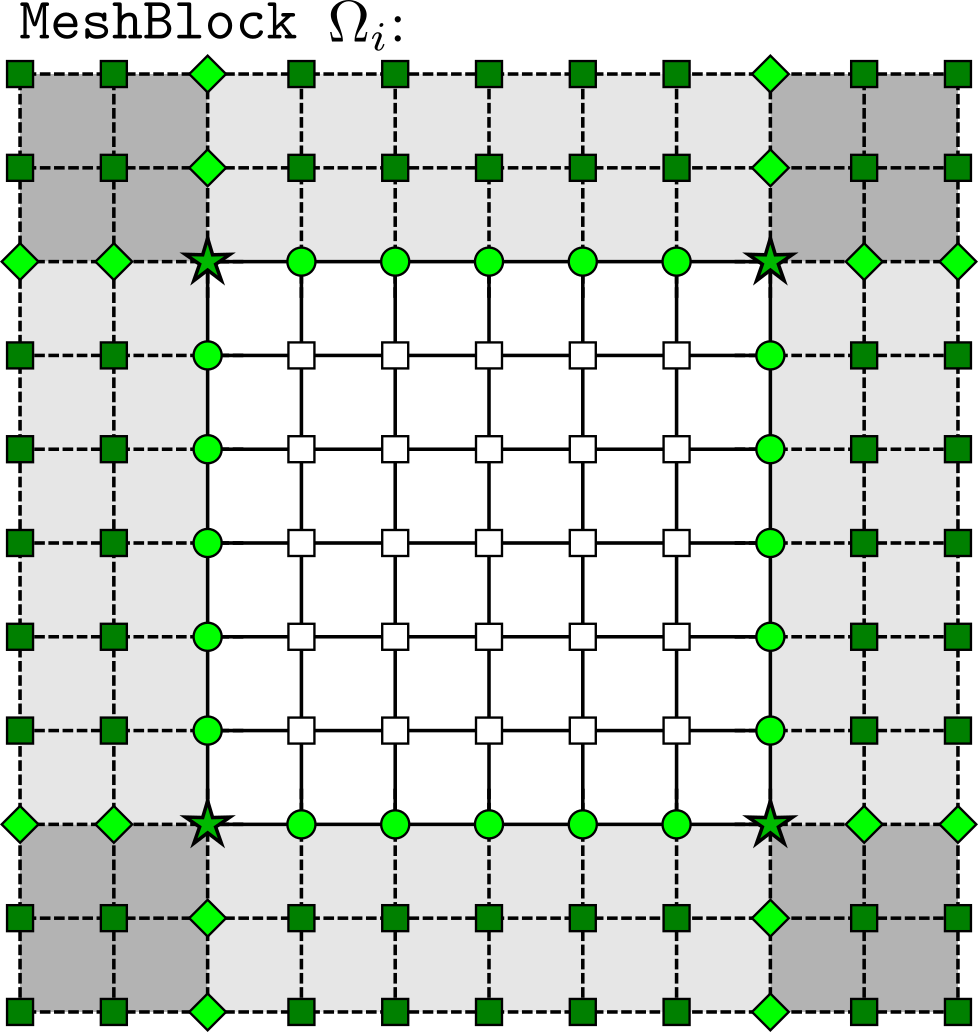}
    \caption{%
    Schematic of (communicated) nodes on a two-dimensional \MeshBlock{} $\Omega_i$.
    The ghost-layer is shaded in gray with alternating shading demarcating differing neighbor
    \MeshBlock{} objects. Nodes marked with ``{\tiny$\square$}'' are interior to $\Omega_i$
    and are unaffected as neighbor data is received -- all other nodes are updated.
    Ghost-layer multiplicities are indicated for ``{\tiny$\blacksquare$}'' and dark-green where $\mu=1$
    whereas nodes in
    ``{\small{$\blacklozenge$}}''
    and light-green have $\mu=2$.
    Interface nodes along edges are marked with ``{\large$\bullet$}'' in light-green and correspond
    to $\mu=2$ whereas corner nodes marked ``{\large$\star$}'' in green correspond to $\mu=4$.
    See text for further discussion.
    }
		\label{fig:communication_schematic_fixed}
\end{figure}

We illustrate this with a two-dimensional example.
Let $\mathcal{V}$ be sampled
on neighboring \MeshBlock{} objects of fixed physical level, where $N_B=6$ is chosen and ghost-zone
layer selected to have $\Ng=2$ nodes. Further, we assume that $\Omega_i$
is not on the physical boundary of the domain.
This entails that $(N_B-1)^2$ nodes are internal
and the remainder require synchronization via data received (i.e. populated) from neighboring
blocks as depicted in
Fig.(\ref{fig:communication_schematic_fixed}).
Note that independent communication requests and buffers are posted for each neighbor.
Communication from neighbors therefore has no preferred order and consequently we follow an
averaging approach to achieve consistency as follows:
All received data is first additively accumulated on the \MeshBlock{}
with node-multiplicity $\mu$ dynamically updated in an auxiliary array of $7^d$ elements based
on the location of the relevant neighbor\footnote{For a given node $\mu$ is uniform in all
VC variables that are to be communicated and hence need only be constructed once in the absence
of \Mesh{} refinement. The choice of $7^d$ elements is made to simultaneously treat communication
over distinct physical levels (see \S\ref{sssec:com_across_lev}).}. After data
from all relevant neighbors has been received, a final division by $\mu$ is performed. This is done
so as to not preferentially weight data from any particular neighbor. In principle, it is possible
to construct $\mu$ a priori, however, we elect to follow this dynamical strategy to facilitate
automatic treatment of boundary conditions and avoid the additional complexity involved during
\Mesh{} refinement.

\subsubsection{Communication: Distinct Physical Levels}\label{sssec:com_across_lev}
We now consider a \Mesh{} featuring refinement. The fundamental description of variables between
neighboring \MeshBlock{} objects may therefore differ by (at most) a single physical level
(see also \ref{sec:tree_structure}).

A \MeshBlock{} at physical level $p$ will be denoted by ${}^p\Omega_j$ and
the corresponding collection of fields
sampled using VC discretization over the \MeshBlock{}
as $\mathcal{F}({}^p\Omega_j)$. In this context
$\mathcal{V}\in\mathcal{F}({}^p\Omega_j)$, has a complementary, coarse analogue
$\mathcal{V}_c$ of $(\lfloor N_B/2 \rfloor+1)^d$ samples further extended by a
coarse ghost-layer comprised of $\Ncg$ nodes. The sampling resolution for
$\mathcal{V}_c$ is thus half that used for $\mathcal{V}$. In order to emphasize the physical
level of a given \MeshBlock{} and not blur the distinction between the types of samplings
we also make use of the
notation $\mathcal{F}_c({}^p\Omega_j)\equiv \mathcal{F}({}^{p-1}\Omega_j)$.

In contrast to CC and FC as implemented in \Athena{} our implementation of VC allows for
$\Ng$ and $\Ncg$ to take odd values and
be independently varied.
For simplicity of discussion we impose $\Ng=\Ncg$.

When a \Mesh{} involves multiple physical levels, prior to any communication
of data, VC variables are initially restricted so as to have a fundamental and coarse description
on each \MeshBlock{} excluding the ghost-layers. For logically Cartesian grids in particular,
this turns out to be an inexpensive
and exact operation (\S\ref{sssec:restr_prol}). With this initial step
neighboring \MeshBlock{} objects at the same physical level have $\mathcal{V}$ and
$\mathcal{V}_c$ communicated using the method
described in \S\ref{sssec:vc_com_fixed}.

\begin{figure}[t]
	\centering
		\includegraphics[width=\columnwidth]{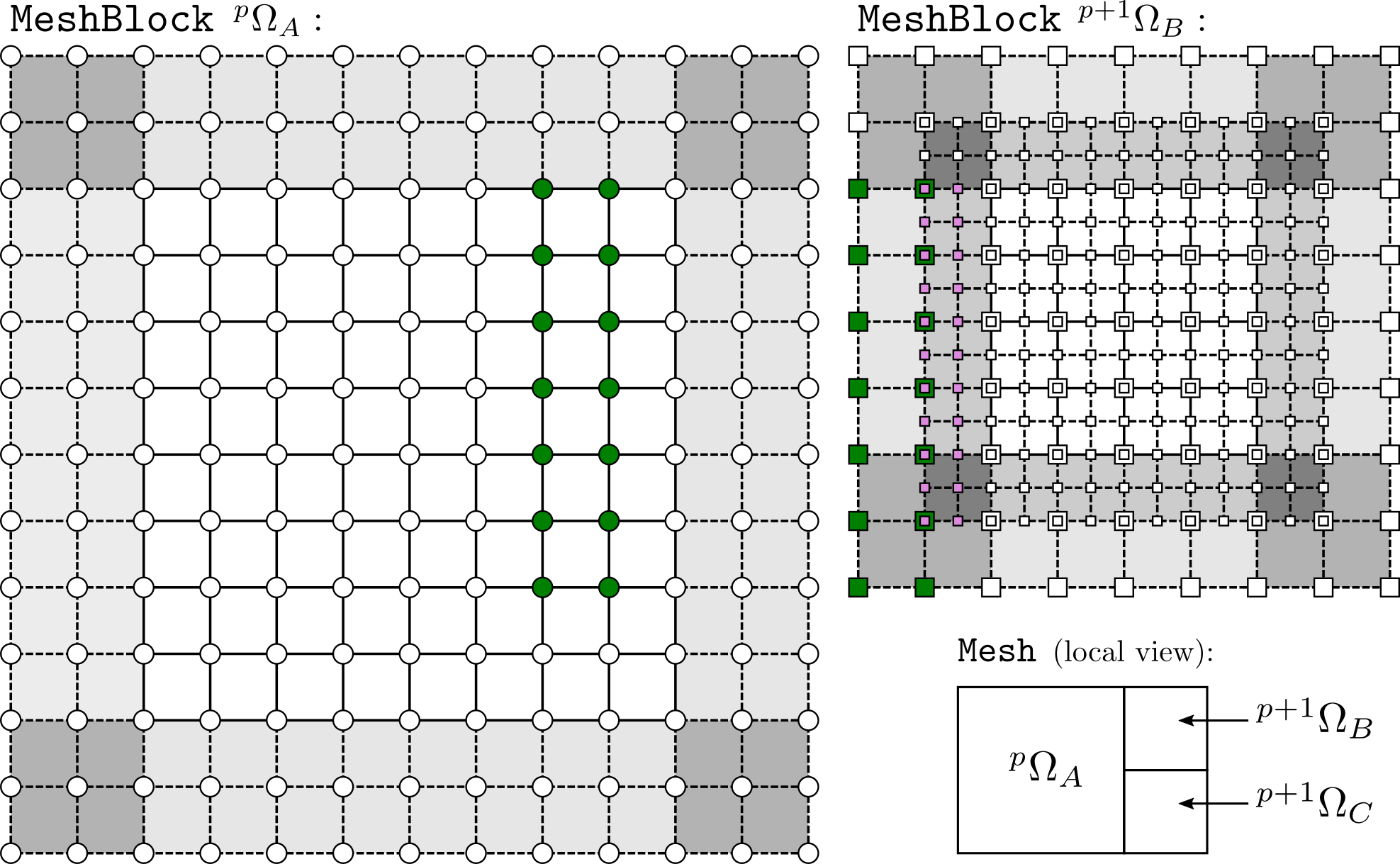}
    \caption{%
    Schematic of two-dimensional \MeshBlock{} ${}^p\Omega{}_A$ used to populate ghost-nodes of
    finer \MeshBlock{} ${}^{p+1}\Omega{}_B$. Local view of the \Mesh{} depicts nearest-neighbor
    \MeshBlock{} connectivity and physical levels. Nodes over ${}^p\Omega_A$,
    and ${}^{p+1}\Omega_B$ together with coarse analogues are shown.
    Sampled values $\mathcal{V}\in\mathcal{F}({}^{p}\Omega_A)$ that are
    to be sent are marked by ``{\Large$\bullet$}'' in dark green; this data is received and
    directly
    populates the ghost-nodes marked by ``{\small$\blacksquare$}'' in dark green, i.e.,
    $\mathcal{V}\in\mathcal{F}_c({}^{p+1}\Omega_B)$. Once the remaining data for
    $\mathcal{F}_c({}^{p+1}\Omega_B)$ -- marked by ``{\small$\square$}'' -- is filled, and
    any multiplicity conditions (here suppressed) are accounted for, prolongation
    $\mathcal{P}:\mathcal{F}_c({}^{p+1}\Omega_B)\mapsto\mathcal{F}({}^{p+1}\Omega_B)$
    can be performed in order to
    populate values at the ghost-nodes of ${}^{p+1}\Omega_B$ marked by
    ``{\tiny$\blacksquare$}'' in purple.
    Notice that for this procedure data at nodes on the neighbor interface remain unchanged.
    See text for further discussion.
    }
		\label{fig:communication_schematic_refined}
\end{figure}

To describe our treatment when neighboring physical levels differ, consider a two-dimensional
\Mesh{} where $N_B=8$ and $\Ng=2$. Once more, we work within a local
portion of the full \Mesh{} where the role of the physical boundary may be ignored. Suppose
${}^p\Omega_A$ is neighbored by ${}^{p+1}\Omega_B$ and ${}^{p+1}\Omega_C$ to the east and the
latter two \MeshBlock{} objects share a common edge.
Figure~\ref{fig:communication_schematic_refined} shows
how the ghost-layer nodes of the finer
${}^{p+1}\Omega_B$ based on the coarser neighbor ${}^p\Omega_A$ are populated.
In this situation data may be freely posted
for communication to the \MeshBlock{} on the finer level whereupon ghost-zones of its coarse
variable are populated. However depending on the details of
$\mathcal{P}$, the prolongation operation over the ghost-layer is blocked in the sense
that the entirety of the coarse ghost-layer of $\mathcal{F}_c({}^{p+1}\Omega_B)$ must
first be populated. Once fully populated, prolongation is carried out on the
target \MeshBlock{}.
During synchronization of data from coarse to fine levels interface nodes are
maintained at the value of the finer level.

\begin{figure}[t]
	\centering
		\includegraphics[width=\columnwidth]{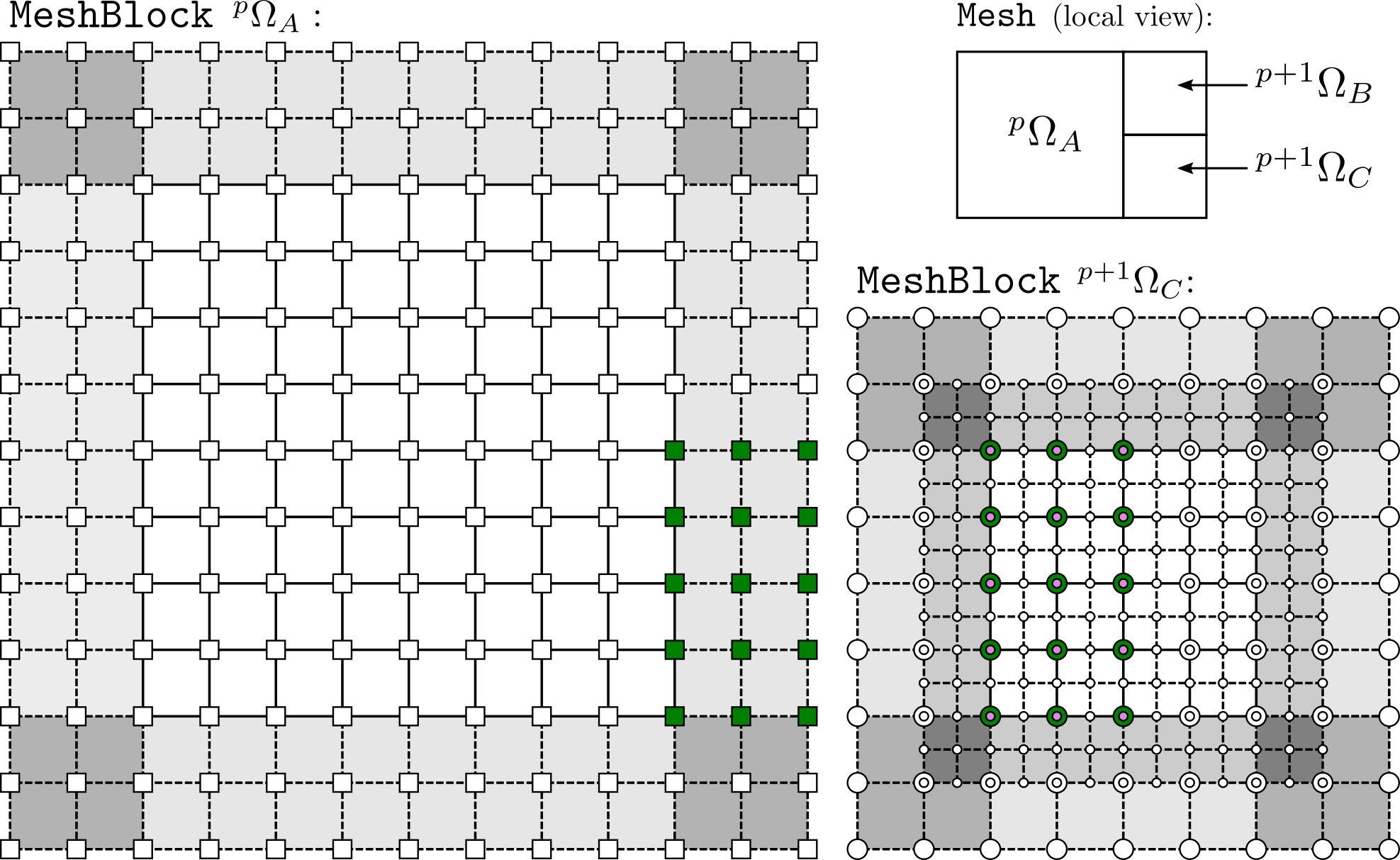}
    \caption{%
    Schematic of two-dimensional \MeshBlock{} ${}^{p+1}\Omega{}_C$ used to populate ghost-layer of
    coarser \MeshBlock{} ${}^{p}\Omega{}_A$. Locally the \Mesh{} has the same structure as in
    Fig.\ref{fig:communication_schematic_refined}. Nodes over ${}^p\Omega_A$,
    and ${}^{p+1}\Omega_C$ together with coarse analogues are shown.
    Prior to communication, sampled data of $\mathcal{V}\in\mathcal{F}({}^{p+1}\Omega_C)$
    at nodes marked by ``{\small$\bullet$}'' in purple must be
    restricted to populate data $\mathcal{V}_c\in\mathcal{F}_c({}^{p+1}\Omega_C)$ at nodes
    marked by ``{\Large$\bullet$}'' in dark green. This is then sent whereupon
    data at the nodes of ${}^p\Omega_A$ marked by ``{\small$\blacksquare$}'' in
    dark green is provided.
    During this procedure data at nodes on the neighbor interface are (additively) updated
    (cf. Fig.\ref{fig:communication_schematic_refined}) and multiplicity conditions
    (here suppressed) dynamically updated in an auxiliary array. See text for further discussion.}
		\label{fig:communication_schematic_coarsen}
\end{figure}

An example of the dual process of populating nodes on a coarser level involving
${}^p\Omega_A$ and ${}^{p+1}\Omega_C$ is depicted in
Fig.\ref{fig:communication_schematic_coarsen}. In this case (previously) restricted data of the
finer \MeshBlock{} interior is communicated, updating the common interface and ghost-layer of
$\mathcal{F}({}^p\Omega_A)$. In this situation no blocking occurs.
However, non-trivial multiplicity conditions arise on the common neighbor interface. Furthermore,
the equivalent operation involving ${}^{p+1}\Omega_B$ instead of ${}^{p+1}\Omega_C$ induces another
edge within the ghost-layer of ${}^{p}\Omega_A$.
Finally, we note that values of $\mathcal{F}_c({}^p\Omega_A)$ must
also be updated.
Thus another restriction of samples of $\mathcal{F}_c({}^{p+1}\Omega_C)$ is also made
and the overall communication process repeated.

The steps for the above communication procedure are summarized in \S\ref{sssec:amr_summary}.

\subsubsection{Restriction and Prolongation}\label{sssec:restr_prol}
When a \Mesh{} is refined restriction $\mathcal{R}$ and prolongation
$\mathcal{P}$ operations are required. In \GRAthena{} these operations for VC variables are
implemented based on univariate Lagrange polynomial interpolation
or products thereof when $\dim(\Omega)>1$ with function data utilized at nodes centered about a
target-point of interest.

For a \Mesh{} sampled according to a Cartesian coordinatization \MeshBlock{} grids are uniformly
spaced in each dimension. This provides immediate simplifications to $\mathcal{R}$ and
$\mathcal{P}$ which may be understood as follows. Consider interpolation of a smooth function
$\mathcal{V}$ on a
one-dimensional interval. A polynomial interpolant $\tilde{\mathcal{V}}$ of degree $2N$ with
samples of the function $\mathcal{V}$ symmetrically and uniformly spaced about $x^*$ that passes
through the $2N+1$ distinct points:
\begin{equation*}
  I_\mathcal{V}:=\left\{
    \left. (x_i,\, \mathcal{V}(x_i)) \right| x_i=x^* + i \sp \wedge i\in\{-N,\,\dots,\,N\}
  \right\},
\end{equation*}
is unique and may be written in Lagrange form \cite{trefethen2013approximation}:
\begin{align}\label{eq:laginterp}
  \tilde{\mathcal{V}}(x) = \sum_{i=-N}^N l_i(x) \mathcal{V}(x_i),
\end{align}
where the Lagrange cardinal polynomials satisfy $l_i(x_j) = \delta_{ij}$ when $x_j$ is a node
used during formation of $I_\mathcal{V}$ and $\delta_{ij}$ is the Kronecker delta.
We use Eq.\eqref{eq:laginterp} (or appropriate product generalizations) in order to specify
$\mathcal{R}$. Given function data on a uniform, VC discretized interval suppose we wish
to construct data on the interior of a coarser overlapping interval that shares the same
end-points and is sampled at twice the spacing.
We find that points in the image of $\mathcal{R}$ (i.e. desired points over the coarse grid)
form a subset of points over the original fine grid. Therefore the desired data may simply be
immediately copied (see Fig.\ref{fig:communication_schematic_coarsen}). This is efficient
and involves no approximation.

Recall that the restriction operator is utilized during transfer of data
from a \MeshBlock{} to
a coarser neighbor. Consider the case of the two-fold coarsened data that must be
provided to the neighbor \MeshBlock{}.
While $\mathcal{R}$ as specified here is an idempotent operation, some care must be taken,
because the spatial extent of the ghost-nodes to be populated is sampled by the non-ghost data
of the source \MeshBlock{}.
To ensure this is possible we impose a constraint relating
\MeshBlock{} sampling and ghost-layer through:
\begin{equation}\label{eq:N_B_restriction}
  N_B \geq \max(4,\,4\Ng-2).
\end{equation}

The above does not place any constraint on whether $\Ng$ is even and therefore a choice of an
odd or even number is allowed.

Interpolation based on Eq.\eqref{eq:laginterp} is also utilized for prolongation. Here function
data is transferred to a finer, uniformly sampled grid of half the spacing and consequently
interspersed nodes coincide (see Fig.\ref{fig:communication_schematic_refined}) offering another
optimization in execution efficiency. Due to the uniform structure of the source and target grids,
interpolation at non-coincident nodes may be implemented through a weighted sum where weight
factors can be precomputed \cite{berrut2004barycentriclagrangeinterpolation}.
In practice the width of the interpolation stencil we utilize is $N=\lfloor \Ncg / 2 \rfloor + 1$.

Tailored, optimized routines for \GRAthena{} incorporate the above simplifications for the case of
logically Cartesian grids.

Finally, for later convenience we note that the $d$-rectangle $[x_L,\,x_R]^d$ as
represented by a \Mesh{} with Cartesian coordinatization, $N_M$ points along each dimension, and
$N_L$ physical levels of refinement has a grid spacing on the finest level of:
\begin{equation}\label{eq:refinedCartRes}
  \sp = \frac{x_R - x_L}{N_M}\frac{1}{2^{N_L - 1}}.
\end{equation}

\subsubsection{Summary}\label{sssec:amr_summary}
We close discussion of VC by providing a compact summary of the overall logic involved during
synchronization of data between \MeshBlock{} objects for a problem involving refinement.

At compile time $\Ng$ and $\Ncg$ are selected and \Cpp templates specify precomputed weights for
any requisite interpolation during a computation (thus fixing $\mathcal{R}$ and $\mathcal{P}$).
A given problem of interest may then be executed for some choice of $N_M$, $N_B$ (subject to
Eq.\eqref{eq:N_B_restriction}), $N_L$, and physical grid.

The following steps are taken when function data from a \MeshBlock{} ${}^p\Omega_i$ is
to be \emph{sent}:
\begin{enumerate}[i.]
  \item{Non-ghost data is restricted populating $\mathcal{F}_c({}^p\Omega_i)$.}
  \item{Neighbor \MeshBlock{} objects are iterated over and treated according to the physical level
  of the target neighbors and the communication buffers are populated from:}
  \begin{enumerate}
    \item[$p-1:$]{Relevant interior (and shared interface) nodes of $\mathcal{F}_c({}^p\Omega_i)$;
    similarly $\mathcal{F}({}^p\Omega_i)$ is twice restricted directly to the communication buffer.}
    \item[$p:$]{Relevant interior (and shared interface) nodes of $\mathcal{F}({}^p\Omega_i)$
    together with $\mathcal{F}_c({}^p\Omega_i)$.}
    \item[$p+1:$]{Relevant interior nodes not on the common interface from
    $\mathcal{F}({}^p\Omega_i)$.}
  \end{enumerate}
\end{enumerate}
The following steps are taken when function data on ${}^p\Omega_i$ is to be \emph{received}:
\begin{enumerate}[i.]
  \item{The ghost-layer of variable data for the given \MeshBlock{} ${}^p\Omega_i$ is set to zero
  and any previously accumulated multiplicites are reset.}
  \item{Non-ghost data is restricted populating $\mathcal{F}_c({}^p\Omega_i)$.}
  \item{Function data is independently received (unordered) from neighbor \MeshBlock{} objects.
  Treatment again splits based on physical level of the salient neighbor with additive updating
  of the following \MeshBlock{}-local function data:
  \begin{enumerate}
    \item[$p-1:$]{Relevant ghost-layer nodes of $\mathcal{F}_c({}^p\Omega_i)$.}
    \item[$p:$]{Relevant ghost-layer and interface nodes of $\mathcal{F}_c({}^p\Omega_i)$
    and $\mathcal{F}({}^p\Omega_i)$.}
    \item[$p+1:$]{Relevant ghost-layer and interface nodes of $\mathcal{F}({}^p\Omega_i)$ and
    $\mathcal{F}_c({}^p\Omega_i)$.}
  \end{enumerate}
  \item{Once all neighbor function data is received, division by multiplicity conditions is
  carried out.}
  \item{Regions of the ghost-layer involving a coarser level neighbor may finally be prolongated.}
  }
\end{enumerate}
For local calculations (in the absence of distributed, MPI communications) operations are
performed locally in memory.
Finally we emphasize that the base \Athena{} CC and FC variables when required continue to
simultaneously function as explained in \cite{stone2020athenamathplusmathplus}.

\subsection{Geodesic spheres}\label{ssec:geodesic_spheres}
\begin{figure}[t]
  \centering
  \includegraphics[width=\columnwidth]{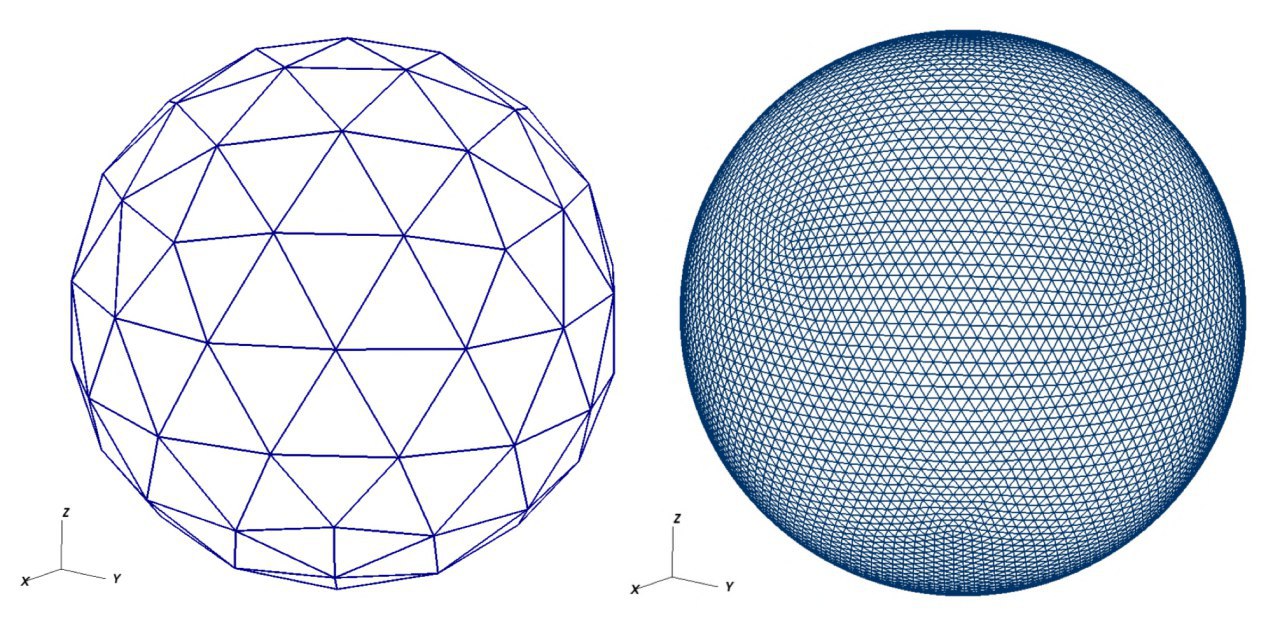}
  \caption{Structure of the geodesic grid used by \GRAthena{}. Left
  panel: an example of a low resolution (92 vertices) geodesic grid highlighting the
  features of the grid. Right panel: a grid used for gravitational wave extraction in
  production simulations (9002 vertices).}
  \label{fig:geogrid}
\end{figure}

Calculating quantities such as the ADM mass, momentum and gravitational
radiation associated with an isolated system typically involves integration over spherical
surfaces, the radii of which are controlled by a limiting procedure.
In practice, a large but finite radius is often selected during numerical work. Denote
the $2$-sphere of fixed radius $R$ by $\mathbb{S}{}^2_R$. The natural choice of spherical
coordinatization for $\mathbb{S}{}^2_R$ involves uniform sampling in the polar and azimuthal
angles $(\vartheta,\,\varphi)$ and it is well-known that problems may arise at the poles
during description of geometric quantities; furthermore, points tend
to cluster there which may be undesirable from the stand-point of efficiency in some
applications.
In \GRAthena{} we avoid these issues by instead working with triangulated
geodesic spheres.
In short, a geodesic sphere of radius
$R$ (denoted $Q_R$) may be viewed
as the boundary of a convex polyhedron embedded in $\mathbb{R}^3$ with triangular faces, i.e.,
a simplicial 2-sphere that is homeomorphic to $\mathbb{S}{}^2_R$.
A sequence of geodesic spheres with an increasing number of vertices (and consequently surface
tiling triangles) thus serves as a sequence of
increasingly accurate approximants to $\mathbb{S}{}^2_R$; see Fig.\ref{fig:geogrid}

To construct the geodesic grid we start from a regular icosahedron with 12
vertices and 20 plane equilateral triangular faces, embedded in a unit sphere.
We refine it using the so called ``non-recursive'' approach described in
\cite{Wang:2011}.  In this approach, each plane equilateral triangle of the
icosahedron is divided into $n_Q^2$ small equilateral triangles (each side of the
triangle is split into $n_Q$ equal segments, where $n_Q$ is called the grid level).
The intersection points are projected onto the unit sphere, and together with
the original 12 vertices of the icosahedron they form the convex polyhedron used
as a grid. The resulting polyhedron has $10n_Q^2+2$ vertices, in which we define
the desired physical quantities. The left panel of Fig.\ref{fig:geogrid} shows
the grid consisting of 92 vertices ($n_Q=3$), while the right panel shows the grid
consisting of 9002 vertices ($n_Q=30$).

Integrals on the sphere are computed with numerical quadratures. To this aim we
associate to each grid point a solid angle in the following way. We construct
cells around each vertex of the grid by connecting the circumcenters of any
pair of triangular faces that share a common edge. The resulting cells are
mostly represented by hexagons, apart from the 12 vertices of the original
icosahedron, which have only five neighbors and therefore correspond to
pentagonal cells.  The solid angles subtended by the cells at the center of the
sphere are used as weighting coefficients when computing the averages. The
logical connection between the neighboring cells is implemented as described in
\cite{Randall:2002}.

Using a geodesic grid ensures more even tiling of the sphere compared to the
uniform latitude-longitude grid of similar resolution. The ratio between the
solid angles corresponding to the largest and smallest cells in the
$n_Q=30$ grid is equal to $2$. For comparison, a grid of comparable resolution with uniform
sampling in the polar and azimuthal angles (say, 67 $\vartheta$ angles and 134
$\varphi$ angles, with the total of 8978 cells), would have the ratio between
the areas of the smallest and largest cells $\simeq
1/\sin(\pi/67)\simeq 21.3$.

\section{$\mathrm{Z}4\mathrm{c}$ system in \GRAthena{}}\label{sec:z4csystem}
In the Cauchy problem for the Einstein field equations (EFE), a globally hyperbolic space-time
$\mathcal{M}$ is foliated by a family of non-intersecting spatial
slices $\{\Sigma_t\}_{t\in\mathbb{R}}$ where the parametrizing time-function $t$ is assumed
globally defined. An initial slice $\Sigma_{t_0}$ is selected and well-posed evolution equations
based on the EFE must be prescribed. A variety of mature approaches exist to this problem such as
BSSNOK \cite{Nakamura:1987zz,Shibata:1995we,Baumgarte:1998te}
or those based on the generalized harmonic gauge (GHG) formulation
\cite{Friedrich:1985,Pretorius:2005gq,Lindblom:2005qh}.
A unifying framework is provided in the \z4{} approach \cite{Bona:2003fj}
where particular cases of both GHG and BSSNOK formulations may be recovered
(see \cite{bona2010actionprinciplenumericalrelativity} and references therein).
In particular \z4c{} \cite{Bernuzzi:2009ex,Ruiz:2010qj,Weyhausen:2011cg,Hilditch:2012fp}
seeks to combine the strengths of these other two approaches \cite{Cao:2011fu}
thus motivating it as the choice of formulation for \GRAthena{}.

In \S\ref{ssec:eqnformgrathena} and \S\ref{ssec:gauge_bc} we describe the overall idea behind
numerical evolution with \z4c{} and implementation within \GRAthena{}.
Details on our method for wave
extraction (i.e., calculation of gravitational radiation) is provided
in \S\ref{ssec:wave_extraction} whereupon
\S\ref{ssec:numerical_technique} closes with a brief description of numerical methods we
utilize.

\subsection{Overview}\label{ssec:eqnformgrathena}
At its core, the \z4{} formulation \cite{Bona:2003fj} seeks to stabilize the
time-evolution problem through direct augmentation of the EFE via suitable introduction of an
auxiliary, dynamical vector field $Z^a$
and first-order covariant
derivatives thereof.
The approach admits natural incorporation of constraint damping via explicit appearance of
(freely chosen) parameters $\kappa_i$ \cite{%
Gundlach:2005eh,%
Weyhausen:2011cg}.

Recall that in the standard method of ADM-decomposition \cite{Arnowitt:1959ah,Baumgarte:2010}
one introduces a future-directed $t{}^a$ satisfying $t{}^a\nabla{}_a[t]=1$ and
considers $t{}^a=\alpha n{}^a+\beta{}^a$ where $n^a$ is a future-directed, time-like,
unit normal $n^a$ to each member of the foliation $\Sigma{}_t$, $\alpha$ is the
lapse and $\beta{}^a$ the shift.
Subsequently geometric projections of ambient fields, to (products of) the tangent and normal
bundle(s) of $\Sigma$ may be considered, which here leads to evolution equations for the augmented
EFE. The evolution equations are written in terms of the variables
$\left(%
\gamma{}_{ij},\,K{}_{ij},\,\Theta,\,\check{Z}_i%
\right)$
where $\gamma{}_{ij}$ is the induced metric and $K{}_{ij}$ the extrinsic curvature
associated with $\Sigma$;
$\Theta:=-n{}_a Z{}^a$ and $\check{Z}_i:=\perp^a_i Z{}_a$ (with $\perp^a_b:=g{}^a{}_b+n{}^a n{}_b$
and $g{}_{ab}$ being the space-time metric)
are the normal and spatial projections of $Z^a$ respectively.
Furthermore Hamiltonian, momentum, and auxiliary vector constraints
must also be satisfied $\mathcal{C}_U:=(\mathcal{H},\,\mathcal{M}_i,\,Z_a)=0$ such that a numerical
space-time is faithful to a solution of the standard EFE.
Importantly, for a space-time without
boundary
if $\mathcal{C}_U=0$ for some element of the
foliation $\Sigma_{t^*}$ then analytically this property extends for all $t$ \cite{Bona:2003fj}.
This compatibility of $\mathcal{C}_U$ with the evolution is one crucial property for numerical
calculations allowing for a choice of free-evolution scheme. In such a scheme equations are
discretized and initial data of interest is prepared so as to satisfy $\mathcal{C}_U=0$ on
$\Sigma_{t_0}$, during the course of the time-evolution $\mathcal{C}_U$ is monitored and it must
be verified that any accumulated numerical error converges away with increased resolution.

In \z4c{} \cite{Bernuzzi:2009ex,Hilditch:2012fp} to fashion an evolution scheme an additional
step is taken wherein a spatial conformal degree of freedom is first factored out via:
\begin{align}\label{eq:confxform}
  \tilde{\gamma}{}_{ij} :=& \psi{}^{-4}\gamma{}_{ij}, &
  \tilde{A}{}_{ij}:=& \psi^{-4}\Big(K{}_{ij} - \frac{1}{3}K\gamma{}_{ij}\Big);
\end{align}
with $K:=K{}_{ij}\gamma{}^{ij}$ and $\psi:=(\gamma/f)^{1/12}$ where $\gamma$ and $f$
are determinants of $\gamma{}_{ij}$ and some spatial reference metric $f{}_{ij}$ respectively.
Here we assume $f{}_{ij}$ is flat and in Cartesian
coordinates which immediately yields the \emph{algebraic constraints}:
\begin{align}\label{eq:algConstr}
  \mathcal{C}_A:=
  \big(
    \ln(\tilde{\gamma}),\,\tilde{\gamma}{}^{ij} \tilde{A}{}_{ij}
  \big)=0.
\end{align}
The expression $\mathcal{C}_A=0$ must be continuously enforced\footnote{From the point
of view of computational efficiency this is trivial to accomplish but strictly speaking doing so
entails a partially-constrained evolution scheme.} during numerical evolution
(see \S\ref{ssec:numerical_technique}) to ensure consistency \cite{Cao:2011fu}.
Additionally we introduce the transformations:
\begin{align}
  &\chi:=\gamma{}^{-1/3}, &&
  &\hat{K}:= K-2\Theta;\\
  &\tilde{\Gamma}{}^i := 2\tilde{\gamma}{}^{ij} Z{}_j
  + \tilde{\gamma}{}^{ij}\tilde{\gamma}{}^{kl} \pd{}_l[\tilde{\gamma}{}_{jk}],&&
  &\defG{}^i := \tilde{\gamma}{}^{jk}\tilde{\Gamma}{}^i{}_{jk};
  \label{eq:defn_transforms}
\end{align}
where the definition of $\chi$ implies that $\chi=\psi^{-4}$.
Collectively the \z4c{} system is comprised of dynamical variables
$\big(%
  \chi,\,\tilde{\gamma}{}_{ij},\,\hat{K},\tilde{A}{}_{ij},\,\Theta,\,\tilde{\Gamma}{}^i%
\big)$ which are governed by the \emph{evolution equations}:
\begin{align}
  \pd{}_t[\chi] =& \frac{2}{3}\chi
  \left(
    \alpha(\hat{K} + 2\Theta) - \partial_i[\beta{}^i]
  \right) + \beta{}^i\partial_i[\chi],
  \label{eq:evo_chi}
\end{align}
\begin{align}\nonumber
  \pd{}_t[\tilde{\gamma}{}_{ij}] =&
  -2\alpha\tilde{A}{}_{ij} + \beta{}^k\partial{}_k[\tilde{\gamma}{}_{ij}]
  -\frac{2}{3}\tilde{\gamma}{}_{ij}\partial{}_k[\beta{}^k]\\
  &+2\tilde{\gamma}{}_{k(i}\partial{}_{j)}[\beta{}^k].
  \label{eq:evo_conf_metr}
\end{align}
\begin{align}\nonumber
  \pd{}_t[\hat{K}] =&
  -\D{}^i[\D{}_i[\alpha]]
  + \alpha\left[
    \tilde{A}{}_{ij}\tilde{A}{}^{ij} + \frac{1}{3}(\hat{K}+2\Theta)^2
  \right]\\
  &
  + \beta{}^i \pd{}_i[\hat{K}]
  +\alpha\kappa{}_1(1-\kappa{}_2)\Theta+4\pi\alpha[S+\rho],
  \label{eq:evo_Khat}
\end{align}
\begin{align}\nonumber
  \pd{}_t[\tilde{A}{}_{ij}] &=
  \chi[-\D{}_i[\D{}_j[\alpha]] + \alpha(R{}_{ij}
  -8\pi S{}_{ij})]^{\mathrm{tf}}
  \\\nonumber
  &\hphantom{=}
  +\alpha[
    (\hat{K} + 2\Theta)\tilde{A}{}_{ij}
    -2\tilde{A}{}^k{}_i\tilde{A}{}_{kj}
  ] + \beta{}^k \pd{}_k[\tilde{A}_{ij}]\\
  &\hphantom{=}
  + 2\tilde{A}{}_{k(i} \pd{}_{j)}[\beta{}^k]
  - \frac{2}{3}\tilde{A}{}_{ij}\pd{}_k[\beta{}^k],
  \label{eq:evo_conf_Atil}
\end{align}
\begin{align}
  \pd{}_t[\Theta] =& \frac{\alpha}{2}
  \left[
    \tilde{\mathcal{H}} -2\kappa{}_1(2+\kappa{}_2) \Theta
  \right] + \beta{}^i \pd{}_i[\Theta],
  \label{eq:evo_Theta}
\end{align}
\begin{align}
  \nonumber
  \pd{}_t[\tilde{\Gamma}{}^i] =&
  -2\tilde{A}{}^{ij} \pd{}_j[\alpha]
  +2\alpha
  \Big[
    \tilde{\Gamma}{}^i{}_{jk} \tilde{A}{}^{jk}
    -\frac{3}{2}\tilde{A}{}^{ij} \pd{}_j[\ln(\chi)]\\
  \nonumber
  &
  -\kappa{}_1(\tilde{\Gamma}{}^i-\defG{}^i)
  -\frac{1}{3}\tilde{\gamma}{}^{ij}\pd{}_j[2\hat{K}+\Theta]
  -8\pi\tilde{\gamma}{}^{ij}S{}_j
  \Big]\\
  \nonumber
  &
  +\tilde{\gamma}{}^{jk}\pd{}_k[\pd{}_j[\beta{}^i]]
  +\frac{1}{3}\tilde{\gamma}{}^{ij}\pd{}_j[\pd{}_k[\beta{}^k]]\\
  &
  +\beta{}^j \pd{}_j[\tilde{\Gamma}{}^i]
  -\defG{}^j \pd{}_j[\beta{}^i]
  +\frac{2}{3}\defG{}^i \pd{}_j[\beta{}^j];
  \label{eq:evo_conf_GamTil}
\end{align}
where in Eq.\eqref{eq:evo_conf_Atil} the trace-free operation is computed with respect to
$\gamma{}_{ij}$ and $\tilde{\mathcal{H}}$ is defined in Eq.\eqref{eq:constr_ham}.
Definitions of matter fields are based on projections of the decomposed space-time,
energy-momentum-stress tensor:
\begin{equation}\label{eq:space_time_ems_dec}
  T{}_{ab} = \rho n{}_a n{}_b + 2 S{}_{(a} n{}_{b)} + S{}_{ab},
\end{equation}
in terms of the
energy density $\rho := T{}_{ab} n{}^a n{}^b$,
momentum $S{}_i := -T{}_{bc} n{}^b \perp{}^c_i$,
and spatial stress $S{}_{ij} := T{}_{cd} \perp{}^c_i \perp{}^d_j$ with associated
traces $T:=g{}^{ab}T{}_{ab}=-\rho + S$ and $S:=\gamma{}^{ij}S{}_{ij}$.
The intrinsic curvature appearing in Eq.\eqref{eq:evo_conf_Atil} is decomposed according to:
\begin{align}
  R{}_{ij} =& \tilde{R}^\chi{}_{ij} + \tilde{R}{}_{ij},
\end{align}
where in terms of the conformal connection $\tilde{\mathrm{D}}{}_i$ compatible
with $\tilde{\gamma}{}_{jk}$:
\begin{align}\nonumber
  \tilde{R}{}^\chi{}_{ij} =&
  \frac{1}{2\chi}\left[
    \tilde{\D}{}_i[\tilde{\D}{}_j[\chi]]
    +\tilde{\gamma}{}_{ij}\tilde{\D}{}^l[\tilde{\D}{}_l[\chi]]
    -\frac{1}{2\chi}\tilde{\D}{}_i[\chi]\tilde{\D}{}_j[\chi]
  \right]\\
  &
  -\frac{3}{4\chi^2} \tilde{\D}{}^l[\chi]\tilde{\D}{}_l[\chi] \tilde{\gamma}{}_{ij},
  \label{eq:defn_RicTilChi}
\end{align}
and:
\begin{align}\nonumber
  \tilde{R}{}_{ij} =&
  -\frac{1}{2}\tilde{\gamma}{}^{lm}
  \pd{}_l[\pd{}_m[\tilde{\gamma}{}_{ij}]]
  +\tilde{\gamma}{}_{k(i}
  \pd{}_{j)}[\tilde{\Gamma}{}^k]
  +\defG{}^k\tilde{\Gamma}{}_{(ij)k}\\
  &
  +\tilde{\gamma}{}^{lm}
  (
    2\tilde{\Gamma}{}^k{}_{l(i}\tilde{\Gamma}{}_{j)km}
    +\tilde{\Gamma}{}^k{}_{im}\tilde{\Gamma}{}_{klj}
  ).
  \label{eq:defn_RicTil}
\end{align}
Furthermore, we emphasize that in
Eq.\eqref{eq:evo_conf_GamTil} and Eq.\eqref{eq:defn_RicTil} it is crucial to impose
$\defG{}^i$ where it appears through the definition of Eq.\eqref{eq:defn_transforms}.

The \emph{dynamical constraints} in terms of transformed variables
$(\tilde{\mathcal{H}},\,\tilde{\mathcal{M}}{}_i,\,\Theta,\,\check{Z}{}^i)$
may be monitored to assess the quality of a numerical calculation:
\begin{equation}
  \tilde{\mathcal{H}} := R - \tilde{A}{}_{ij} \tilde{A}{}^{ij}
  +\frac{2}{3}\big(
  \hat{K}+2\Theta
  \big)^2 - 16\pi \rho = 0,
  \label{eq:constr_ham}
\end{equation}
\begin{align}\nonumber
  \tilde{\mathcal{M}}{}_j :=&
  \tilde{\mathrm{D}}_i[\tilde{A}{}^{i}{}_j]
  - \frac{3}{2}\tilde{A}{}^i{}_j\partial{}_i[\ln(\chi)]\\
  &-\frac{2}{3}\pd{}_j[\hat{K}+2\Theta] - 8\pi S_j =0,
\end{align}
\begin{align}
  \Theta=&0, &
  \check{Z}{}^i=& \tilde{\Gamma}{}^i - \defG{}^i =0.
\end{align}
Depending on the quantities of interest we may alternatively monitor the original
non-rescaled constraints $\mathcal{C}_U$. Furthermore, we introduce for later convenience a single,
scalar-valued collective constraint monitor:
\begin{align}\label{eq:collective_constraint}
  \mathcal{C}:= \sqrt{
    \mathcal{H}^2 + \gamma{}_{ij}\mathcal{M}{}^i\mathcal{M}{}^j
  + \Theta^2 + 4\gamma{}_{ij}\check{Z}{}^i\check{Z}{}^j
  }.
\end{align}
Finally we note that we have
made use of the freedom to adjust the system by non-principal parts prior to conformal
decomposition so as to have a result closer to BSSNOK (which may be obtained by taking the formal
limit $\Theta\rightarrow 0$ in Eqs.(\ref{eq:evo_chi}--\ref{eq:evo_conf_GamTil})).

\subsection{Gauge choice and boundary conditions}\label{ssec:gauge_bc}
To close the \z4c{} system it must be further supplemented by gauge conditions
(i.e., conditions on $\alpha$ and $\beta{}^i$)
that specify how the
various elements $\Sigma{}_t$ of the foliation piece together. Furthermore in this work the
computational domain does not extend to spatial infinity and consequently boundary conditions (BC)
on $\partial\Omega$ must also be imposed.

In \GRAthena{}
we make use of the puncture gauge condition which consists of the Bona-M\'asso lapse
\cite{Bona:1994b} and the gamma-driver shift \cite{Alcubierre:2002kk}:
\begin{gather}\label{eq:gaugeBGDRV}
\begin{aligned}
  \partial_t[\alpha] &= -\mu{}_L \alpha^2 \hat{K} +
    \beta{}^i \partial_i[\alpha],\\
  \partial_t[\beta{}^i] &= \mu{}_S \alpha^2 \tilde{\Gamma}{}^i
    -\eta \beta{}^i + \beta{}^j \partial{}_j[\beta{}^i].
\end{aligned}
\end{gather}
In specification of
Eq.\eqref{eq:gaugeBGDRV} we employ the $1+\log$ lapse variant
$\mu{}_L=2/\alpha$ together with $\mu{}_S=1/\alpha{}^2$. Initially
a ``precollapsed'' lapse and zero-shift is set:
\begin{align}\label{eq:Gprecoll}
  \left.\alpha\right|_{t=0} &= \left.\psi^{-2}\right|_{t=0}, &
  \left.\beta{}^i\right|_{t=0} &= 0;
\end{align}
where the choice is motivated by a resulting reduction in initial gauge
dynamics \cite{Campanelli:2005dd}. The shift damping parameter $\eta$ appearing in
Eq.\eqref{eq:gaugeBGDRV}
reduces long-term drifts in the metric variables \cite{Alcubierre:2002kk} and serves to
magnify the effective spatial resolution near a massive feature, which in turn reduces noise in its
local motion and extracted gravitational waveforms \cite{Brugmann:2008zz} (see also
\S\ref{sec:puncture_tests}). We adopt a fixed choice $\eta=2/M$ where
$M$ is the total ADM mass \cite{Arnowitt:1962hi} of the system throughout this work
as it is known to lead to successful time evolution of binary black holes (BBH) of comparable
masses \cite{Brugmann:2008zz} and improves stability more broadly \cite{Cao:2011fu}.
With a view towards potential investigations of high mass
ratio binaries we have also incorporated $\eta$ damping conditions within \GRAthena{}
based on BBH location as a function of time \cite{Purrer:2012wy}
together with the conformal factor based approach
of \cite{Nakano:2011pb,Lousto:2010qx,Muller:2009jx}.

When coupled to the puncture gauge with the choices made above \z4c{} forms a PDE system that is
strongly hyperbolic \cite{Cao:2011fu} and consequently the initial value problem is
well-posed \cite{Bernuzzi:2009ex}.
Artificial introduction of a boundary at finite distance complicates the analysis of numerical
stability significantly \cite{Hilditch:2016xos}.
For the initial boundary value problem the analysis benefits from symmetric hyperbolicity of
the underlying evolutionary system however for \z4c{} starting in fully second order form
this property does not appear to exist
within a large class of symmetrizers \cite{Cao:2011fu}. We do not seek to address this issue
further here.
Our boundary treatment follows an approach due to \cite{Hilditch:2012fp}.
We consider a Cartesian
coordinatization of the \Mesh{} $\Omega$ as a compactly contained domain within $\Sigma{}_t$
capturing the physics of interest. On $\partial \Omega$ Sommerfeld BC are imposed
on the subset of dynamical fields $\{\hat{K},\,\tilde{\Gamma},\,\Theta,\,\tilde{A}{}_{ij}\}$.
Though this choice is not optimal, as it is not constraint-preserving,
we have not experienced issues on account of this.
An interesting consideration for future work would be to incorporate within \GRAthena{} the
constraint-preserving BC of e.g. \cite{Ruiz:2010qj,Hilditch:2012fp} (see also \cite{Rinne:2008vn}).

\subsection{Wave extraction}\label{ssec:wave_extraction}
To obtain the gravitational wave content of the space-time, we calculate the Weyl scalar $\Psi_4$,
the projection of the Weyl tensor onto an appropriately chosen null tetrad $k,l,m, \bar{m}$. We use
the same definition of $\Psi_4$ here as \cite{Brugmann:2008zz}:
\begin{eqnarray}
  \Psi_4 &=& - R_{abcd} k^a\bar{m}^b k^c \bar{m}^d,
\end{eqnarray}
where we have exchanged the Weyl tensor for the Riemann tensor since we extract the gravitational waves
in vacuum. The $4$ dimensional Riemann
tensor is constructed from $3+1$ split ADM variables using the
Gauss-Codazzi relations as detailed in \cite{Brugmann:2008zz}.
To construct the null tetrad we start from a spatial coordinate basis:
\begin{align}
  \phi^i =& (-y,x,0), &
  r^i =& (x,y,z), &
  \theta^i =& \epsilon^i_{kl}\phi^k r^l\,;
\end{align}
which is then Gram-Schmidt orthonormalized. The newly formed orthonormal, spatial triad
is extended to space-time with $0^{\mathrm{th}}$ components set to $0$.
With this we construct the null tetrad:
 \begin{align}\nonumber
  {k} &= \frac{1}{\sqrt{2}}({n} - \hat{{r}}), &
  {l} &= \frac{1}{\sqrt{2}}({n} + \hat{{r}}); \\
  {m} &= \frac{1}{\sqrt{2}}(\hat{{\theta}} + i \hat{{\phi}}), &
  \bar{{m}} &= \frac{1}{\sqrt{2}}(\hat{{\theta}} - i \hat{{\phi}});
\end{align}
where $n{}^a=(1/\alpha,\,-\beta{}^i/\alpha)$.
Once the Weyl scalar is obtained we perform a multipolar decomposition onto spherical harmonics of spin-weight $s=-2$, defined as
follows\footnote{Convention here is that of \cite{Goldberg:1966uu} up to a Condon-Shortley phase factor of $(-1)^m$. }:

\begin{eqnarray}
\psi_{\ell m} &=& \int^{2\pi}_0\int^\pi_0 \Psi_4 \bar{{}_{-2}Y_{\ell m}}\sin \vartheta \mathrm{d}\vartheta \mathrm{d} \varphi,\label{eq:psilm}\\
{}_{-2}Y_{\ell m} &=& \sqrt{\frac{2\ell+1}{4\pi}} d_{\ell m}(\vartheta)e^{im\varphi},\\
d_{\ell m}(\vartheta) &=& \sum_{k=k_1}^{k_2} \frac{(-1)^k((\ell+m)!(\ell-m)!(\ell-2)!(\ell+2)!)^{1/2}}{(\ell+m-k)!(\ell-k-2)!k!(k-m+2)!}\nonumber\\
&&\times\left(\cos\left(\frac{\vartheta}{2}\right)\right)^{2\ell+m-2-2k}\left(\sin\left(\frac{\vartheta}{2}\right)\right)^{2k-m+2}\\
k_1 &=& \max(0,m-2),\\
k_2 & =& \min(\ell+m, \ell-2).\label{eq:k2}
\end{eqnarray}

To obtain the multipolar decomposition, $\Psi_4$ is first calculated at
all grid points throughout the \Mesh{}. This is then interpolated onto a set of
geodesic spheres $Q_R$ (see \S\ref{ssec:geodesic_spheres}) at given extraction radii
$R_Q$, over which the integral in Eq.\eqref{eq:psilm} is performed.
Recall that the grid level parameter $n_Q$ controls the total number of samples on
on $Q_R$ as $10 n_Q^2 + 2$.
We select $n_Q$ through local matching to an area element of a \MeshBlock{}:
\begin{equation}\label{eq:etaSpec}
  n_Q = \left \lceil
  \sqrt{\frac{\pi R_Q^2}{\sp^2} - 2}
  \right \rceil.
\end{equation}

The modes of the gravitational wave strain $h$ are computed
from the projected Weyl scalar by integrating twice in time
\begin{eqnarray}
\label{eq:strain}
\psi_{\ell m}=\ddot{h}_{\ell m}\,,\label{eq:hlm}
\end{eqnarray}
The strain is then given by the mode-sum:
\begin{equation}
  R\left(h_+ - i h_\times\right) = \sum_{\ell=2}^{\infty}\sum_{m=-\l}^\l h_{\ell m}(t)\; {}_{-2}Y_{\ell m}(\vartheta,\varphi)\,.
\end{equation}
Following the convention of the LIGO algorithms library \cite{lalsuite} we set:
\begin{equation}\label{eq:strain_complex}
R h{}_{\ell m} = A{}_{\ell m} \exp(-i\phi{}_{\ell m}),
\end{equation}
and the gravitational-wave frequency is:
\begin{equation}
\omega{}_{\ell m}= \frac{d}{dt}\phi{}_{\ell m}.
\end{equation}

\subsection{Numerical technique}\label{ssec:numerical_technique}
We implement the \z4c{} system described in \S\ref{ssec:eqnformgrathena} in \GRAthena{} based on
the method of lines approach where field variables may be chosen to obey a VC or CC discretization
at compile time and time-evolution is performed using the fourth order in time, four stage,
low-storage RK$4()4[2S]$ method of \cite{ketcheson2010rungekuttamethods}.

Generic spatial field
derivatives in the bulk (away from $\partial\Omega$) are computed with high-order,
centered, finite difference
(FD) stencils whereas shift advection terms use stencils lopsided by one grid point
\cite{zlochower2005accurateblackhole,%
Husa:2007hp,Brugmann:2008zz,chirvasa2010finitedifferencemethods}.
The implementation is based on \cite{Alfieri:2018a} and utilizes \Cpp templates to offer
flexibility in problem-specific accuracy demands without performance penalties.
A similar approach is taken for implementation of the $\mathcal{R}$ and $\mathcal{P}$ operators
discussed in \S\ref{sssec:restr_prol}. With this a consistent, overall, formal order throughout
the bulk of the computational domain is maintained during calculations by compile-time
specification of the ghost-layer through choice of $\Ng$ together with $\Ncg$. Throughout this
work we take $\Ng=\Ncg$ though for a VC discretization this is not a requirement within
\GRAthena{} and may be tuned to the demands of the desired \Mesh{} refinement strategy.
In the case of \z4c{} and VC this translates to an order for spatial discretization in the
bulk of $2(\Ng-1)$.

We emphasize that during calculation of FD, $\mathcal{R}$ and $\mathcal{P}$ approximants,
special care has been taken in the ordering and grouping of arithmetical operations
so as to reduce accumulation of small floating-point differences. This is a particularly
important consideration in the presence of physical symmetries where linear instabilities may
amplify unwanted features present in the operator approximants and lead to resultant, spurious
appearance of asymmetry during late-time solutions \cite{stone2020athenamathplusmathplus}.

For most calculations involving \z4c{} the treatment of the physical boundary is non-trivial.
\GRAthena{} extends \Athena{} by providing the Sommerfeld BC motivated in \S\ref{ssec:gauge_bc}.
To accomplish this, within every time-integrator substep an initial Lagrange extrapolation is
performed so as to populate the ghost-layer at $\partial\Omega$. Order is again controlled at
compile-time and we typically select $\Ng+1$ points for the extrapolation, albeit numerical
experiments did not indicate significant changes when this choice was varied.
The dynamical equations of
\S\ref{ssec:eqnformgrathena} and gauge conditions of \S\ref{ssec:gauge_bc}
populate the subset of fields
$\{\chi,\,\tilde{\gamma}{}_{ij},\,\alpha,\,\beta{}^i\}$ on nodes of $\partial\Omega$
whereas for
$\{\hat{K},\,\tilde{\Gamma},\,\Theta,\,\tilde{A}{}_{ij}\}$
Sommerfeld BC are imposed as in \cite{Hilditch:2012fp} where first order spatial derivatives are
approximated through second order accurate, centered FD; we have found this to be
crucial for numerical stability.

As observed in \cite{Cao:2011fu} in the absence of algebraic constraint $\mathcal{C}_A$ projection
\z4c{} is only weak-hyperbolic. Therefore in \GRAthena{} we enforce $\mathcal{C}_A$ at each
time-integrator substep. On the other hand a coarse indicator on the overall error during the
course of a calculation is provided through inspection of the constraints $\mathcal{C}_U$ together
with $\mathcal{C}$ (of Eq.\eqref{eq:collective_constraint}). Note that these latter constraints
are \emph{not} enforced.

In order to have confidence in implementation details we have replicated a
subset of tests from the ``Apples with Apples'' test-bed
suite \cite{Alcubierre:2003pc,Babiuc:2007vr,Cao:2011fu,Daverio:2018tjf}
a discussion of which is provided in the appendix \S\ref{sec:awa_tests}.

The \z4c{} system does not strictly impose any particular underlying \Mesh{} structure or
refinement strategy and consequently we use this freedom to improve efficiency and accuracy by
raising resolution only where it is required. During evolution the Courant-Friedrich-Lewy (CFL)
condition must be satisfied. To achieve this in the context of refinement, spatial resolution on
the most refined level and the choice of $\CFL$ factor itself determines the global
time-step that is applied on each \MeshBlock{}. Finally, in order to suppress high-frequency
numerical
artifacts generated at \MeshBlock{} boundaries and not present in the physical solution we make use
of high-order Kreiss-Oliger (KO) dissipation \cite{Kreiss:1973,Gustafsson:2013td} of uniform
factor $\KO$ over all levels. In particular given a system of time-evolution equations for a vector
of variables $\mathbf{u}$ the replacement
$\partial{}_t[\mathbf{u}] \leftarrow \partial{}_t[\mathbf{u}] + \sigma \mathcal{D}[\mathbf{u}]$
is made where $\mathcal{D}[\cdot]$ is proportional to a spatial derivative of order $2\Ng-2$.

\section{Mesh refinement for punctures}\label{sec:amr_criterion}
Black holes in \GRAthena are modeled as in \BAM making use of the puncture formalism~
\cite{Brugmann:2008zz}.
In numerical relativity, BH can be treated by adopting the Brill-Lindquist wormhole
topology which consists of considering N black holes with N+1 asymptotically
flat ends for the initial geometry. These flat ends are compactified and
identified with points on $\mathbb{R}^3$ and the coordinate singularities at these
points are called \emph{punctures}.
This allows one to produce black hole initial data associating masses, momenta and
spins to any number of black holes.
The main application of this formalism is binary black hole evolution.

The adaptive mesh refinement criterion implemented in \GRAthena for puncture
evolution mimics the classic box-in-box refinement (used in e.g. \BAM, \Cactus),
within the \Athena infrastructure. The main idea is to follow the punctures' position
during the evolution and refine the grid depending on the distance from each puncture.

\subsection{Punctures' initial data}\label{ssec:amr_criterion}
Black holes initial data are constructed following~\cite{Brugmann:2008zz}.
We consider as our initial data the positive-definite metric and extrinsic curvature
$(\gamma{}_{ij},~K{}_{ij})$ on a spatial hypersurface $\Sigma$ with time-like
unit normal $n{}^i$ such that $n{}^i n{}_i = -1$.
Such initial data are constructed by means of the conformal,
transverse-traceless decomposition of the initial-value equations~\cite{York:1979}.
We can use the map of Eq.\eqref{eq:confxform} and freely choose an initially conformally flat
background $\tilde{\gamma}{}_{ij}=\delta{}_{ij}$ and take a maximal slice, i.e. set $K=0$.
Doing so, the momentum constraint
becomes $\partial{}_j \left(\psi^6\tilde{A}_{ij}\right) =0$ and admits Bowen-York
solutions~\cite{Bowen:1980yu} for an arbitrary number of black holes.

The Hamiltonian constraint reduces to an elliptic equation
for $\psi$, with solution (for $N$ black holes with centers at $r_i$):
\be\label{eq:init_psi0}
	\psi_0 = 1 + \sum_{i=1}^{N} \frac{m_i}{2r_i} + u.
\ee
The variable $\psi_0$ represents the initial value of $\psi$, which, based on its relation
to $\chi$, is evolved according to Eq.\eqref{eq:evo_chi}.
In this equation the function $u$ can be determined by an elliptic
equation on $\mathbb{R}^3$
and is $C^2$ at the punctures and $C^{\infty}$ elsewhere.
The variable $m_i$ is called the \emph{bare mass} of a black hole and it coincides with the
actual mass only in the Schwarzschild case. The total ADM mass of each black hole at the puncture
is given by:
\be
	M_i = m_i\left(1+u_i+\sum_{i\neq j} \frac{m_j}{2d_{ij}}\right),
\ee
where $u_i$ is the value of $u$ at each puncture and
$d_{ij}$ is the coordinate distance between each pair of punctures.
Ultimately, we denote the total mass of the system with $M$, which represents
the physical mass scale of the problem and thus all results will be reported
accordingly.

To produce BBH initial data following the above description, we make use of an
external \texttt{C} library based on the pseudo-spectral approach
of~\cite{Ansorg:2004ds}, which is also used in the \TwoPunctures\footnote{We adapted
the public code into a stand-alone library that may be found at
the URL \url{https://bitbucket.org/bernuzzi/twopuncturesc/}.}
thorn of \Cactus.

\subsection{Puncture tracker}\label{ssec:punc_tracker}
To follow the punctures' position we need to solve an additional ODE, which is
not coupled to the Z4c system. Since the conformal factor $\chi$ vanishes at the
puncture,
Eq.\eqref{eq:evo_chi} implies that \cite{Campanelli:2005dd}:
\be
\dot{\mathbf{x}}_\mathrm{p} (t) = -\mathbf{\beta}_{|\mathbf{x}_\mathrm{p}}(t),
\label{eq:tracker}
\ee
where $\mathbf{\beta}_{|\mathbf{x}_\mathrm{p}}$ is the shift function evaluated at
the puncture position.
We solve this vectorial equation at every timestep using an explicit Euler solver.
Though \BAM implements higher order methods to solve
this equation (Crank–Nicolson method), the solution obtained with the first order Euler
solver agrees with that of \BAM where a comparison is made for
two trajectories in the left panel of Fig.\ref{fig:comparison}.

\subsection{Oct-tree box-in-box}\label{sec:oct-tree_bnb}
\begin{figure}[thp]
	\centering
		\includegraphics[width=\columnwidth]{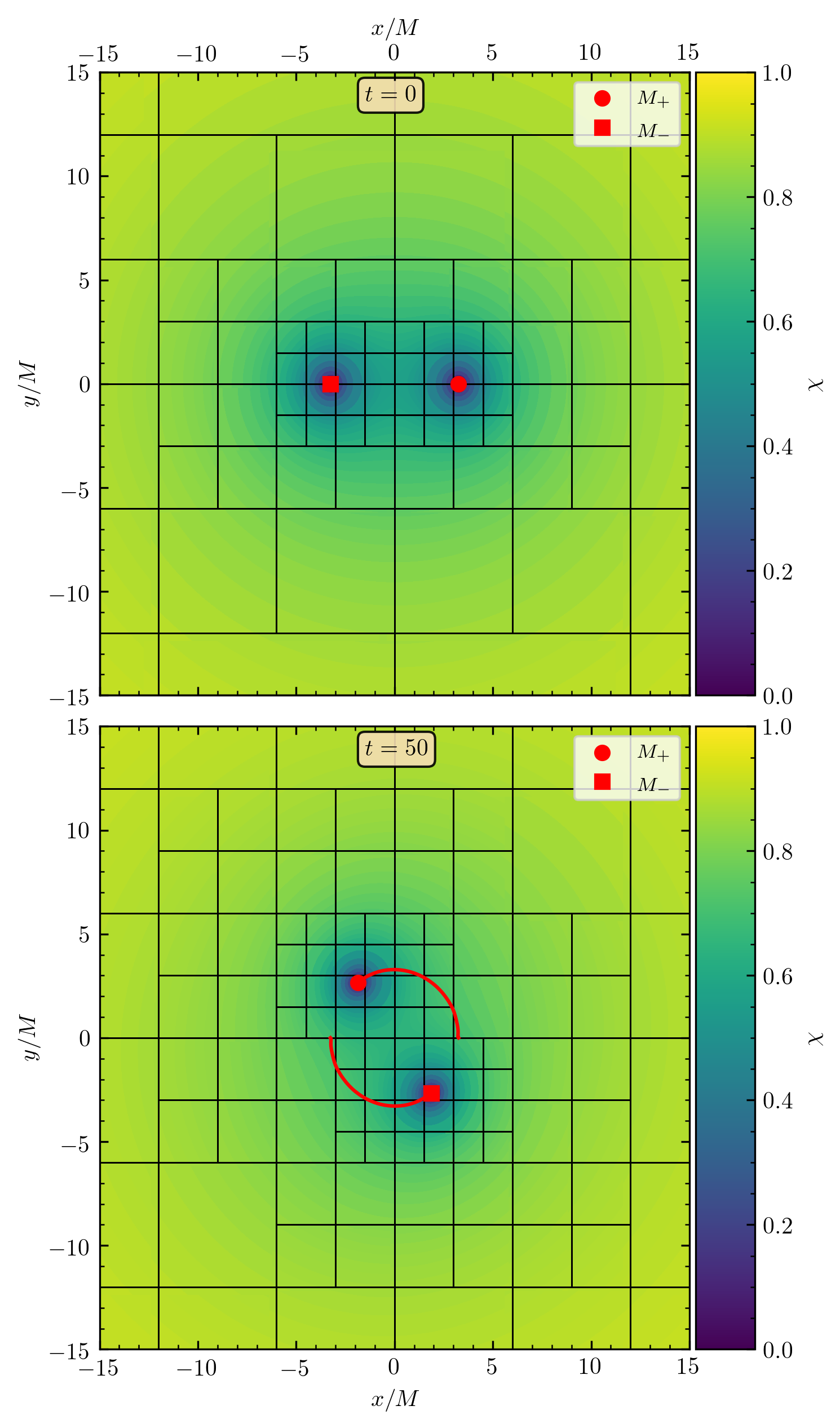}
		\caption{2D slice at $z=0$ of a mesh grid produced with \GRAthena, setting 11
		         refinement levels. Top panel: black holes at initial position
		         $\textbf{x}_\mathrm{\pm} = (\pm 3.257,0,0)~M$. Bottom panel:
		         snapshot at $50~M$ in which trackers are shown (red lines).
		         The color code refers to the value of the conformal factor $\chi$.
				 For clarity, in the figure only a subset of the total slice is shown,
				 therefore only highest levels are visible.
				 It is also possible to see the underlying box-in-box structure, in
				 which boxes are made up of \MeshBlock{}s.
		}
		\label{fig:mesh}
\end{figure}

In \Athena, adaptive mesh refinement (AMR) is implemented as follows:
during the evolution a certain condition is evaluated on each \MeshBlock{} and
consequently the code refines the particular \MeshBlock{}, de-refines it or does
nothing.
In the punctures' case the condition we employ relies on the punctures'
position.
For a given \MeshBlock{} we first calculate the distance
$\min\limits_{i}||{\textbf{x}}^i_\mathrm{p} - \textbf{x}_\mathrm{MB}||_\infty$, where
$\textbf{x}_\mathrm{p},
~\textbf{x}_\text{MB}$
denote the puncture and \MeshBlock{} positions\footnote{For implementation reasons,
$\textbf{x}_\text{MB}$ is defined as follows: we consider a cube with same center,
the edge of which is $1/4$ of the edge of the original~\MeshBlock{}.
$\textbf{x}_\text{MB}$ are the coordinates of the corner
of this cube which is closer to the closest puncture.}
respectively, and $i$ labels each
puncture that is present. This allows one to assess the \textit{theoretical}
refinement level the \MeshBlock{} should be in.
If the refinement level of the considered \MeshBlock{} is
not the same as the theoretical one just calculated, then the block is either
refined or de-refined according to its current level.
The theoretical refinement level is determined by considering a classic box-in-box
structure of the grid, in which each puncture is enclosed in a series of nested boxes
centered on the puncture, all with the same number of points but with increasing
physical extent, i.e. with decreasing resolution. In particular, each box has half of
the resolution of the next one it contains.
The presence
of punctures and their position
determines a structure of nested boxes, in such a way that
the smallest (and finest) imaginary box around a puncture defines the highest
refinement level. The box containing it corresponds to a lower refinement level
and so on up to the $0^\text{th}$ level which corresponds to the initial mesh 
itself.
Practically, to define a grid in \GRoAthena one needs to specify the number of
points of the initial mesh grid $N_M$, the number of points per \MeshBlock{} 
constituting the mesh $N_B$, and the number of total refinement levels $N_L$ up to 
which the grid has to be refined.
Following the procedure above, \GRAthena will refine each \MeshBlock{},
producing an \emph{oct-tree box-in-box} grid structure. A visualization of this can be
seen in Fig.\ref{fig:mesh}, in which the initial configuration of two coalescing
black holes and a snapshot at later time are shown.
Here $N_M = 64$ and $N_B = 16$, thus the initial mesh is divided into
$4^3$ \MeshBlock{}s (level 0, see \S\ref{sec:tree_structure}).
Following the procedure described above, \MeshBlock{}s are sub-divided up to level 10,
which corresponds to the smallest \MeshBlock{}s containing the two black holes.
The final grid is composed of the initial \MeshBlock{}s, which are
those far from the punctures and thus untouched, and increasingly
smaller blocks (in terms of physical extent) for each refinement level.
Note that in the top panel
levels 7, 8, 9, 10 are visible, while due to regridding at later times
in the bottom panel \MeshBlock{}s of level 7 (the ones closest to the edges
of the plot) have been subdivided and their children belong to level 8.
\subsection{Grid configurations}\label{sec:grid_conf}
In order to \emph{accurately} and \emph{efficiently} perform a BBH merger simulation it is crucial
to optimize the grid configuration for a given problem.
Thus to attain accuracy at reduced computational cost a balance
must be struck such that the strong-field dynamics are well-resolved
and their effects propagate cleanly into the wave-zone for extraction. The former can be directly
controlled based on the oct-tree box-in-box refinement criterion and fixing a target puncture
resolution $\sp_p$ or equivalently, maximum number of refinement levels $N_L$. For the
wave-zone an extraction radius $R$ must be selected and the underlying refinement strategy
together with choices of $N_M$ and $N_B$ induce\footnote{Direct control on $\sp_w$ is
offered in \GRAthena{} through optional introduction of a minimum refinement level maintained over
a ball of radius $R$
centered at $C$
but for the results presented it was not found to be necessary to utilize.}
a resolution $\sp_w$.
Finally, the maximum spatial extent $x_M$ of the computational domain $\Omega$ must be chosen to be
sufficiently large so as to mitigate any potential spurious effects due to imposed approximate
boundary conditions.

Unless otherwise stated we select $\Omega$ as the Cartesian coordinatized cube
$[-x_M,\,x_M]^3$
which results in a resolution on the most refined level in the vicinity of a puncture of:
\be\label{eq:spacing}
	\sp_p=\frac{2x_M}{N_M 2^{N_L - 1}}.
\ee
Wave-zone resolution $\sp_w$ may similarly be computed taking
$N_L=\lceil\log_2\left(\frac{2x_M}{R}\right)\rceil$ in the above where
$R$ is the extraction radius.
For the calculations presented in this work,
we typically select $N_B=16$, which allows for up to $\6th$ order accuracy for
approximants to
operators pertaining to quantities appearing during spatial
discretization. The constraint on maximum approximant order that can be selected for a
choice of $N_B$ is given in Eq.\eqref{eq:N_B_restriction} which
arises on account of the double restriction operation
described in \S\ref{sssec:restr_prol}.
Natively, \GRAthena{}
supports up to $\8th$ order which may be further extended
through simple modification of the relevant \Cpp templates.

We have observed that a simple approach to further optimize for \emph{efficiency} once
convergence properties are established is to modify $N_M$ and $N_B$.

\section{Puncture tests}\label{sec:puncture_tests}
In this section we present several tests of puncture evolutions to validate $\GRAthena$.
We compare our results against \BAM code and {\tt TEOBResumS}, used as benchmarks.
We also demonstrate the convergence properties of our code for these tests.

Unless otherwise stated, throughout this section we adopt tortoise coordinates,
in which evolution time $t$ is mapped as $t \rightarrow u \equiv t - r^*$, where
$r^* = r + 2M\log\left|\frac{r}{2M}-1\right|$ and $M,~r$ are the total mass of the
binary system and the Schwarzschild coordinate, respectively. In waveform plots
quantities are suitably rescaled by $M$ (see~\S\ref{ssec:amr_criterion}) and
by the symmetric mass ratio $\nu:=\frac{M_1M_2}{\left(M_1+M_2\right)^2}$.
The \emph{merger} or \emph{time of merger} is defined as the time corresponding to
the peak of the $(\ell = 2, m = 2)$-mode of $A_{\lm}$\footnote{Hereafter the (2,2)
mode, and similarly for other $(\ell,m)$ modes and for all related quantities.} (defined as in
Eq.\eqref{eq:strain_complex}).

\subsection{Single Spinning Puncture}\label{ssec:single_spin_punc}
In order to verify the evolution of a single black hole puncture, as well as the
gravitational wave signal calculated by
\GRAthena{} we perform a direct comparison with the established \BAM
code.
Using initial data generated by the \TwoPunctures library
for both codes, as used in similar tests of the \BAM code in \cite{Hilditch:2012fp}, we simulate
the evolution of a single spinning
puncture, representing a Kerr black hole with dimensionless spin
parameter $a=0.5$. To this end we initialize two black holes, one with the
target mass, $1M$, and spin $a=0.5$, and another with negligible mass,
$10^{-12}M$, and zero spin, with a small initial separation, $10^{-5}M$.
These black holes merge soon after the simulation begins, and the resulting
single black hole can be treated as our target Kerr BH. We use the static
mesh refinement of \GRAthena{} to construct a refined grid around the puncture
that matches the resolution of the \BAM evolution both at the puncture ($\sp = 0.08333 M$) and
in the wave zone ($\sp = 0.66667 M$).

To compare the two wave signals, we calculate the dominant $(2,0)$ mode
of the strain from the expressions above in
Eqs.(\ref{eq:psilm}--\ref{eq:k2}, \ref{eq:hlm}).
In doing so, we perform two integrations in the time domain (note this is
different to the frequency domain integration performed for the $(2,2)$
mode studied below for the black hole binary, as here we have no
well-motivated cut-off frequency available \cite{Dietrich:2014wja}). These
integrations may add an arbitrary quadratic polynomial in time onto the
strain as constants of integration \cite{Damour:2008te,Baiotti:2008nf} and so,
in the results presented here, we fit for, and then subtract, this quadratic.
We further note this reconstruction has been shown to introduce errors in the
waveform ring-down \cite{Dietrich:2014wja}.

\begin{figure}[t]
	\centering
		\includegraphics[width=\columnwidth]{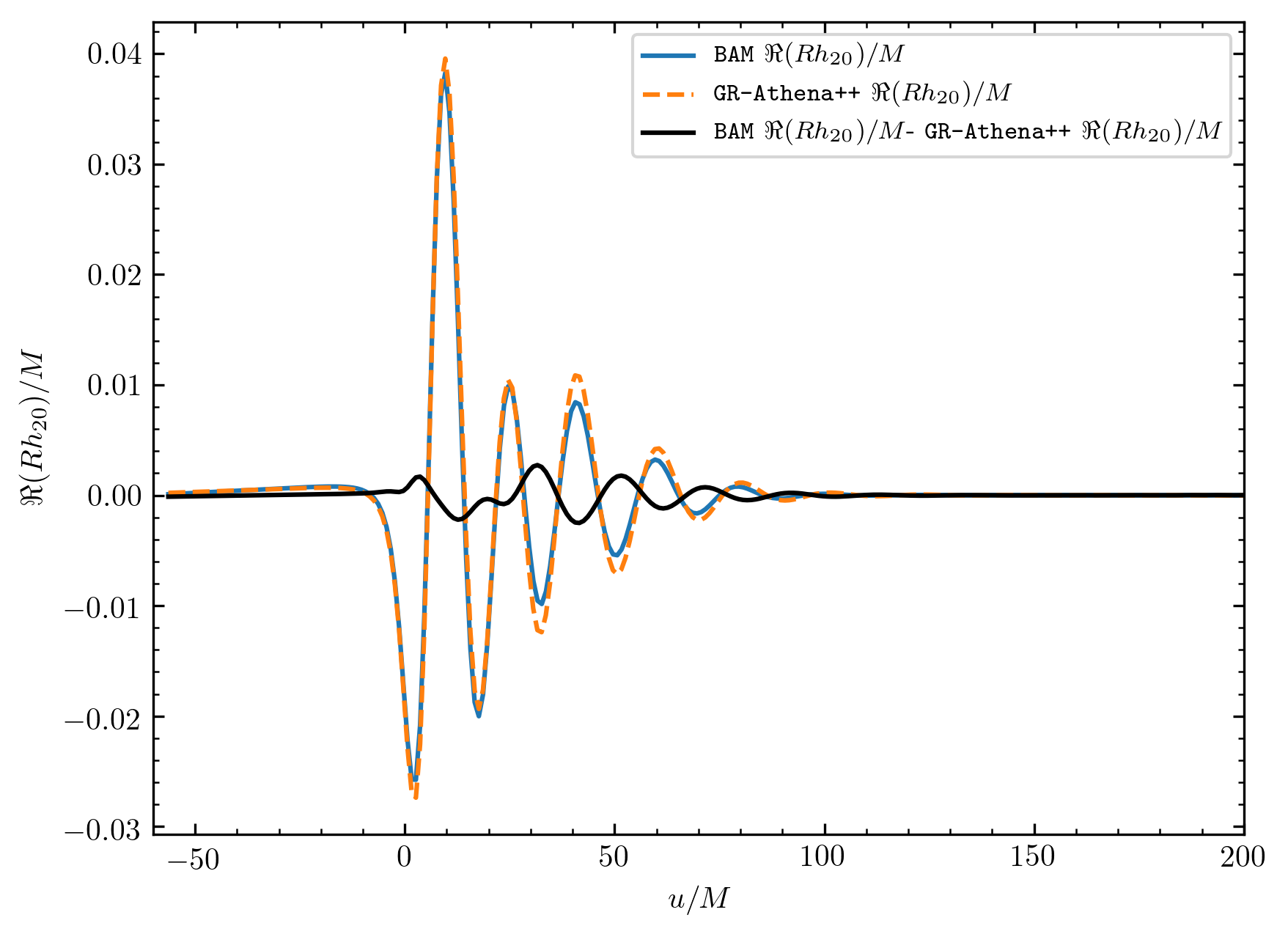}
		\caption{$\Re (Rh_{20}/M)$ for a single spinning puncture in \GRAthena and \BAM with difference shown in black as a function of Schwarzschild tortoise coordinate, $u$. Wave extracted at $R=50M$.}
		\label{fig:spcomp}
\end{figure}

In Fig.\ref{fig:spcomp} we show the match between
the two calculated signals for the real part of the $(2,0)$ mode of the
gravitational wave strain.
These show consistency in the phasing of the signal, with slight
discrepancies in the amplitude of the strain.

We also demonstrate the convergence properties of the waveforms in \GRAthena{}
for this test.
We perform the same simulation at a coarse, medium and fine resolution
(with finest grid spacings $\sp_c = 0.025M, \sp_m = 0.02083M, \sp_f = 0.01563M$),
 and show that the difference between the medium and fine resolution waveform
matches the difference between the coarse and medium waveform, when rescaled by
 the factor $Q_n$ for $n^{\mathrm{th}}$ order convergence, defined as:
\begin{eqnarray}
  \label{eq:conv_fact}
  Q_n = \frac{\sp_c^n - \sp_m^n}{\sp_m^n - \sp_f^n}.
\end{eqnarray}

\begin{figure}[t]
	\centering
	\includegraphics[width=\columnwidth]{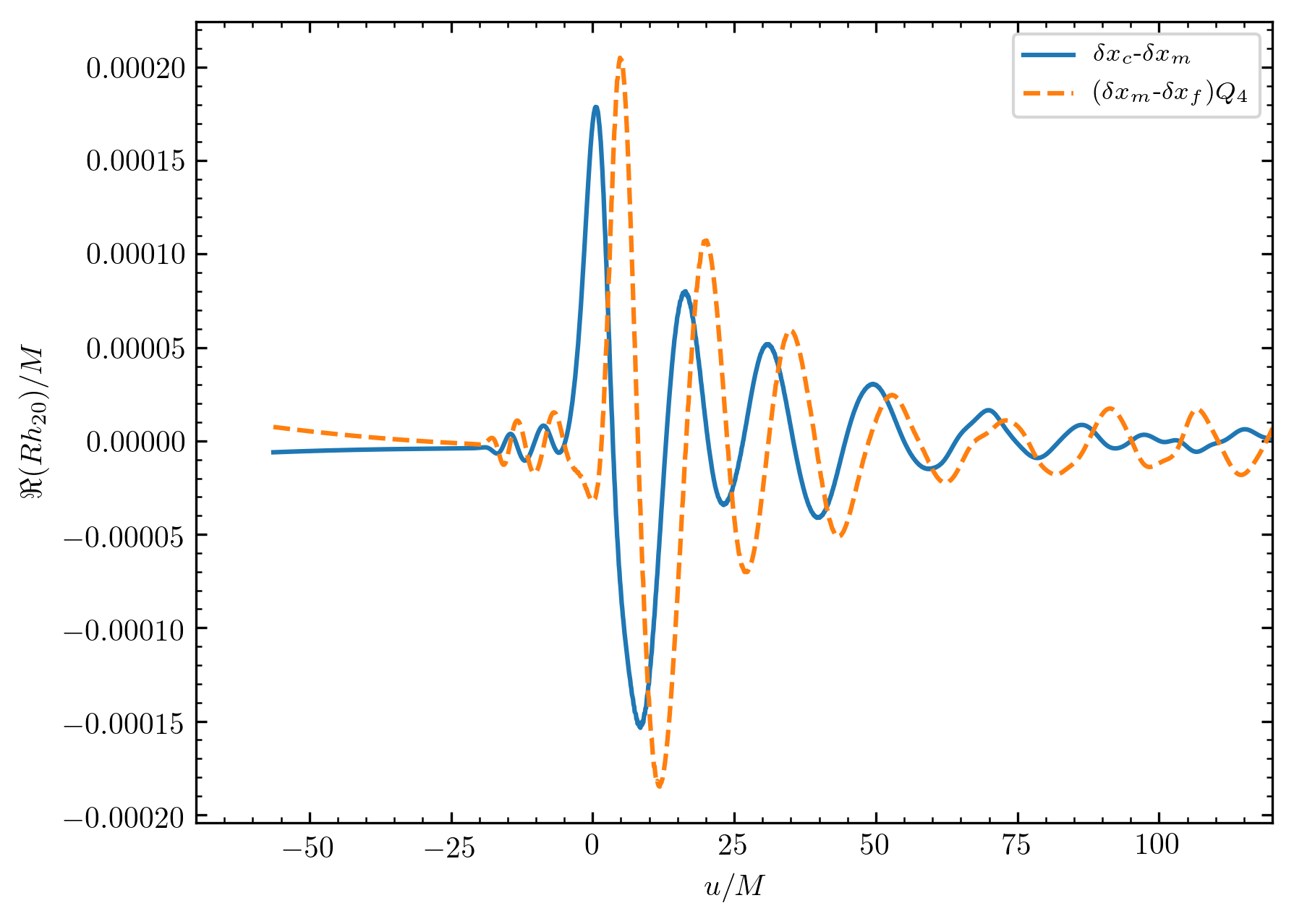}
	\caption{The difference between the waveform at coarse and medium resolution ($\sp_c-\sp_m$) is consistent with the difference between the waveform at medium and fine resolution ($\sp_m-\sp_f$) when rescaled by the appropriate factor for $\4th$ order convergence $Q_4$. Waves extracted at $R=50M$}
	\label{fig:spconv}
\end{figure}

In Fig.\ref{fig:spconv} we show convergence properties for a simulation with $\4th$
order accurate finite differencing operators.

We see here that, rescaling assuming $\4th$ order convergence,
 \GRAthena{} demonstrates under-convergence at initial times and a
 consistent order of convergence at later times, although without a
 precise point-wise scaling.  We note that for similar tests performed
 with the \BAM code in \cite{Hilditch:2012fp} using
 initial data constructed in the same manner, these same
 convergence properties are observed for the waveform of the spinning
 puncture problem.

\subsection{Calibration evolution of two punctures}\label{ssec:calibration}
We validate \GRAthena against binary black
hole evolutions by comparing with \BAM and performing convergence tests.
For these tests the two initial non-spinning black holes, with bare-mass
$m_\pm = 0.483~M$, are located on the $x-$axis, with
$x^1_\mathrm{p,\pm}(t=0) = \pm 3.257~M$, and
initial momenta directed along the $y-$axis,
$p^2_\mathrm{\pm}(t=0) = \mp 0.133~M$.
The gauge is chosen as explained in \S\ref{ssec:gauge_bc}.
For the \GRAthena vs. \BAM comparison and for convergence tests several runs at
different resolutions have been performed. The grid configuration for both codes
is described in detail below.
This initial setup results in a $\sim 2.5$ orbits evolution of the two black holes
before merger, which happens at evolution time $t\sim 170~M$, as can be seen in
Fig.\ref{fig:comparison}.

In the next two subsections the $(2,2)$ mode of gravitational
wave strain is calculated according to Eq.\eqref{eq:strain}.

\subsubsection{\GRAthena vs. \BAM comparison}\label{ssec:grathena_vs_bam}

 \begin{figure*}[t]
   \centering
     \includegraphics[width=0.45\textwidth]{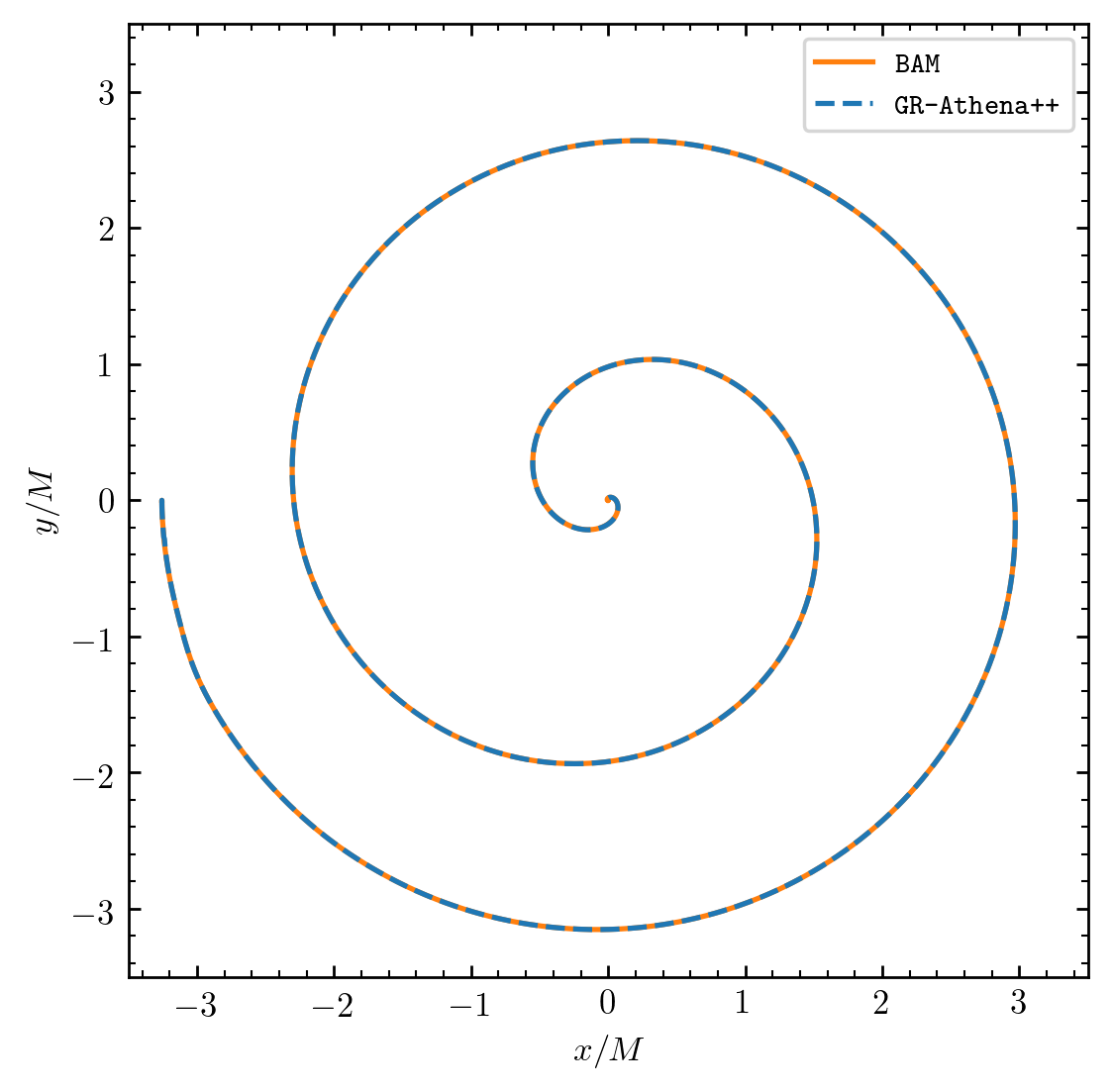}
     \includegraphics[width=0.45\textwidth]{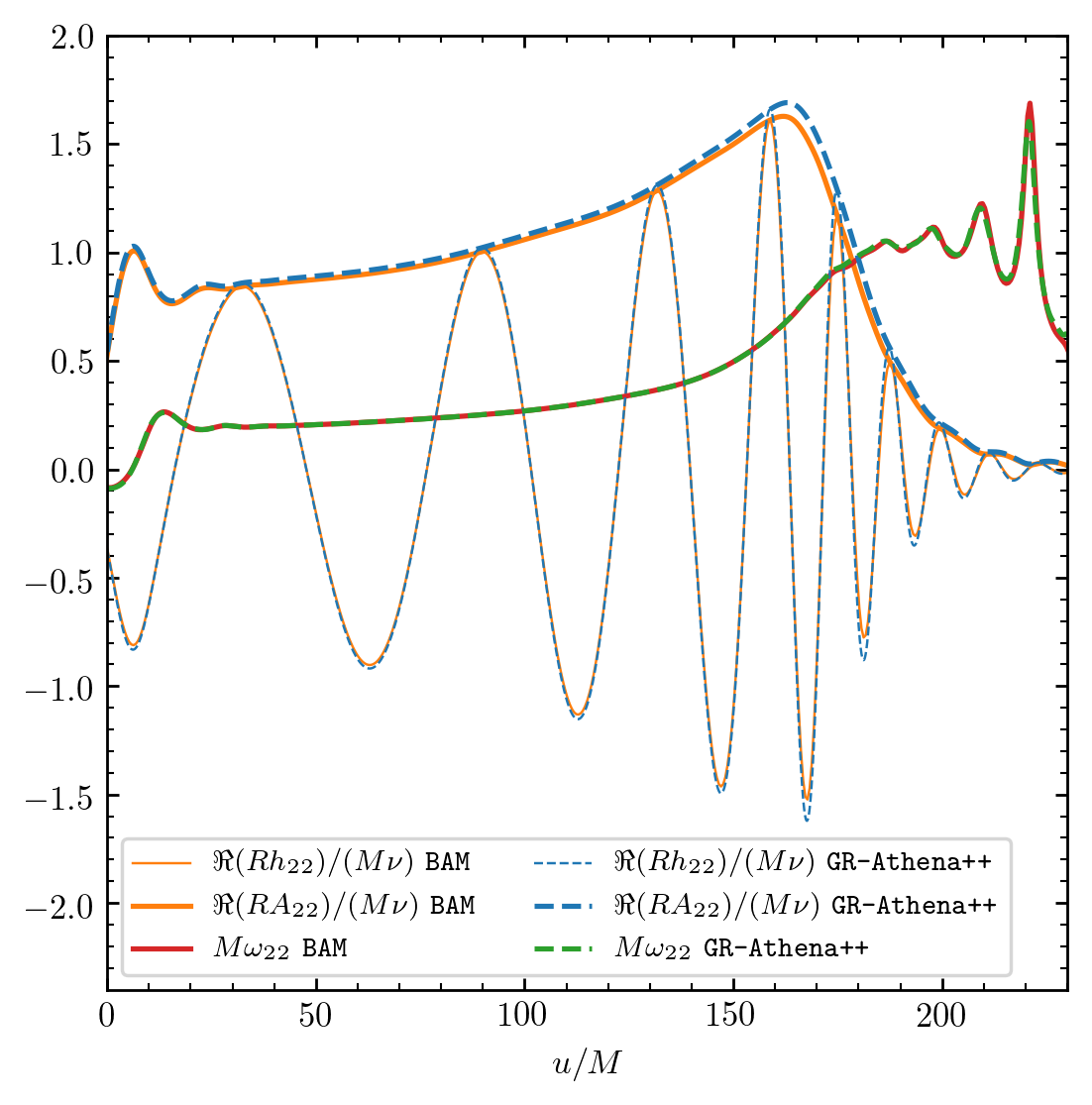}
     \caption{Comparison between \BAM and \GRAthena of the trajectories of
     		  puncture $-$ (left panel) and of gravitational waveforms (right panel) for
     		  resolution $N_M = 192$. Waveforms are extracted at a
     		  representative radius $R = 120~M$ and merger time is defined as the
     		  amplitude peak time. Discrepancy between the two amplitudes
     		  is $\lesssim 2\%$.}
  \label{fig:comparison}
 \end{figure*}

\Athena and \BAM implement completely different grid structures.
To compare the two codes, we try to generate grids as similar
as possible aiming to match the resolution at the puncture and
the physical extent of the grid.
In the case of \GRAthena we choose grid parameters $N_B = 16$,
$N_L = 11$, $x_M = 3072~M$ and various resolutions
$N_M = [96,\,128,\,192,\,256]$. This results in
resolutions at the puncture of
$\sp_p = [\,1.5625,\,1.171875,\,0.78125,\,0.5859375] \times 10^{-2}~M$ (see
Eq.\eqref{eq:spacing}).
For \BAM the same is achieved in each corresponding
simulation by considering 6 nested boxes of
$N_M$ points and 10 smaller boxes (5 per puncture, centered at each one) of $N_M/2$
points and maximum spacing in the outermost grids $\Delta x =
[96,\,64,\,48,\,32,\,24]~M$ respectively.
In both cases all simulations are performed with $\4th$ order finite 
differences stencils for derivatives.
Fig.\ref{fig:comparison} shows very good agreement between the two codes regarding
black hole trajectories (left panel). This can be also seen looking at the right
panel, in which the two GW frequencies perfectly match. 
There is a discrepancy of 2\% between GW amplitudes that converges away with
increasing resolution.

\subsubsection{Convergence tests for \GRAthena}\label{ssec:calib_conv}

\begin{figure*}[t]
   \centering
     \includegraphics[width=0.48\textwidth]{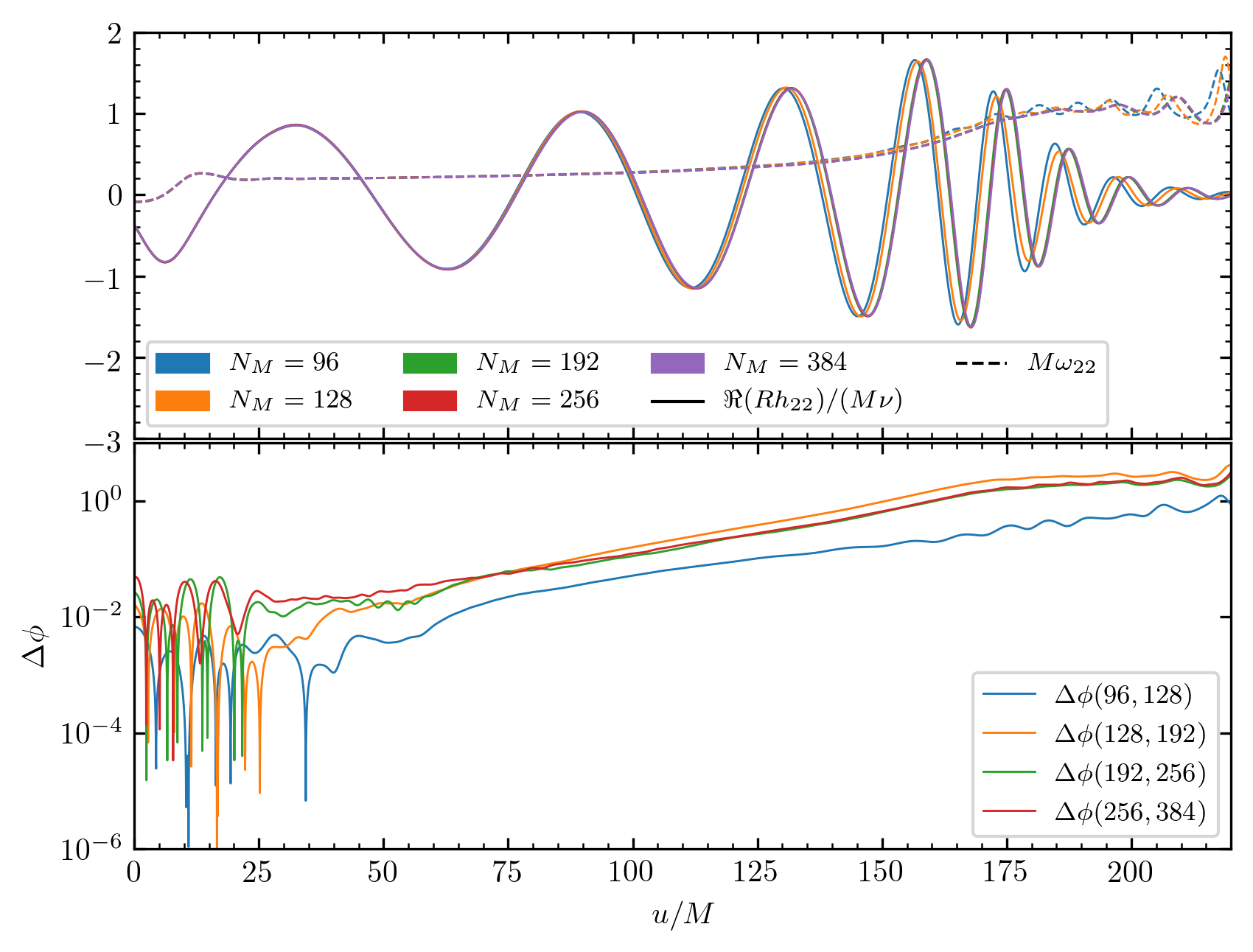}
     \includegraphics[width=0.48\textwidth]{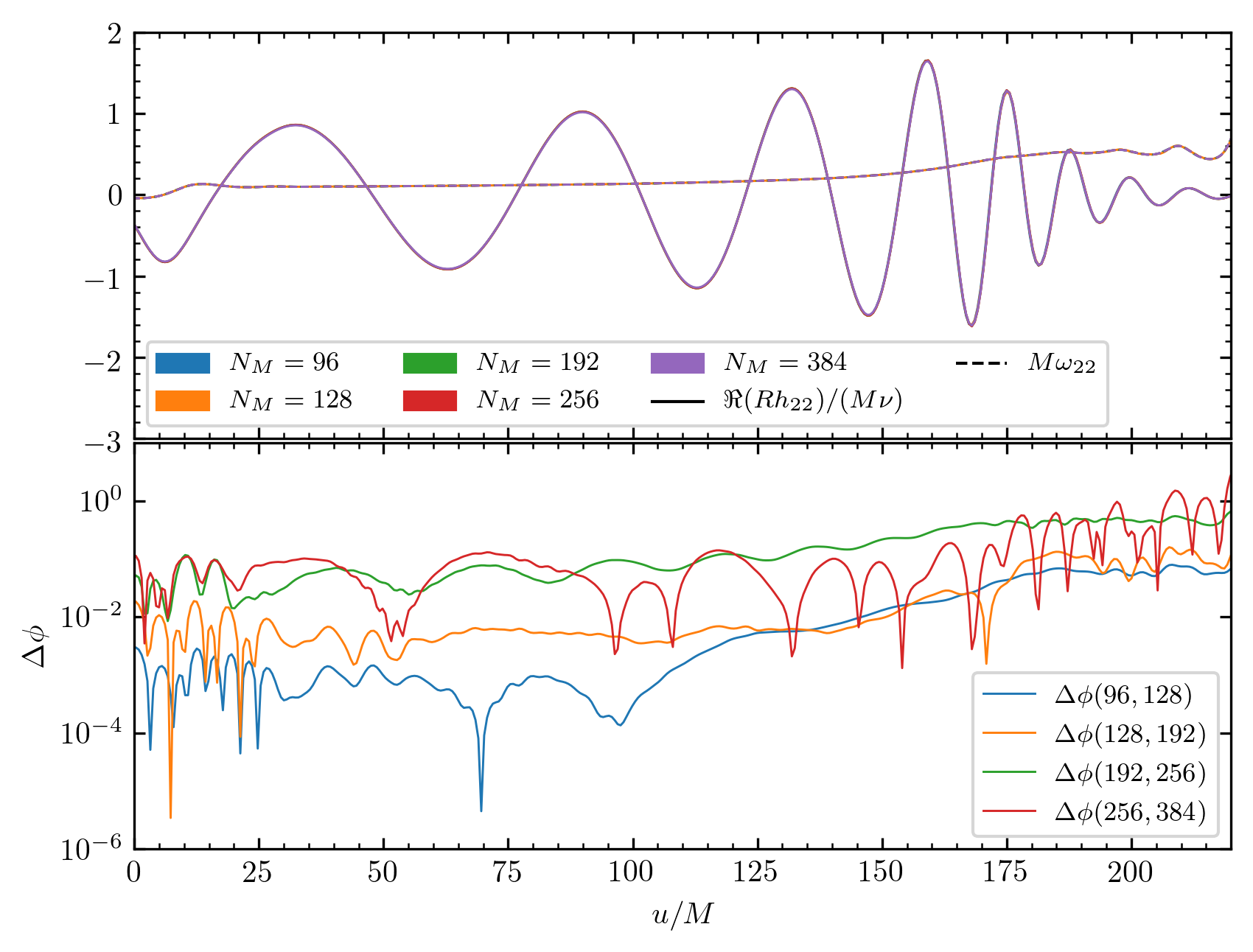}
     \caption{Convergence plots for calibration BBH evolution. Left and right
     plots correspond respectively to $\4th$ order and $\6th$ order finite
     differencing. In both case waves are extracted at $R=120 M$. In bottom panels
     phase differences between resolutions are rescaled according to
     Eq.\eqref{eq:conv_fact} with respect to the blue line (corresponding to
     lowest resolutions).
     }
  \label{fig:convergence_test}
\end{figure*}

Convergence tests are performed on the same runs as the previous section plus an
additional run
made at resolution $N_M = 384$, with $N_B = 24$ resulting in $\sp_p = 0.390625\times
10^{-2}~M$. Moreover, we consider another set of runs employing $\6th$ order finite
differences, with the same grid setup as in the previous section but halving the
maximum extent of
the physical grid, namely in this case $x_M = 1536~M$. This has the effect of
doubling the resolution at the puncture.
In the top panels of Fig.\ref{fig:convergence_test} we compare the gravitational
wave strain extracted at $R = 120~M$ for every resolution, both for $\4th$ and
$\6th$ order.

In order to quantitatively investigate the effect of resolution on phase error we
inspect differences in phase between runs in the bottom panels of
Fig.\ref{fig:convergence_test}.
Inverting Eq.\eqref{eq:strain_complex} allows us to write the phase difference as:
\begin{equation}
  |\Delta \phi(\alpha,\,\beta)|:=\left|
   \left.\phi[h{}_{22}]\right|_\alpha
  -\left.\phi[h{}_{22}]\right|_\beta
  \right|.
  \label{eq:h22_phase_diffs}
\end{equation}
In the bottom panel of the $\4th$ order plot (Fig.\ref{fig:convergence_test}
left) the red and green lines match, demonstrating $\4th$ order convergence for the
highest resolutions.
In the $\6th$ order case (Fig.\ref{fig:convergence_test} right) the
waveforms (and corresponding frequencies) lay on top of each other, so as to
be indistinguishable, and this translates into smaller phase differences comparing
the bottom panels of the two plots. Even though the red line, corresponding to
the phase difference between the two highest resolution, is quite noisy, $\6th$
order convergence can be seen here for the three highest resolutions as in
the previous case. This behavior is present
in every extraction radius. In all cases, the plots show a convergent behavior
with respect to
the phase differences, i.e., the differences between each pair of lines decreases
with increasing resolution. Additionally, we check the accuracy of our convergence
tests by evaluating the error on the phase differences,
estimated for each line as the difference between the phase given by the Richardson
extrapolation formula and the phase corresponding to the highest resolution used to
calculate the Richardson extrapolated phase,
similarly to what is done in~\cite{Bernuzzi:2016pie}.
For each corresponding phase difference line, we find that this error is always at
least $\sim 50\%$ smaller than $\Delta \phi$.

\begin{figure}[t]
   \centering
     \includegraphics[width=0.48\textwidth]{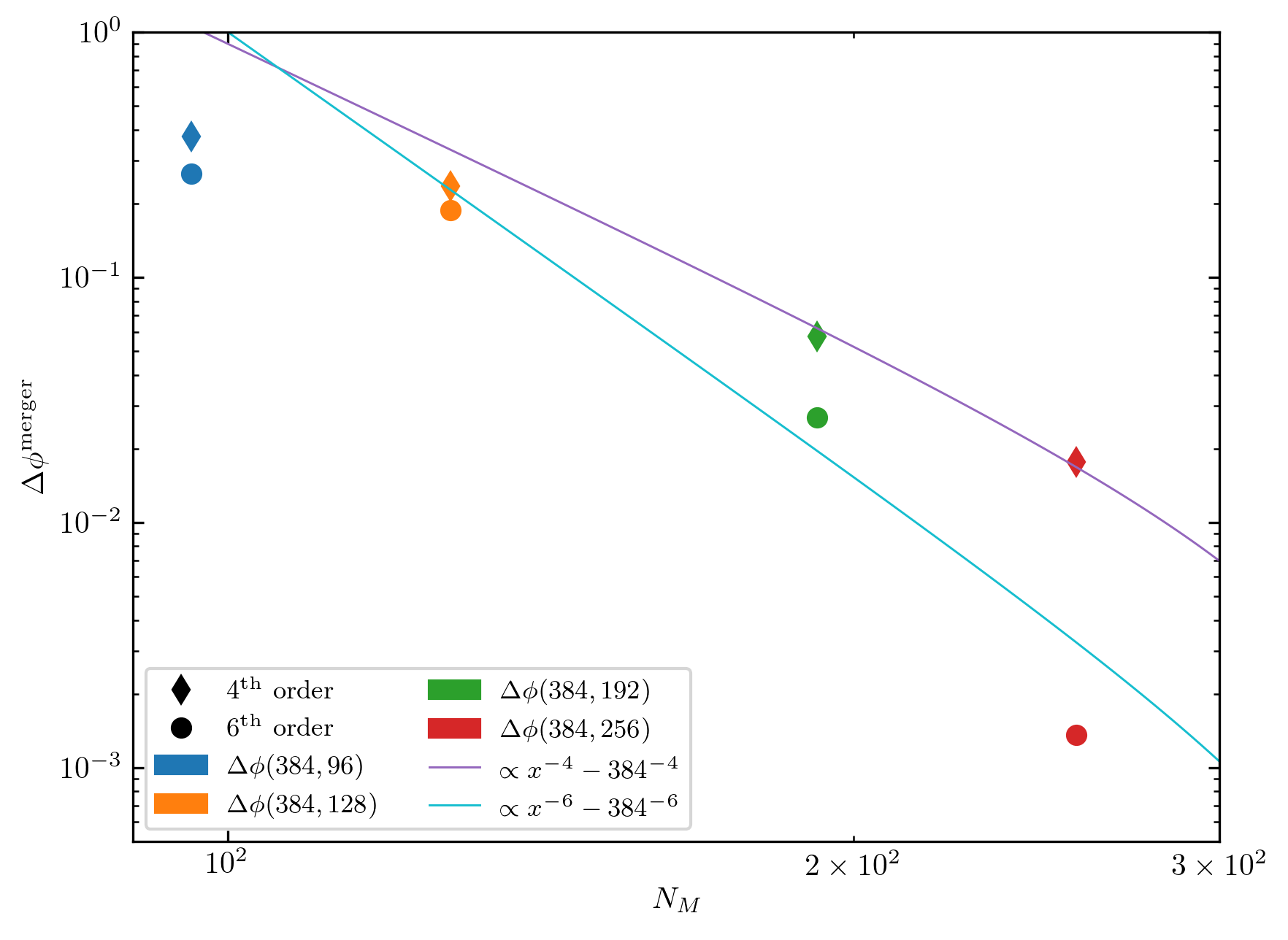}
     \caption{Self-convergence test for the calibration BBH evolution.
     		  In the plot phase difference at merger with respect to
              the highest resolution available is reported on $y$-axis, against
     		  the resolution $N_M$. Diamonds correspond to $\4th$ order series,
     		  while dots refer to $\6th$ order series.
     		  Purple and cyan lines represent the theoretical convergence
     		  for both cases.
              }
  \label{fig:conv_merger}
\end{figure}

Additionally, we show in Fig.\ref{fig:conv_merger} a convergence plot
in which phase differences are calculated at merger (see beginning of
\S\ref{sec:puncture_tests}).
This figure further confirms the clean $\4th$ order convergence of
$\GRAthena$ for the highest resolution cases. For the $\6th$ order
case, comparing with the other, phase differences are smaller
and the convergence is faster. However,
the noise in the phase differences with respect to the highest resolution
(right plot in Fig.\ref{fig:convergence_test}, bottom panel) makes the
convergence assessment less clean.

\subsection{Two punctures evolution of ten orbits}\label{ssec:twopuncture_ten_orbit}
Physically one anticipates that inspiral of astrophysical binary systems is
well-described by a significant duration of co-orbit on a
quasi-circular trajectory \cite{Peters:1963ux,Peters:1964zz}.
This assumption is consistent with the events detected by the
LIGO and Virgo
collaborations \cite{Abbott:2016blz,TheLIGOScientific:2016wfe,TheLIGOScientific:2017qsa}.
Consequently it is of considerable interest to also test the performance of \GRAthena{} in this
scenario.
To this end we evolve non-spinning, equal-mass, low eccentricity initial data based
on \cite{Hannam:2010ec} where bare-mass parameters are
$m_\pm = 0.488479\,M$ and the punctures are initially on-axis at
$x_\pm=\pm 6.10679\,M$ with instantaneous momenta
$\mathbf{p}_\pm=(\mp 5.10846\times 10^{-4},\,\pm 8.41746\times 10^{-2},\,0)\,M$.
This choice of parameters results in
${\sim}10$ orbits prior to merger at $t{\sim} 2145\,M$. In comparison to the calibration evolution
this evolution is of significantly longer duration and therefore it is of interest
to investigate how waveform accuracy is affected for a selection of \Mesh{} parameters that reduce
computational resource requirements.
In order to provide another comparison that is independent of \BAM{} here we
provide a final assessment on the quality of waveforms computed with \GRAthena{}
based on the NR informed, effective-one-body model {\tt TEOBResumS} \cite{Nagar:2018zoe}.

\subsubsection{Setup}\label{ssec:setup_disc}
The convergence studies performed for the calibration BBH merger problem provide a guide as to
how to choose resolution at the puncture $\sp_p$. Here we fix the
\MeshBlock{} sampling to $N_B=16$ and work at $\6th$ order in the spatial discretization. For the
\Mesh{} sampling we select $N_M = 64$ and construct a sequence of grid configurations
where each value of $\sp_p$ is reduced by a factor of $3/2$ compared to the previous
in Tab.\ref{tab:ERq1_gridtab}.

\begin{table}[ht!]
  \centering
  \setlength\tabcolsep{2pt}
  \begin{tabular}{| c || r | r || r || c |}
    \hline
    $\rho{}_{(\cdot)}$ & $N_L$ & $x_M$  & $\sp_p \times 10^{-2} [M]$ & $\#\mathrm{MB}$\\
    \hline
    $\mathrm{vvl}$     & $10$  & $768$  & $4.6875$       & $1072$ \\
    $\mathrm{vl}$      & $11$  & $1152$ & $3.515625$     & $1352$ \\
    $\mathrm{l}$       & $11$  & $768$  & $2.34375$      & $1184$ \\
    $\mathrm{ml}$      & $12$  & $1152$ & $1.757812$     & $1464$ \\
    $\mathrm{m}$       & $12$  & $768$  & $1.171875$     & $1296$ \\
    $\mathrm{mh}$      & $13$  & $1152$ & $0.878906$     & $1576$ \\
    $\mathrm{h}$       & $13$  & $768$  & $0.585938$     & $1744$ \\
    \hline
  \end{tabular}
  \caption{%
    A distinct label $\rho{}_{(\cdot)}$ is assigned to each run with corresponding
    maximum number of refinement levels $N_L$
    and fixed physical extent $x_M$ of the \Mesh{} (see \S\ref{sec:grid_conf}).
    Resultant puncture resolutions $\sp_p$ and total number of \MeshBlock{} objects
    initially partitioning a \Mesh{} are provided.
  }
  \label{tab:ERq1_gridtab}
\end{table}

That the choice of parameters in Tab.\ref{tab:ERq1_gridtab} reduce overall computational
resource requirements can be understood as follows.
Consider the choice of parameters made in $\rho{}_\mathrm{h}$ and suppose $N_M$ and
$N_L$ are varied while maintaining $\sp_p$ fixed. With this, the number of \MeshBlock{}
objects required to partition the initial \Mesh{} changes. For example, taking $N_M=128$ and
$N_L=12$ resulted in an initial number of \MeshBlock{} objects of $\#\mathrm{MB}=8352$.
Whereas selecting $N_M=256$ and $N_L=11$ leads to $\#\mathrm{MB}=58752$ initially.
Generally, we found an approximate cubic scaling in $\#\mathrm{MB}$ as $N_M$ is scaled which
is related to the dimensionality of the problem.

In this section an extraction radius
of $R=100\,M$ is used. The CFL condition is $0.25$ and a KO dissipation of $\sigma=0.5$ is
chosen. Constraint damping parameters are selected as $\kappa_1=0.02$ and $\kappa_2=0$.

\begin{figure}[t]
  \centering
  \includegraphics[width=0.95\columnwidth]{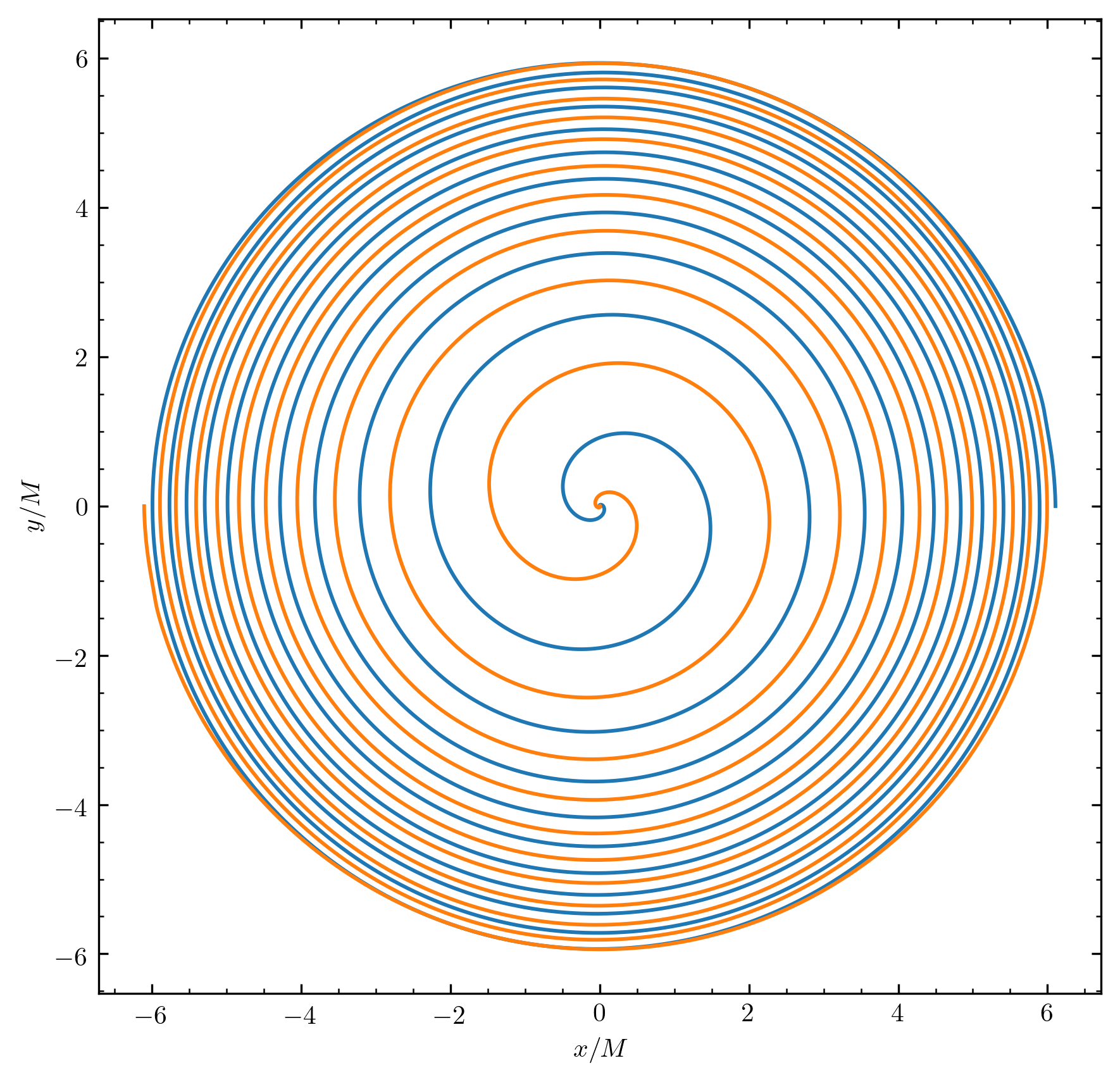}
  \caption{%
  Coordinate trajectories of both punctures ($\mathbf{x}{}_+(t)$ in blue and $\mathbf{x}{}_-(t)$ in orange)
  for parameter choice $\rho_{\mathrm{h}}$.
  }%
  \label{fig:q001tracker}%
\end{figure}

\begin{figure}[t]
  \centering
  \includegraphics[width=\columnwidth]{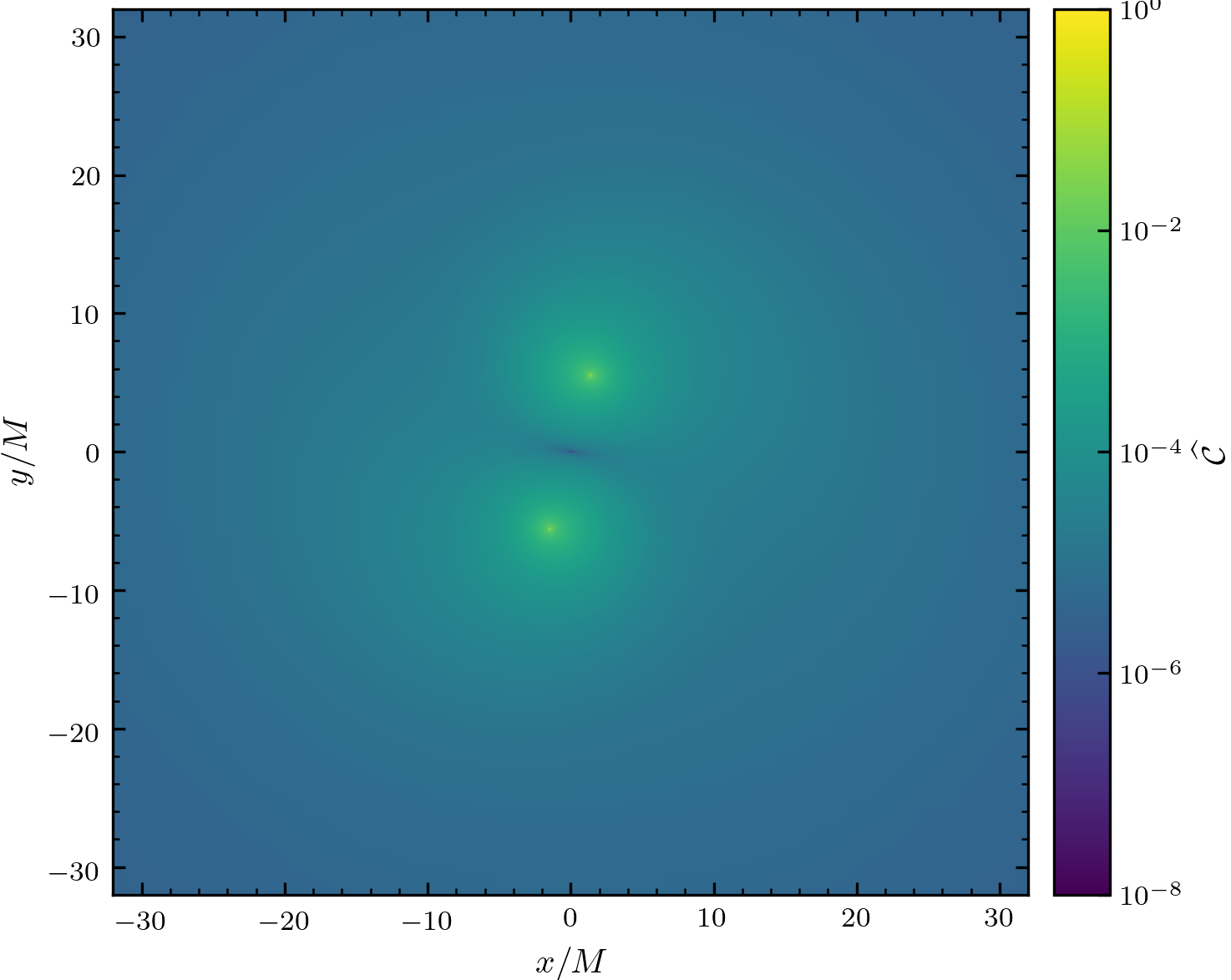}
  \includegraphics[width=\columnwidth]{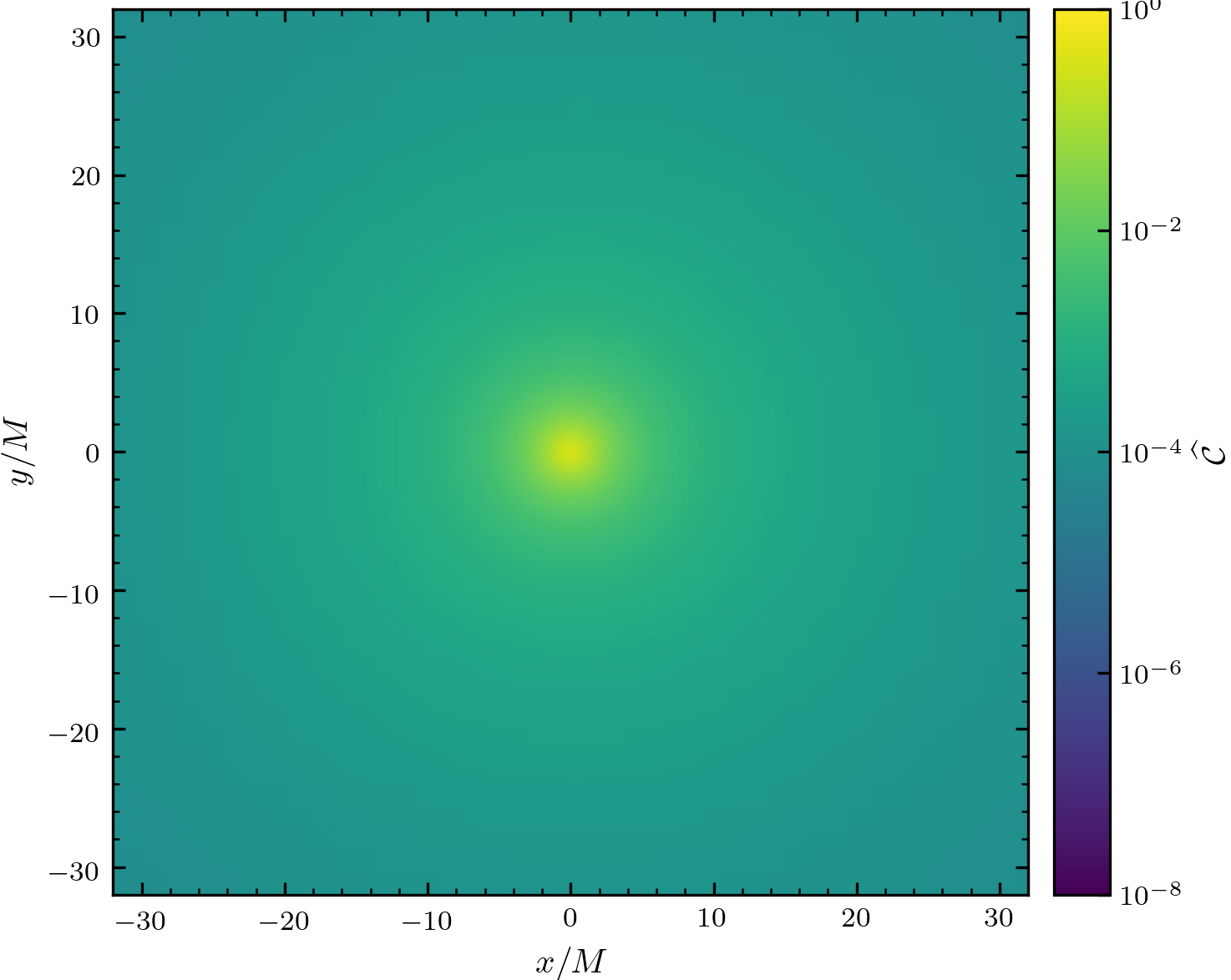}
  \caption{%
  Values of the (normalized) collective constraint
  $\widehat{\mathcal{C}}(x,\,y,\,z):=\mathcal{C}(x,\,y,\,0) / \max{}_{x,\,y}\mathcal{C}(x,\,y,\,0)$ over the
  orbital plane $z=0$ for a simulation with $\rho{}_{\mathrm{h}}$ of Tab.\ref{tab:ERq1_gridtab}.
  Upper panel: Evolution time is $t=500\,M$
  where $\max{}_{x,\,y}\,\mathcal{C}(x,\,y,\,0)\simeq 111.3$.
  Lower panel: Evolution time is $t=2100\,M$
  where $\max{}_{x,\,y}\,\mathcal{C}(x,\,y,\,0)\simeq 3.3$.
  As can be seen in both cases constraint violation is greatest directly in the vicinity of the
  punctures. See text for further discussion.
  }%
  \label{fig:q001colconstr}%
\end{figure}

The coordinate trajectories of the punctures for a calculation utilizing grid parameters
$\rho_{\mathrm{h}}$ of Tab.\ref{tab:ERq1_gridtab} can be seen to satisfy ten orbits
in Fig.\ref{fig:q001tracker}. This provides an initial verification of expected qualitative
properties \cite{Hannam:2010ec} of the BBH inspiral and merger.
In order to investigate the behavior of the constraints we focus attention on the collective
constraint $\mathcal{C}$ of Eq.\eqref{eq:collective_constraint}. We display
values of $\mathcal{C}$ in the orbital plane ($z=0$) at fixed times $t=500\,M$ and $t=2100\, M$ in
Fig.\ref{fig:q001colconstr}.
The general properties of $\mathcal{C}$ discussed here we found to be shared between other
simulations utilizing parameters from Tab.\ref{tab:ERq1_gridtab}. Crucially, this means that
increasing refinement in the vicinity of the puncture does not contaminate the rest of the
physical domain. In all cases we found that away from the punctures
values of $\mathcal{C}$ decrease on average as the boundary of the computational domain is
approached. In particular, for the calculation involving parameters $\rho{}_{\mathrm{h}}$ and
during $500\,M\leq t \leq 2200 \,M$
as $\varrho:=\sqrt{x^2+y^2}\rightarrow 100\,M$ we found
$\mathcal{C}\sim 10^{-8}$ which continues to decrease as $\varrho\rightarrow 300\,M$
to $\mathcal{C}\sim 10^{-10}$ thereafter plateauing at $\mathcal{C}\sim 10^{-11}$
towards the boundary. We found qualitatively similar behavior when inspecting the Hamiltonian
constraint. We remark that, even in the
continuum limit, constraints are not expected to converge to zero in the
entire domain for this solution because punctures are excluded from $\mathbb{R}^3$.

\begin{figure}[t]
  \centering
    \includegraphics[width=\columnwidth]{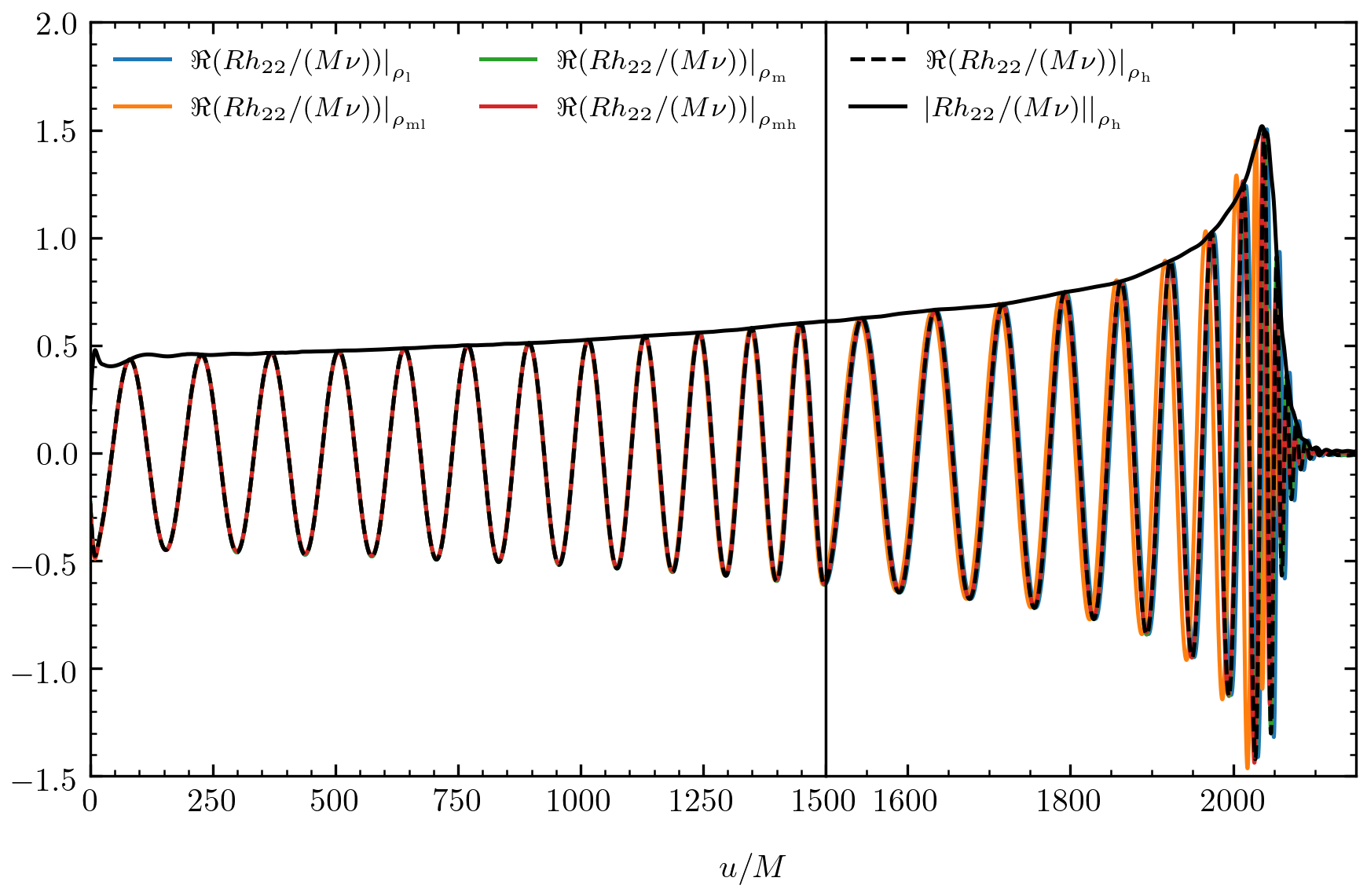}
    \caption{%
    The $(2,2)$ multipole of the GW strain normalized to the
    symmetric mass ratio $\nu=1/4$ computed for simulations based on parameters
    of Tab.\ref{tab:ERq1_gridtab}.
    Peak amplitude for a choice of $\rho_{\mathrm{h}}$
    occurs at $u/M=2037.5$ which indicates the end of the inspiral \cite{Bernuzzi:2014kca}.
    Dephasing as merger-time is approached reduces rapidly with increased resolution
    (see also Fig.\ref{fig:q001hphidiff} though note that the legend differs there).
    Note: horizontal axis scale
    changes at $u/M=1500$.
    }
    \label{fig:q001rRehcmp}
\end{figure}

Of principal interest for gravitational wave detection is the strain. To this end we solve
Eq.\eqref{eq:strain} for $h{}_{22}$ in the frequency domain making use of
the FFI method of \cite{Reisswig:2010di}. A cut-off frequency
of $f_0=1/300 \times 3/4$ is chosen which is physically motivated by inspecting
the early time puncture trajectories of the inspiral. We display the resulting
$(2,\,2)$ mode of the strain
for calculations using the parameters
of Tab.\ref{tab:ERq1_gridtab} in Fig.\ref{fig:q001rRehcmp}.

\begin{figure}[t]
  \centering
  \includegraphics[width=\columnwidth]{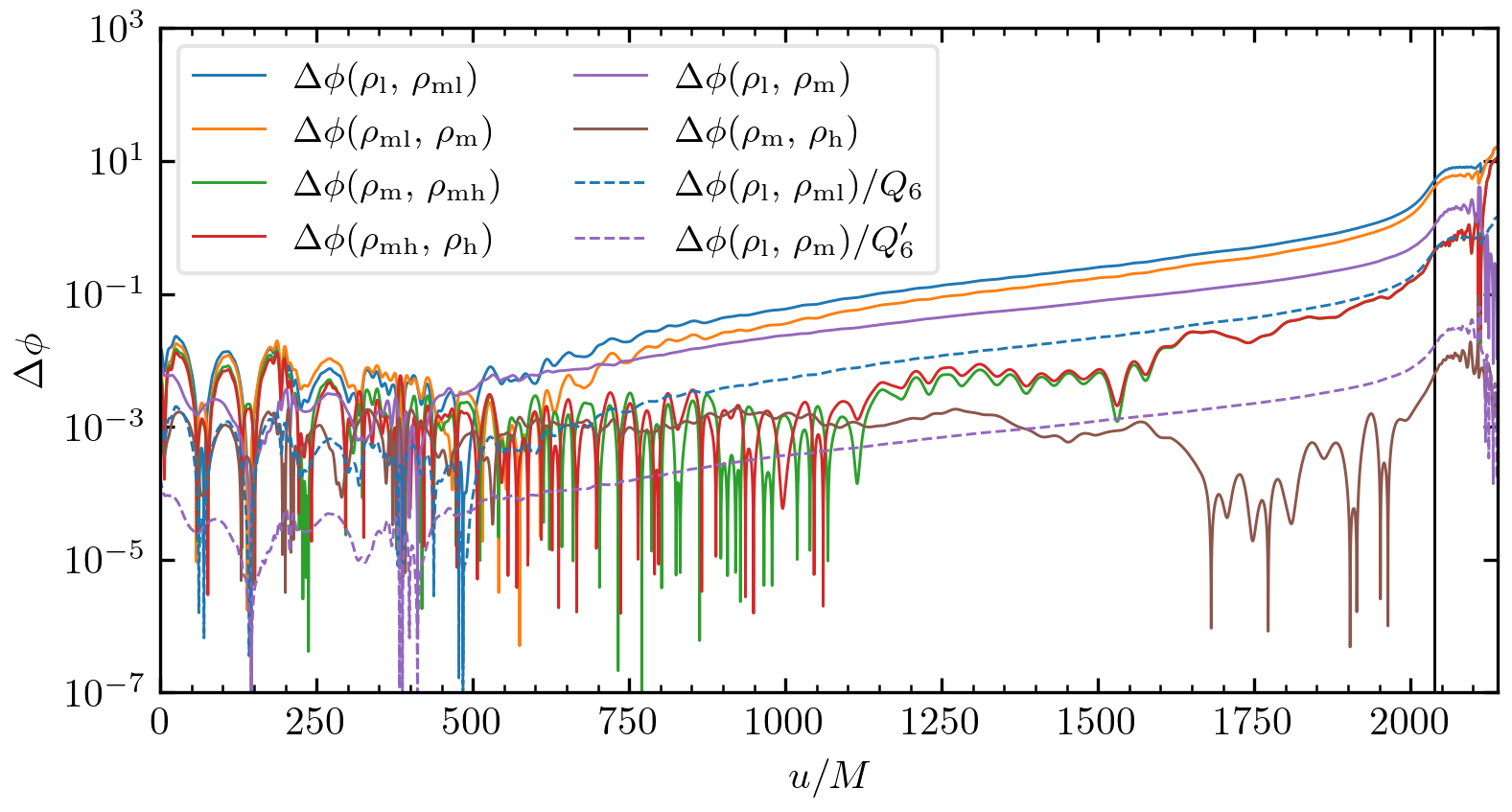}
  \includegraphics[width=\columnwidth]{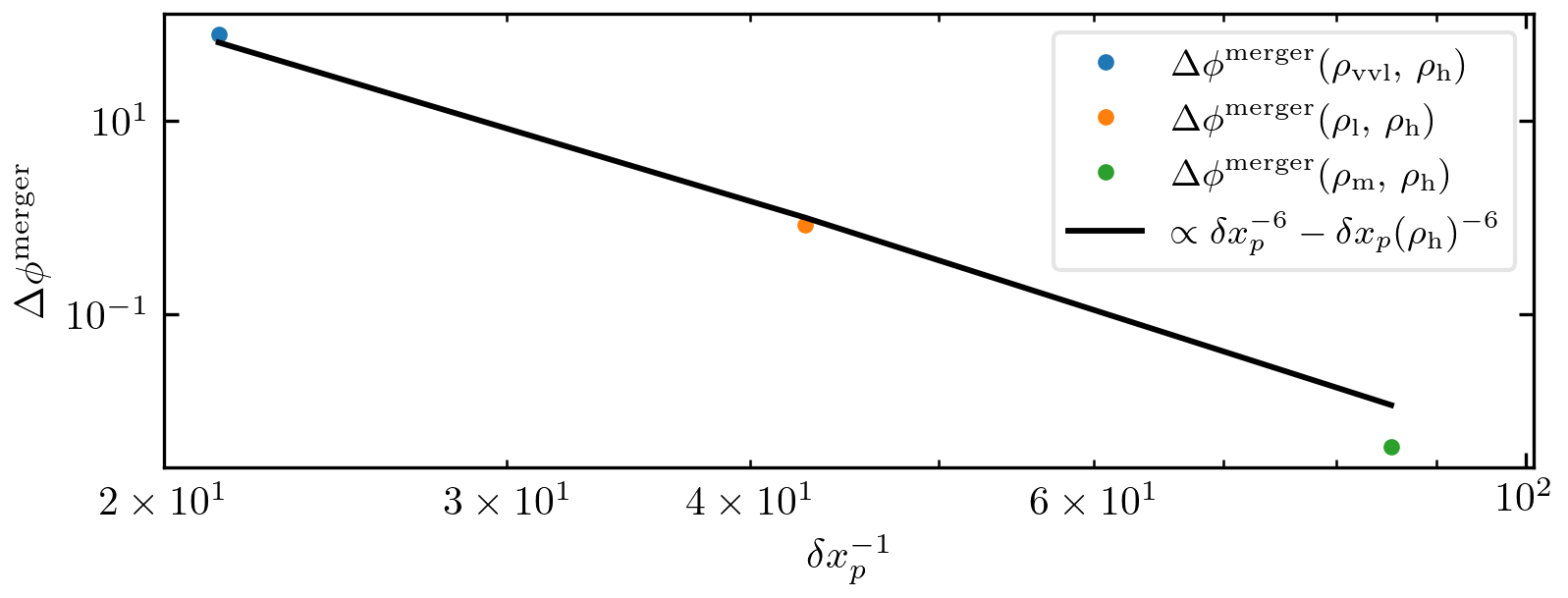}
  \caption{
  Phase differences $\Delta \phi$ between simulations involving parameters of
  Tab.\ref{tab:ERq1_gridtab}.
  Upper panel: A trend of $\Delta \phi$ accumulating with time is present. Merger
  time corresponding to $\rho{}_{\mathrm{h}}$ is indicated with a vertical black line
  at $u/M=2037.5$.
  A decrease in $\Delta \phi$ occurs as $\delta x_p$ pairs of decreasing values
  are compared. In order to mitigate a systematic effect of varied
  spatial extent in the computational domain we also compute phase differences at fixed $x_M$ such
  as $\Delta\phi(\rho{}_{\mathrm{l}},\,\rho{}_{\mathrm{m}})$
  and $\Delta\phi(\rho{}_{\mathrm{m}},\,\rho{}_{\mathrm{h}})$.
  These two differences are also shown rescaled with $Q_6$ and $Q_6'$ as computed using
  Eq.\eqref{eq:conv_fact} under the assumption of $\6th$ order spatial discretization.
  Lower panel:
  Phase differences at merger computed with reference data taken from the
  $\rho{}_{\mathrm{h}}$ run
  are depicted as a function of puncture
  resolution. Data on the black reference curve would obey a $\6th$ order convergence trend.
  See text for further discussion.
  }
  \label{fig:q001hphidiff}
\end{figure}

The peak amplitude of $h{}_{22}$ indicates the end of the inspiral \cite{Bernuzzi:2014kca}
and for a grid parameter choice of $\rho_{\mathrm{h}}$ occurs at $u=2037.5\,M$. The maximum
deviation from this value for parameters investigated in Fig.\ref{fig:q001rRehcmp}
occurs when $\rho_{\mathrm{ml}}$ is utilized resulting in
$\Delta u=10.3\,M$. In order to directly quantify how the choice of $\delta x_p$ affects the
phase error in the strain waveform as merger time is approached we compute $\Delta\phi$ using
Eq.\eqref{eq:h22_phase_diffs} and show the result in Fig.\ref{fig:q001hphidiff}.

As we have not modified resolution globally over the computational domain but rather considered the
effect of introducing additional refinement levels in the vicinity of the punctures it is not clear
what sort of convergence should be expected. Furthermore the extent to which a
time-integrator order below the order of the spatial discretization
affects GW waveform quality can also be somewhat
delicate (see e.g. the super-convergence discussion
of \cite{Reisswig:2010di}).
In Fig.\ref{fig:q001hphidiff}
(upper panel) clean $\6th$ order convergence in $\Delta \phi$ is not found for all $u$
upon rescaling with the appropriate factors determined through
Eq.\eqref{eq:conv_fact}. An
additional issue that complicates the discussion here
is that the parameters of Tab.\ref{tab:ERq1_gridtab} also vary the spatial extent of
the computational domain potentially introducing a source of systematic error. For example, at
merger $\Delta\phi(\rho{}_{\mathrm{mh}},\,\rho{}_{\mathrm{h}})\simeq4\times 10^{-1}$ and
$\Delta\phi(\rho{}_{\mathrm{m}},\,\rho{}_{\mathrm{h}})\simeq6\times 10^{-3}$ though
$\delta x_p(\rho{}_{\mathrm{m}}) > \delta x_p(\rho{}_{\mathrm{mh}})$. In order to compensate for
this effect we consider phase differences at fixed $x{}_M$. In particular the lower panel of
Fig.\ref{fig:q001hphidiff} displays phase differences at merger where $\rho{}_{\mathrm{h}}$ is
taken as the reference value for all comparisons. While displayed $\Delta \phi$ are compatible
with a $\6th$ order trend the $\rho{}_{\mathrm{vvl}}$ choice of parameters is likely of too low
resolution to make a robust claim on convergence order with respect to varying $\delta x{}_p$.
Nonetheless it is clear that judicious choice of refinement level (and hence
resolution local to the punctures through $\delta x{}_p$)
reduces GW phase error.

\subsubsection{EOB comparison}\label{sssec:eob_cmp}
We compare the gravitational waveform from the $10$ orbit
simulation to the state-of-the-art EOB model {\tt TEOBResumS} \cite{Nagar:2018zoe}.
The latter is informed by several existing NR datasets and faithfully models
the two-body dynamics and radiation of spin-aligned BBH multipolar
waveforms for a wide variety of mass ratio and spins magnitudes. We focus again on the
$(2,2)$ mode of the gravitational wave strain.
The \GRAthena $\psi_{22}$ output mode is first extrapolated to
null infinity using the asymptotic extrapolation formula \cite{Lousto:2010qx,Nakano_2015}:
\begin{align}\label{eq:extrap_wvf}
  \lim_{r\rightarrow\infty} r \psi_{22} \sim
  A\Big(r\psi_{22}
  - \frac{(l-1)(l+2)}{2r}\int r \psi_{22} \,\mathrm{d}t\Big),
\end{align}
where $A(r):=1-2M/r$ and $r:=R(1+M/(2R))^2$ with $R$ being the (finite) GW extraction radius
of an NR simulation.
The extrapolated result
is successively integrated twice in time (Eq.\eqref{eq:strain}) using the
FFI method \cite{Reisswig:2010di} to obtain the strain mode $h_{22}$.
The waveform comparison is performed by suitably aligning the two
waveforms; the time and phase relative shifts are determined by minimizing the $L^2$ norm
of the phase differences \cite{Bernuzzi:2011aq}.

\begin{figure*}[t]
    \centering
     \includegraphics[width=0.9\textwidth]{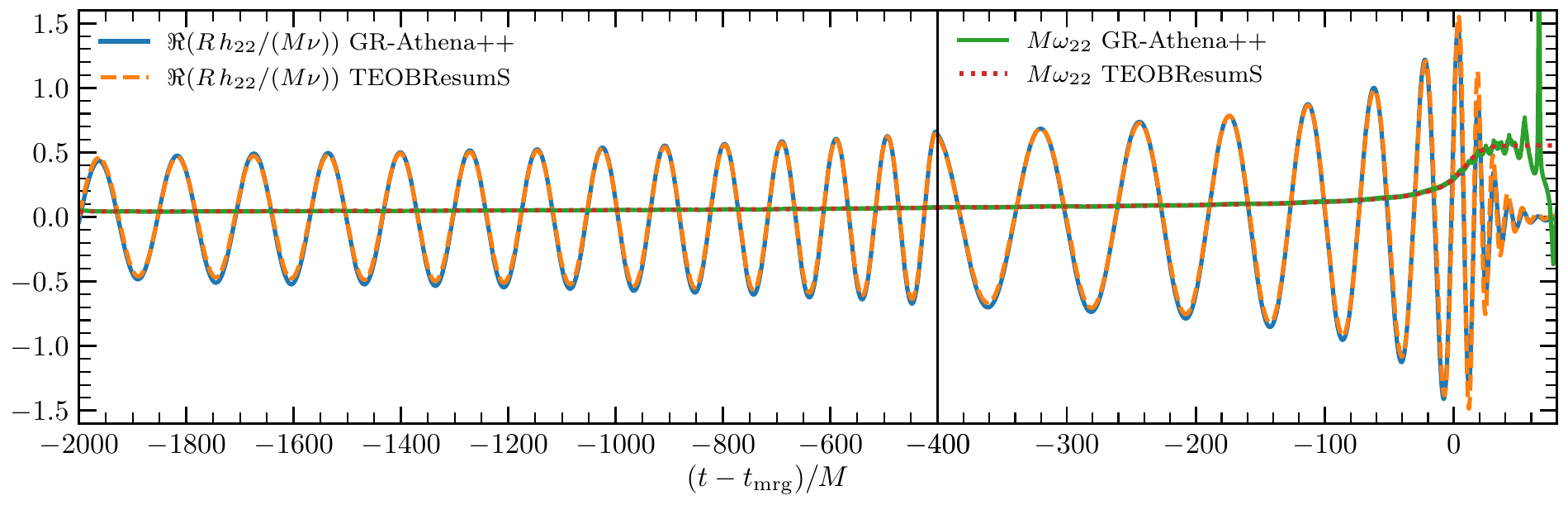}
     \caption{Comparison between the \GRAthena BBH $q=1$ waveform and the
       semi-analytical effective-one-body model {\tt TEOBResumS}. The
       plot shows the $(2,2)$ multipole of the GW strain normalized to the
       symmetric mass ratio $\nu=1/4$ and the instantaneous GW
       frequency. The time is shifted to the mode amplitude peak that
       approximately defines the merger time. The \GRAthena waveform is
       from the highest resolution simulation ($\rho_{\mathrm{h}}$
       of Tab.\ref{tab:ERq1_gridtab}), extracted at
       coordinate radius $R=100\,M$ and extrapolated to
       null infinity using Eq.\eqref{eq:extrap_wvf}.
       Note: horizontal axis scale
       changes at $(t-t{}_{\mathrm{mgr}})/M=-400$.
    }
\label{fig:eobnr}
\end{figure*}

Figure~\ref{fig:eobnr} shows the two waveforms are compatible within
the NR errors. The accumulated EOBNR phase differences
are of order ${\simeq}0.1\,$rad to merger and ${\simeq}0.4\,$rad
to the ring-down for the highest resolution \GRAthena simulation.
The larger inaccuracy of the ring-down part is a resolution effect
related to the higher frequency of the wave; it can potentially be improved by
adding a refinement level so as to better resolve the black hole remnant.
The maximum relative amplitude difference is of order
${\simeq}{0.01}$.
The same comparison using the lowest resolution gives
${\simeq}{0.4}\,$rad at merger (${\simeq}{1}\,$rad during the
ring-down) and maximum relative amplitude differences of ${\simeq}{0.01}$.
Overall, this analysis demonstrates that \GRAthena can produce high-quality data
for waveform modeling.

\section{Scaling tests}\label{sec:scaling}
To check the performance of \GRAthena{} and make sure it maintains the
scalability properties of \Athena,
we conduct weak and strong scaling tests with the same problem setup as in \S
\ref{ssec:calibration}.
In these tests BBH evolutions of 20 Runge-Kutta time-steps are performed, with
full AMR and full production grids, in which $N_B = 16$, $N_L = 11$,
$x_M = 1536~M$ are fixed and making use of hybrid MPI and OMP parallelization.
These tests are performed on the cluster SuperMUC-NG at LRZ.
Specifically, on each node of the cluster (48 CPUs per node)
8 MPI tasks with 6 OMP threads are launched, thus filling up the node.
We find very good results using up to 2048 nodes ($\sim 10^5$)
CPUs. With respect to scaling tests presented in \Dendrogr~\cite{Fernando:2018mov}, 
our results favorably compare both in the case of strong and weak scaling tests.

\subsection{Strong scaling tests}\label{ssec:strong_scaling}
\begin{figure*}[t]
   \centering
     \includegraphics[width=0.48\textwidth]{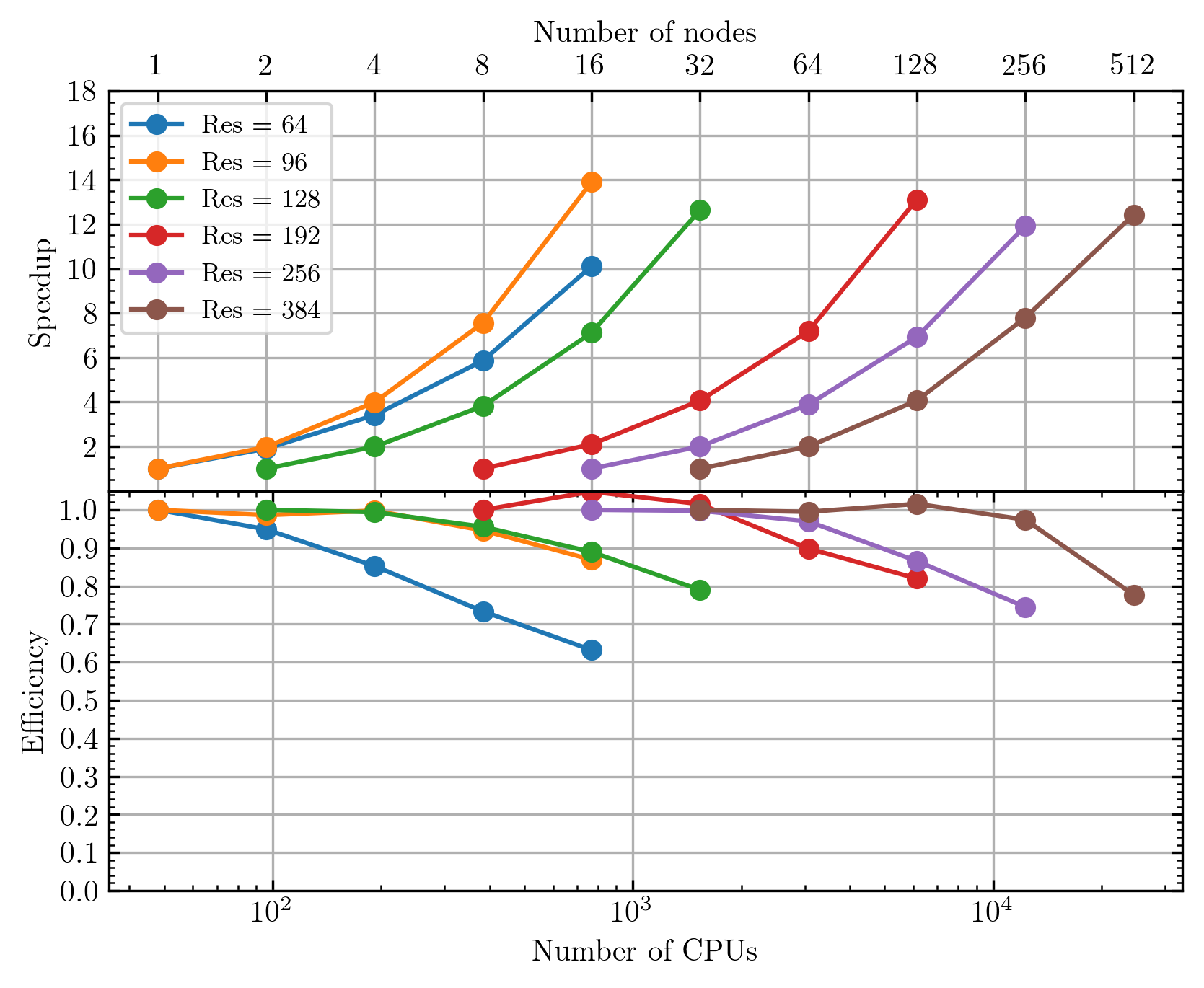}
     \includegraphics[width=0.48\textwidth]{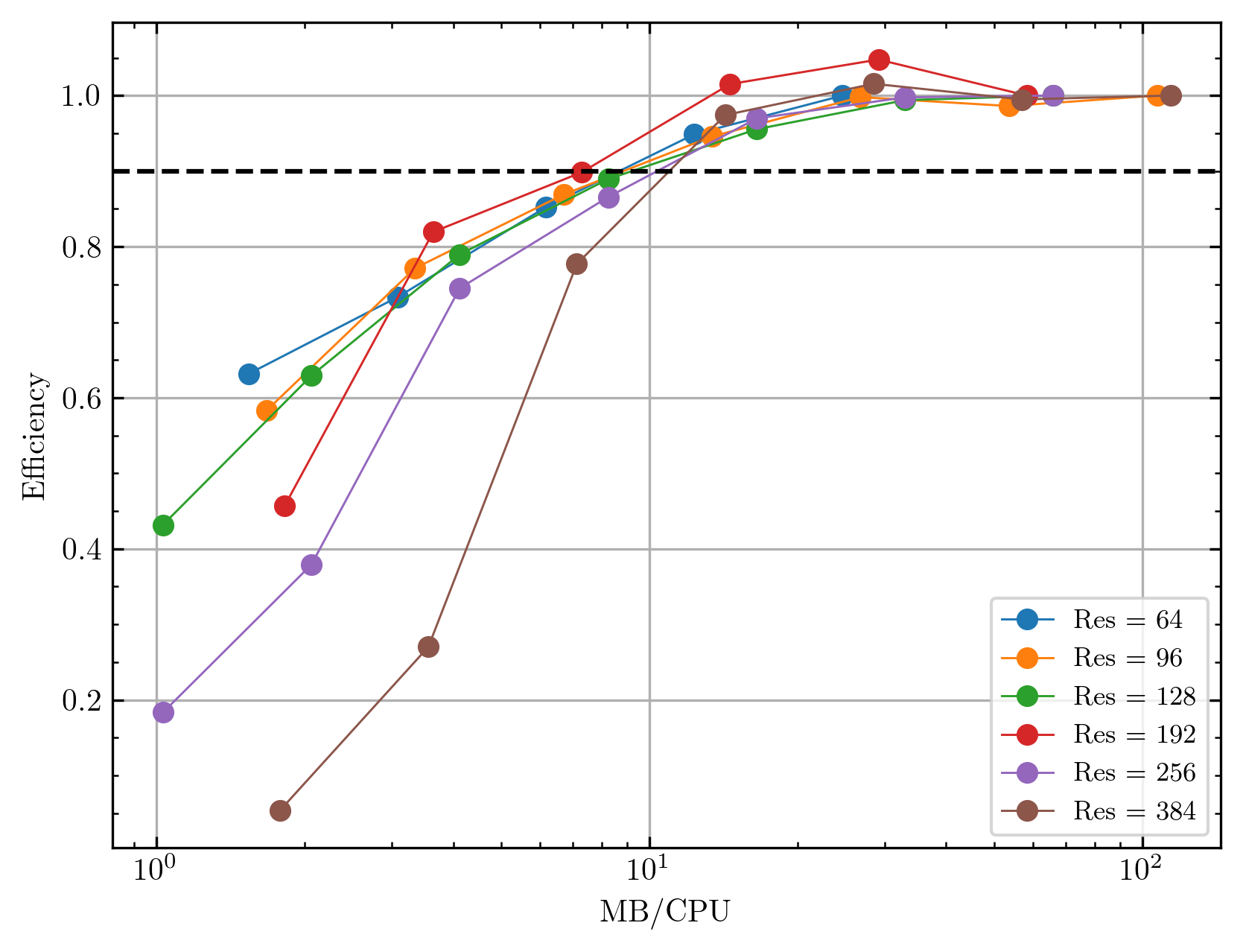}
     \caption{Strong scaling tests for \GRAthena for several CPU number regimes.
     Left plot: speed-up (top panel) and efficiency (bottom panel) calculated
     with respect to the first point of the series. In each series the second point corresponds
     to a theoretical speed-up by a factor of 2, the third point a factor of 4 and so on.
     Right plot: efficiency as a function of the \MeshBlock{} load each CPU carries.
     }
  \label{fig:strong_scaling}
 \end{figure*}

To test the strong scaling behavior of the code in different regimes of
CPU numbers, we consider 6 resolutions, namely
$N_M = [64,\,96,\,128,\,192,\,256,\,384]$. For each of them
we perform an evolution on $i = N_\text{min}^\text{nodes},\dots,
N_\text{max}^\text{nodes}$ nodes,
where $N_\text{min/max}^\text{nodes}$ represent
some limits on the possible number of nodes that can be used for each resolution.
The presence of these boundaries is due to the fact that for a certain resolution a
specific number of \MeshBlock{}s is produced and consequently on the one hand a
sufficient number of CPUs
is required to
handle those \MeshBlock{}s and, on the other hand, \Athena's parallelization strategy
is based on \MeshBlock{}s and so is not possible to use more OMP threads (and
consequently CPUs) than \MeshBlock{}s. Fig.\ref{fig:strong_scaling} (left) shows
that excellent strong scalability
is obtained up to $\sim1.5\times10^4$ CPUs, with efficiency above $90\%$. For the
aforementioned reasons, for a given resolution it is not possible to achieve high
efficiency when increasing CPU number above a certain point.
Fig.\ref{fig:strong_scaling} (right) shows that efficiency strongly
depends on the ratio between \MeshBlock{} number and CPUs. In particular, for each
resolution, an efficiency of above $90\%$ is obtained when there are at least 10
\MeshBlock{}s/CPU. By contrast, the efficiency shown in the strong scaling 
plot of~\cite{Fernando:2018mov} appears to decrease faster when comparing
to the brown line in Fig.\ref{fig:strong_scaling} (left), although the two
results are obtained in slightly different regimes of CPU number.

\subsection{Weak scaling tests}\label{ssec:weak_scaling}
\begin{figure*}[t]
   \centering
     \includegraphics[width=0.98\textwidth]{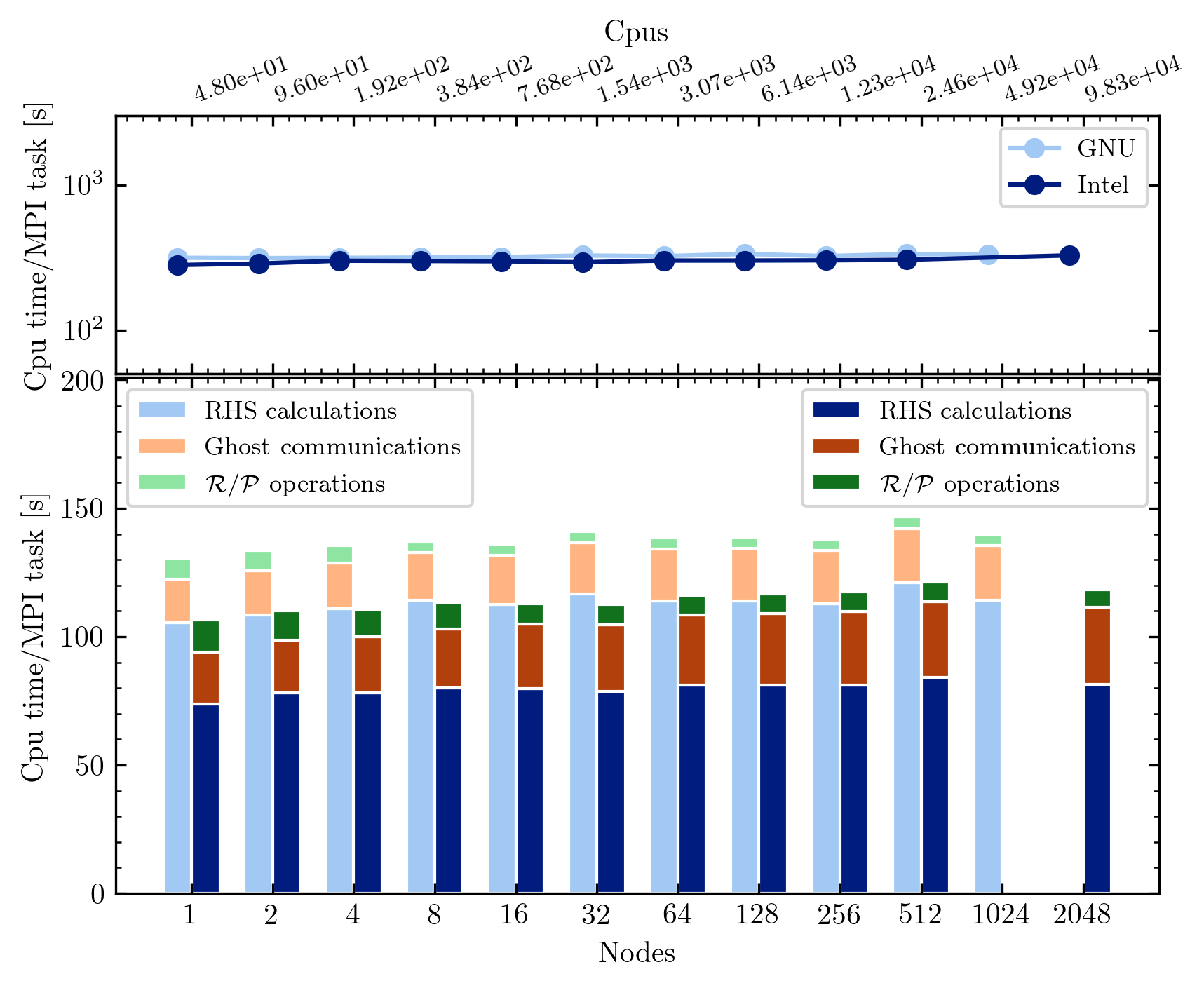}
     \caption{Weak scaling tests for \Athena, performed on SuperMUC-NG at LRZ compiling the code
     		  with GNU compiler and Intel compiler.
     		  Top panel reports total CPU times measured for rank 0 directly in
     		  the code with the \Cpp function $\texttt{clock\textunderscore t}$.
     		  CPU time in the bottom panel is measured instead using the profiling
     		   tool
			  \texttt{gprof}; here, times which count less than 2\% of the total
			  CPU time are neglected. Light (left) bars represent results obtained
			  with GNU compiler,
			  while dark (right) ones are for Intel compiler.
     		  }
  \label{fig:weak_scaling}
 \end{figure*}

For weak scaling tests we use
asymmetric grids in terms of $N_M$ in each direction, while keeping \MeshBlock{}s
of a constant size $N_B=16$. In particular,
we start with a run on a single node with a grid $N_{M}^x = 128,~N_M^y = 64,
~N_M^z = 64$. Then for 2 nodes we double $N_M^y$. For 4 nodes $N_M^z$ is doubled
as well and for 8 nodes $N_M^x$ is also doubled. We continue this up to 2048 nodes.
In this way, we are able to
double the resources together with the required computations, and we manage
to keep a ratio
of $\sim 33~\MeshBlock{}$s/CPUs in each different run.
The results are displayed in Fig.\ref{fig:weak_scaling}. We performed these tests
twice, once with the code compiled using the Intel compiler and once with the code 
compiled using the GNU compiler.
The top panel of Fig.\ref{fig:weak_scaling} shows that the
total CPU time per MPI task remains constant up to $~10^5$ CPUs employed, thus
demonstrating excellent scalability in an unprecedented CPU number regime
for a numerical relativity code.
The bottom panel displays the same result, but focuses on how the execution time is
distributed between the main computational kernels.
Notably, most of the computation time is spent in the calculation of the right
hand side of the equations, which is indeed the expected behavior in absence of race
conditions or other bottlenecks elsewhere in the code.
The discrepancy between the height of each bar in the bottom panel of the plot
 and the dotted line in the top panel give an estimate on the communication time 
 among all MPI processes and OMP threads
and the comparison between the two plots suggests that this also has good scaling
behavior.

\section{Summary and conclusion}\label{sec:final_word}
In this work we have presented \GRAthena{}; a
vertex-centered extension of the
block-based AMR framework of \Athena{} for numerical relativity (NR) calculations.

To this end we described our introduction of vertex-centered (VC)
discretization for field variables which may be utilized for general problems.
A principle advantage of VC is that restriction of
sampled function data from fine to coarse grids that are interspersed and the coarser of which
has fully coincident grid-points is efficiently achieved by copying of data. This procedure is
formally exact. The dual operation of prolongation via interpolation of data from coarse to finer
grids also takes advantage of the aforementioned grid structure where possible.

Another novel feature is our introduction of geodesic spheres in the sense of highly refined,
triangulated convex, spherical polyhedra. Placement on a \Mesh{} may be arbitrarily chosen
without restriction on underlying coordinatization. An advantage of this is that the associated
vertices defining a geodesic sphere achieve a more uniform spatial sampling distribution when
contrasted against traditional uniform spherical latitude-longitude grids at comparable
resolution. Furthermore, the potential introduction of coordinate singularities is avoided.
We demonstrated the utility of this approach when quantities need to be extracted based on
spherical quadratures.

Our code implements the \z4c{} formulation of NR of the $\chi$ moving-punctures variety.
The overall spatial order of the scheme may be selected at compile time where easily extensible
\Cpp templates define the formal order for desired finite difference operators and VC
restriction and prolongation operators. Time-evolution is performed with a
low-storage $\4th$ order Runge-Kutta method.
Our primary analysis tool is based on extraction of gravitational
radiation waveforms within \GRAthena{} through the use of the $\Psi_4$ Weyl scalar
with numerical quadrature thereof based on the aforementioned geodesic spheres. Through a
post-processing step we also analyze the gravitational strain $h$.

In order to assess the numerical properties of our implementation we have repeated a subset of the
standard Apples with Apples test-bed suite
\cite{Alcubierre:2003pc,Babiuc:2007vr,Cao:2011fu,Daverio:2018tjf} and performed a variety of
cross-code validation tests against \BAM{}. To accomplish the latter within the
oct-tree AMR approach
of \GRAthena{} we constructed a refinement criterion so as to closely emulate a nested
box-in-box grid structure.
Test problems in the cross-code comparison involved a single spinning puncture and the two puncture
binary black hole (BBH) inspiral calibration problem of \cite{Brugmann:2008zz}.
For overall $\4th$ order spatial scheme selection a commensurate $\4th$ order
convergence was cleanly observed. For $\6th$ order, convergence was also achieved for the
same problem, though less cleanly.
As another external validation of \GRAthena{} and a demonstration of utility for
potential future studies of high mass ratio BBH that involve significant evolution duration and
where accurate phase extraction from the gravitational strain is crucial, we investigated a
quasi-circular ten orbit inspiral problem based on parameters from \cite{Hannam:2010ec}. Here
the evolution was performed with $\6th$ order spatial discretization and comparison was made
against the NR informed EOB model of {\tt TEOBResumS} \cite{Nagar:2018zoe}.
Using a \Mesh{} tailored to moderate computational resources we found accumulated
EOBNR phase differences of order $\simeq 0.1$rad to merger and $\simeq0.4$rad to the ring-down
for the highest resolution calculation.
These tests highlight that \GRAthena{} is accurate and robust for BBH inspiral
calculations and a strong contender for construction of high-quality data for waveform modeling.

In addition to accuracy, efficiency of computational resource utilization is crucial.
The task-based computational model for distribution of calculations within
 \Athena{} and concomitant
impressive scalability properties as available resources are increased we have found to
readily extend to \GRAthena{} and the \z4c{} system with VC discretization for a
wide variety of problem specifications. Indeed during strong scaling tests it was
found that efficiency above $95\%$ is reached up to $\sim 1.2\times10^4$ CPUs,
whereas in weak scaling tests almost perfect scaling is achieved up to
$\sim 10^5$ CPUs.
This indicates that for the high resolutions and consequently resources required for a
potential calculation describing an intermediate mass ratio BBH inspiral \GRAthena{}
compares favorably with the code-base of \Dendrogr \cite{Fernando:2018mov}
in terms of scalability.

Finally, as \GRAthena{} builds upon the modular framework of \Athena{} we inherit all its extant
infrastructure which will enable incorporating an NR treatment of the matter sector in
future work. It is our intention to make \GRAthena{} code developments publicly available
in future in coordination with the \Athena{} team.

\acknowledgments
The authors thank Bernd Br\"ugmann, Alessandro Nagar, and Jim Stone for discussions,
and Nestor Ortiz for initial work on \AwA{} tests.
B.D., F.Z. and S.B. acknowledge support by the European Union's
H2020 under ERC Starting Grant, grant
agreement no. BinGraSp-714626.
D.R. acknowledges support from the U.S. Department of Energy, Office
of Science, Division of Nuclear Physics under Award Number(s)
DE-SC0021177 and from the National Science Foundation under Grant No.
PHY-2011725.
Computations were performed on the ARA cluster at Friedrich
Schiller University Jena, on the supercomputer SuperMUC-NG at the
Leibniz-Rechenzentrum (LRZ, \url{www.lrz.de}) Munich, and on the
HPE Apollo Hawk at the High Performance Computing
Center Stuttgart (HLRS).
The ARA cluster is funded in part by DFG grants INST
275/334-1 FUGG and INST 275/363-1 FUGG, and ERC Starting Grant, grant
agreement no. BinGraSp-714626.
The authors acknowledge the Gauss Centre for Supercomputing
e.V. (\url{www.gauss-centre.eu}) for funding this project by providing
computing time on the GCS Supercomputer SuperMUC-NG at LRZ
(allocations {\tt pn56zo}, {\tt pn68wi} and {\tt pn98bu}).
The authors acknowledge HLRS for funding this project by providing
test account access on the supercomputer HPE Apollo Hawk under the grant
number {\tt ACID 44191, GRAthenaBBH}.

\appendix

\section{Apples with Apples Test-beds}\label{sec:awa_tests}
In order to provide a series of computationally inexpensive and standard tests of differing
formulations in numerical relativity (tailored for the vacuum sector) a suite of
so-called ``Apples with Apples'' test-bed problems (hereafter \AwA{}) have been
proposed \cite{Alcubierre:2003pc,Babiuc:2007vr}.

Our goal here is to ensure that the implementation of \z4c{} within \GRAthena{} reflects the
overall properties observed in prior tests made based on the same formulation
\cite{Cao:2011fu,Daverio:2018tjf} with a particular focus on elements directly relevant to
gravitational wave propagation and extraction. Specifically, we examine the AwA{}:
robust stability \S\ref{ssec:awa_robust_stab},
linearized wave \S\ref{ssec:awa_lin_wave}
and gauge-wave \S\ref{ssec:awa_gauge_wave} tests.

Generically the \AwA{} tests are specified for three dimensional periodic spatial grids
(i.e. of $\mathbb{T}^3$ topology) where the
effective dynamics occur over one (or two) spatial dimensions. Dynamics in ``trivial'' directions
are reduced to a thin layer which in \GRAthena{} is achieved by
selecting the relevant component(s) of $N_M$ of the \Mesh{} sampling to be equal to $4$ (selected
chosen as to probe potential ``checker-board instability'' \cite{Babiuc:2007vr}). This entails the
full \z4c{} equations and implementation of (\S\ref{ssec:eqnformgrathena}) are active during a
calculation.
Directions involving non-trivial dynamics are sampled and tested on both
cell-centered $\mathcal{G}_{\mathrm{CC}}$ and vertex-centered $\mathcal{G}_{\mathrm{VC}}$
grids:
\begin{gather}\label{eq:awagrids}
\begin{aligned}
  \mathcal{G}_{\mathrm{CC}} :=& \Big\{-\frac{1}{2} + \Big(n + \frac{1}{2}\Big) \sp\,\Big|\,
  n \in \{0,\,\dots N-1\}\Big\},\\
  \mathcal{G}_{\mathrm{VC}} :=& \Big\{-\frac{1}{2} + n \sp\,\Big|\,
    n \in \{0,\,\dots N\}\Big\};
\end{aligned}
\end{gather}
where $\sp =1/N$ and
$N=50\rho$ with $\rho\in\mathbb{N}$ serving to adjust resolution as required.
Each direction is further extended by ghost-zones as described in \S\ref{sec:method}.
No refinement is present in this section though \MeshBlock{} objects have
$N_B$ chosen so as to partition the domain and allow for a further consistency check on the
MPI-OMP hybrid parallelism. Selection of \z4c{} parameters is as follows:
Constraint damping $\kappa{}_1=0.02$ and $\kappa{}_2=0$ with shift-damping typically selected
as $\eta=2$.
Kreiss-Oliger (KO) dissipation is taken as $\sigma=0.02$ and the Courant-Friedrich-Lewy (CFL)
condition as $1/2$.
These choices are motivated by those made in prior work \cite{Cao:2011fu}
and we comment on behavior upon deviation from them as tests are presented.

For the tests performed here \z4c{} is always coupled to the puncture gauge
described in \S\ref{ssec:gauge_bc}
as this is of primary interest in this work. We take initial gauge conditions to be:
\begin{align*}
  \left.\alpha\right|_{t=0} &= 1, &
  \left.\beta{}^i\right|_{t=0} &= 0.
\end{align*}
Furthermore, in order to facilitate comparison we set
$\Ng=2$ throughout such that spatial discretization is of $\2nd$ order, though
numerical experiments with $\Ng=3$ reveal similar properties. Time-evolution is
performed using the $\4th$ order RK$4()4[2S]$ method of \cite{ketcheson2010rungekuttamethods}.

\subsection{Robust stability}\label{ssec:awa_robust_stab}
\begin{figure}[t]
  \centering
  \includegraphics[width=\columnwidth]{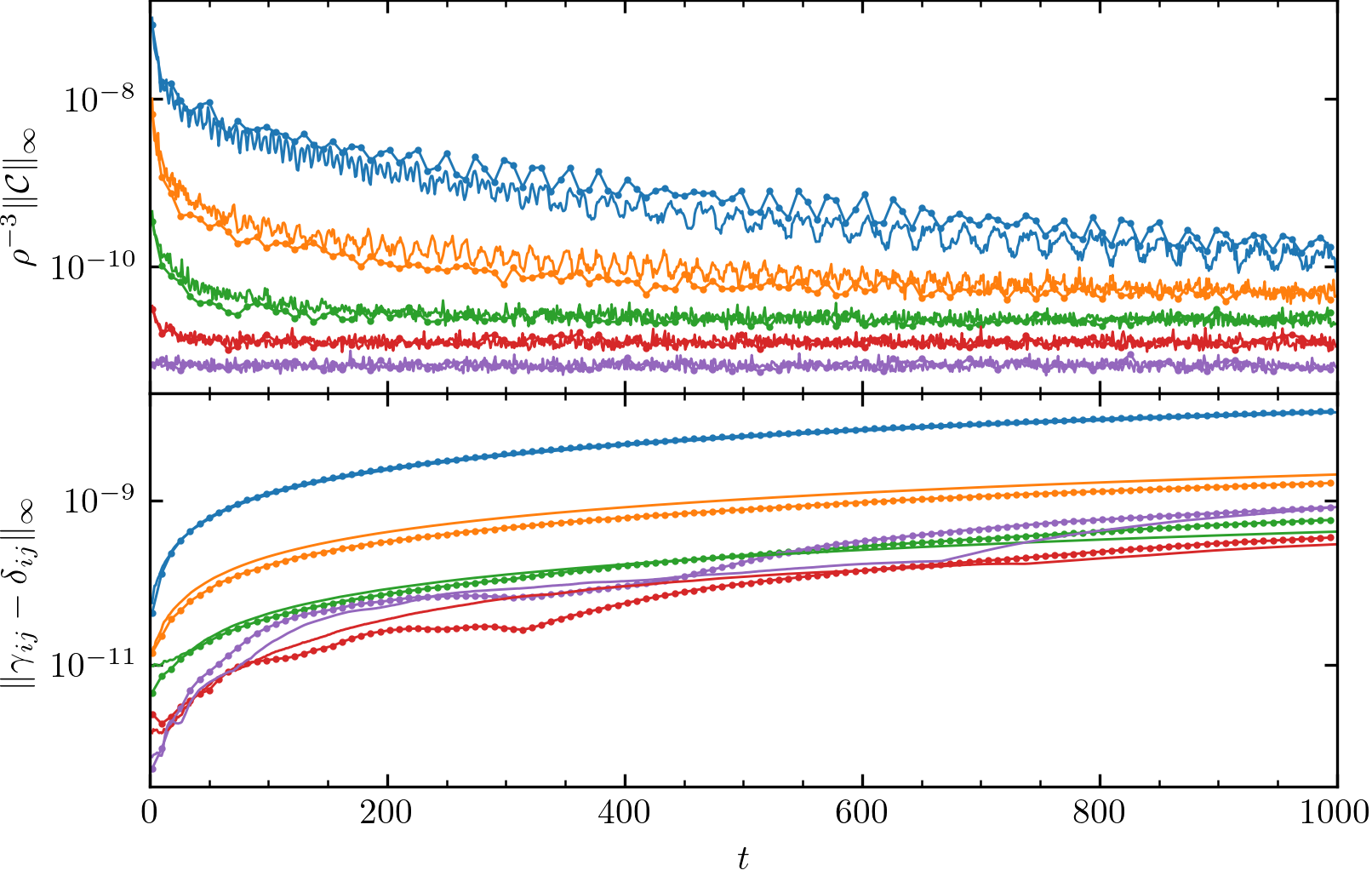}
  \caption{\label{fig:AwA_rs_g2}%
  Robust stability test.
  Errors are displayed for $\mathcal{G}_{\mathrm{VC}}$ with solid lines and
  $\mathcal{G}_{\mathrm{CC}}$ with solid lines marked ``{\large$\bullet$}''.
  Different selections of parameter entering the \Mesh{} sampling are:
  $\rho=1$ in blue; $\rho=2$ in orange;
  $\rho=4$ in green; $\rho=8$ in red; $\rho=16$ in purple.
  Top panel: Rescaled $L_\infty$ norm of collective constraint monitor.
  Bottom panel: Deviation from flat metric.
  Consistent behavior between CC and VC discretization is observed at all resolutions.
  See text for further discussion.
  }
\end{figure}

The robust stability test is performed with one effective spatial dimension and is
designed to efficiently detect instability within numerical algorithms
affecting the principal part of the evolution system. An initial spatial
slice of Minkowski is made where all \z4c{} field components
(\S\ref{ssec:eqnformgrathena}) have to each sampled grid point
a resolution dependent, distinct (i.e. independent) uniform random value added:
\begin{equation}
  \varepsilon \in (-10^{-10}/\rho^2,\,10^{-10}/\rho^2).
\end{equation}
Values are selected such that $\varepsilon^2$ is below round-off in double precision
arithmetic. The variable $\varepsilon$ models the effect of finite machine precision and for a code
to pass the test stable evolution must be observed.
A code that cannot pass this test would potentially have severe issues with any evolution of
smooth initial data.
We evolve to a final time $T=1000$ and
monitor $\Vert \mathcal{C}\Vert_\infty$ where $\mathcal{C}$ is defined in
Eq.\eqref{eq:collective_constraint} together with
$\Vert \gamma{}_{ij} - \delta{}_{ij} \Vert_\infty$ as suggested in
\cite{Daverio:2018tjf}. Results are displayed in Fig.\ref{fig:AwA_rs_g2}.

We find that over the duration of the time-evolution $\Vert \mathcal{C} \Vert_\infty$
plateaus for $\rho\geq 2$ with $\rho=1$ appearing also to tend towards a plateau. Similar behavior
is found for $\Vert \mathcal{H} \Vert_\infty$.
This indicates stability in the sense that no spurious exponential
modes are excited. Reducing $\eta\rightarrow0$ is observed to lead to qualitatively similar
general behavior. This was also observed in \cite{Cao:2011fu}. As errors decrease as resolution is
increased we consider \GRAthena{} to pass this test.
\subsection{Linearized wave}\label{ssec:awa_lin_wave}
The purpose of the linearized wave test is to check whether a code can propagate a linearized
gravitational wave, a minimal necessity for reliable wave extraction from strong-field
sources \cite{Babiuc:2007vr}.

An effective one-dimensional test with dynamics aligned along the $x$-axis is specified through
spatial slicing of:
\begin{equation}\label{eq:awa_linmetr}
  \mathrm{d}s^2 = -\mathrm{d}t^2 + \mathrm{d}x^2 +(1+H) \mathrm{d}y^2 +(1-H) \mathrm{d}z^2,
\end{equation}
where:
\begin{equation}\label{eq:awa_linmetrH}
  H(x,\,t) = A\sin\Bigg(\frac{2\pi(x-t)}{d}\Bigg),
\end{equation}
and $d=1$ is set to match the periodicity of the underlying computational domain and $A=10^{-8}$
selected such that quadratic terms are on the order of numerical round-off in double precision
arithmetic.
Consequently, initially we have $\alpha=1$ and $\beta{}^i=0$ with non-trivial extrinsic curvature
components:
\begin{align}\label{eq:awa_linKextr}
  K{}_{yy} =& -\frac{1}{2}\partial{}_t[H(x,\,t)], &
  K{}_{zz} =& \frac{1}{2}\partial{}_t[H(x,\,t)].
\end{align}
Time evolution is performed to a final time of $T=1000$.

Rather than displaying approximate sinusoidal profiles at some final time as in
\cite{Alcubierre:2003pc,Babiuc:2007vr,Cao:2011fu} we follow a suggestion of \cite{Daverio:2018tjf}
to instead consider the spectra of data. To this end we compute the discrete Fourier transform as:
\begin{equation*}
  F_k(t):= \frac{1}{N} \sum_{n=1}^{N}
  \big(
    \gamma_{zz}(t,\,x_n) - 1
  \big)
  \exp(-2\pi i k (x_n-t)),
\end{equation*}
where $x_n\in \mathcal{G}_{(\cdot)}$ (see Eq.\eqref{eq:awagrids}). In the case of VC
discretization the last point of the (periodic) grid is identified with the first and therefore
dropped from the summation. Thus comparing Eq.\eqref{eq:awa_linmetrH} a spectral measure of the
relative error in the travelling wave amplitude is provided through
$\epsilon_a(t):=||F_1(t)|-A|/A$. The absolute phase
error may be directly inspected through $\epsilon_p(t):=|\arg(F_1(t)) - \pi/2|$.
We also compute the offset of the numerical waveform
relative to the amplitude $\epsilon_o(t)=|F_0(t)|/A$.

\begin{figure}[t]
  \centering
  \includegraphics[width=\columnwidth]{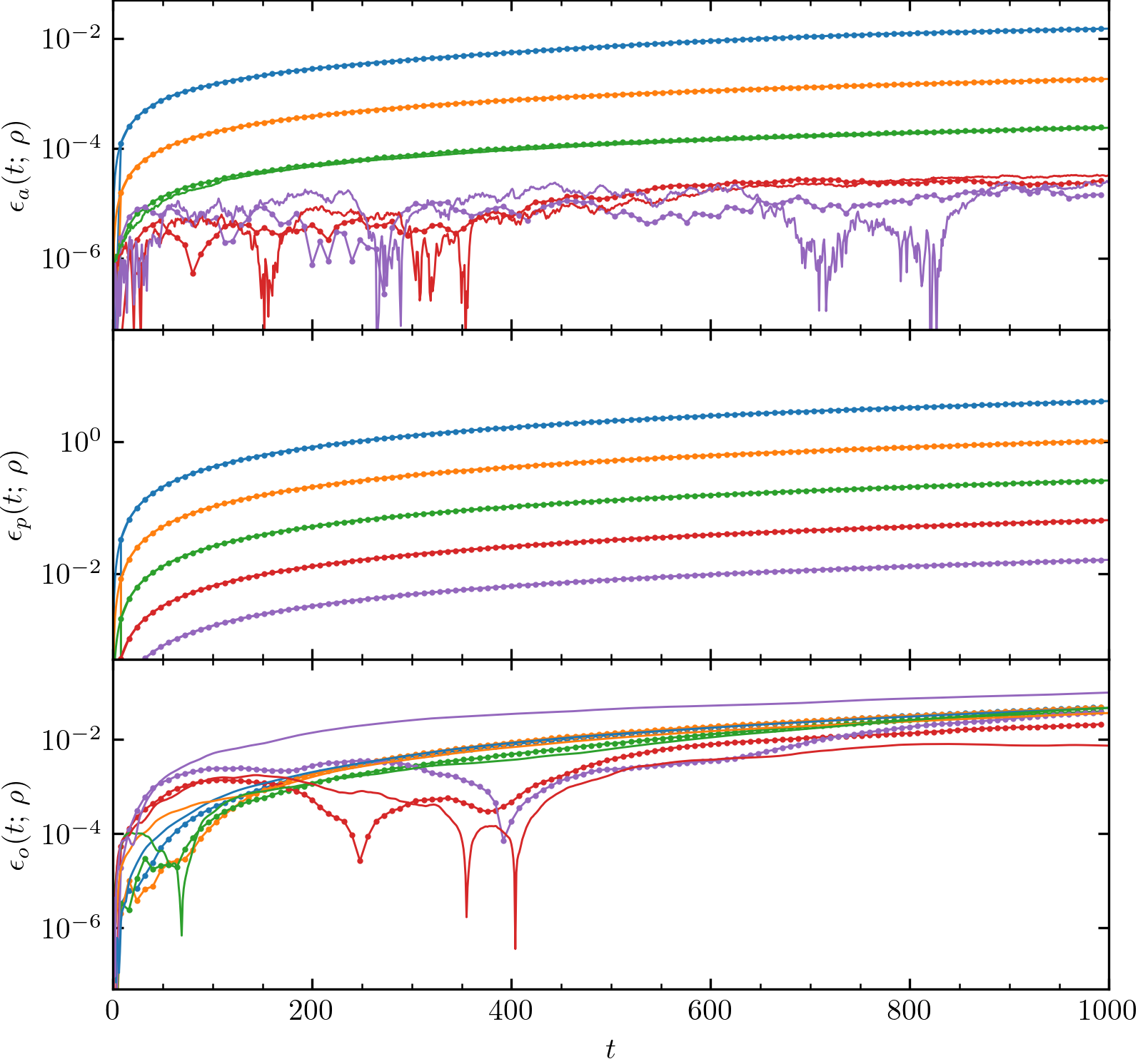}
  \caption{\label{fig:AwA_lw1_g2}%
  Linearized wave test with parameters and legend of
  Fig.\ref{fig:AwA_rs_g2}.
  Top panel: relative error in the travelling wave amplitude $\epsilon_a(t;\,\rho)$.
  Middle panel: phase error $\epsilon_p(t;\,\rho)$.
  Bottom panel: offset of the wave relative to amplitude $\epsilon_o(t;\,\rho)$.
  The CC and VC discretizations display mutually consistent behavior in all variables
  for all $t$ apart from $\epsilon_o(t;\,16)$ though the overall trend is recovered
  as $t\rightarrow T$.
  See text for discussion.
  }
\end{figure}

The initial data are constraint violating \cite{Cao:2011fu} and the puncture gauge is not
necessarily compatible with simple advection,
nonetheless we find that to an excellent approximation the solution is a simple travelling wave
with results of the analysis described above shown in Fig.\ref{fig:AwA_lw1_g2}.

In agreement with \cite{Cao:2011fu,Daverio:2018tjf} we find the dominant source of error to be
in the phase of the propagating waveform where the coarsest sampling $\rho=1$ leads to a final
absolute phase error of $\simeq 3.9\,\mathrm{rad}$ cf., the finest sampling $\rho=16$ yielding
$\simeq 1.5\times 10^{-2}\,\mathrm{rad}$. For $\epsilon_a(t)$ and $\epsilon_p(t)$ we observe
convergence with increasing resolution. For $\epsilon_o(T)$ at $T=1000$ increasing $\rho$
tended to increase error albeit overall this is acceptably within $[7.4,\,97]\times 10^{-3}$;
this is compatible with the general behavior found in \cite{Daverio:2018tjf}.

As observed in \cite{Cao:2011fu} reducing $\kappa{}_1\rightarrow0$ or $\eta\rightarrow0$ leads
to qualitatively very similar results. We thus consider \GRAthena{} to pass this test.

\subsection{Gauge wave}\label{ssec:awa_gauge_wave}
A gauge transformation of Minkowski space-time defines this test with parameters selected so as to
involve the full non-linear dynamics. One takes
$\eta{}_{ab}\dot{=}\mathrm{diag}(-1,\,1,\,1,\,1)$ in Cartesian coordinates $x'{}^a$ and transforms:
\begin{equation*}
  (t',\,x',\,y',\,z')\rightarrow
  \big(t+G_+(x,\,t),\,
  x-G_\pm(x,\,t),\,y,\,z\big),
\end{equation*}
where $G_\pm(x,\,t):= \pm\partial{}_t[H(x,\,t)] / (8\pi^2)$
with $H$ defined in Eq.\eqref{eq:awa_linmetrH}.
Two test cases are defined: \emph{shifted} where $G_-$ is selected for the transformation
on the $x'$ component whereas for \emph{unshifted} $G_+$ is chosen in both components.
For the latter the induced metric is:
\begin{align}
  \gamma{}_{xx} =& 1-H, &
  \gamma{}_{yy}=&\gamma{}_{zz}=1;
\end{align}
resulting in non-trivial extrinsic curvature component:
\begin{equation}
  K{}_{xx} = \frac{\partial{}_t[H(x,\,t)]}{2\sqrt{1-H(x,\,t)}}.
\end{equation}

\begin{figure}[t]
  \centering
  \includegraphics[width=\columnwidth]{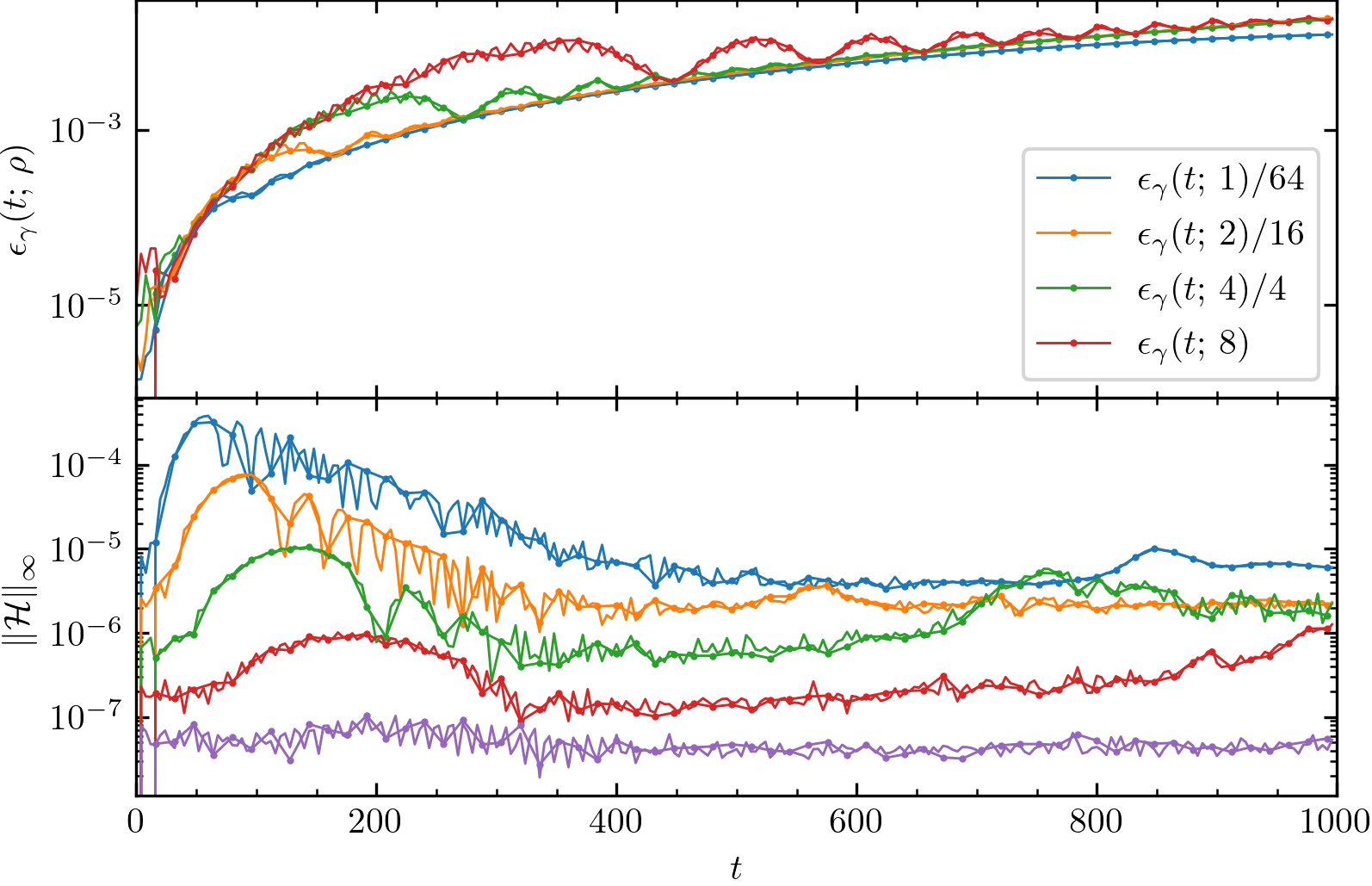}
  \caption{\label{fig:AwA_gw1s_g2}%
  Gauge-wave test.
  Top panel: Error of metric components based on Eq.\eqref{eq:err_rms}. Legend indicates
  scaling applied (based on an assumed $\2nd$ order spatial scheme). Rescaling indicates
  anticipated convergence is well-obeyed.
  Bottom panel: Hamiltonian constraint with displayed data following the legend of
  Fig.\ref{fig:AwA_rs_g2}. It is clear that with increased resolution constraint violation
  converges away.
  Consistent behavior between CC and VC discretizations is observed at all samplings.
  See text for discussion.
  }
\end{figure}

During calculations we evolve to a final time of $T=1000$.
In contrast to the standard \AwA{} specification \cite{Alcubierre:2003pc,Babiuc:2007vr} we
make use of the puncture gauge where we found
it crucial to select a non-zero shift-damping of $\eta=2$.
In addition to inspection of the constraints through
$\Vert\mathcal{H}\Vert_\infty$ we also consider convergence of the aggregate, induced metric
quantity:
\begin{equation}\label{eq:err_rms}
  \epsilon_\gamma(t;\,\rho)=
  \sqrt{\sum_{i,j=1}^3
  \left\langle
  \left.\gamma{}_{ij}(t)\right|_{\rho} - \left.\gamma{}_{ij}(t)\right|_{2\rho}
  \right\rangle_{\mathrm{RMS}}
  },
\end{equation}
where the root-mean-square (RMS) value is computed at fixed times over a \Mesh{} sampled with
$N=50$. In Fig.\ref{fig:AwA_gw1s_g2} we plot
$\epsilon_\gamma(t;\,\rho)$ for a choice of $A=1/100$ and $d=1$ in
Eq.\eqref{eq:awa_linmetrH}
and find a $\2nd$ order rate of convergence as is expected for $\Ng=2$.

The behavior of the shifted case is identical for the stated parameters and therefore we
do not show it.

A two-dimensional variant of the gauge wave test may also be considered where an initial spatial
rotation of $\pi/4$ is made. In particular, coordinates are mapped according to
$SO(2)\ni R:(x,\,y)\mapsto(\hat{x},\,\hat{y})=(x+y,\,x-y)/\sqrt{2}$.
This results in $x$-aligned propagation mapped to
a periodic diagonal trajectory in the $\hat{x}$-$\hat{y}$ plane.
Here as per \AwA{} specification we evolve to a final time of $T=100$ and
selected resolutions based on $\rho\in\{1,\,2,\,4,\,8\}$. For an amplitude
of $A=1/100$ we similarly found a $\2nd$ order rate of convergence with $\Ng=2$ in
rescaling of $\epsilon_\gamma(\rho)$.
For the case of $A=1/10$ however we did not observe clean $\2nd$ order convergence. Indeed the \AwA{}
specification suggestion to use an even higher amplitude $A=1/2$ is well-known to cause issues with
stability in a variety of formulations and regardless of puncture or harmonic
gauge choice \cite{Daverio:2018tjf,Cao:2011fu,boyle2007testingaccuracystability}.

We consider \GRAthena{} to pass this test in both
the one-dimensional (un)-shifted cases and in the two-dimensional unshifted case with the caveat
that initial amplitude must be reduced.

\subsection{\AwA{} summary}\label{ssec:awa_summary}
We have demonstrated that \GRAthena{} with \z4c{} coupled to the moving puncture gauge
passes the \AwA{} robust stability (\S\ref{ssec:awa_robust_stab}) and the one-dimensional
linearized wave (\S\ref{ssec:awa_lin_wave}) tests. For the gauge wave
tests (\S\ref{ssec:awa_gauge_wave}) we find that \GRAthena{} passes for a choice of reduced
initial amplitude of the propagated wave.

\bibliography{refs20210120}

\begin{thebibliography}{}
\expandafter\ifx\csname natexlab\endcsname\relax\def\natexlab#1{#1}\fi
\providecommand{\url}[1]{\href{#1}{#1}}
\providecommand{\dodoi}[1]{doi:~\href{http://doi.org/#1}{\nolinkurl{#1}}}
\providecommand{\doeprint}[1]{\href{http://ascl.net/#1}{\nolinkurl{http://ascl.net/#1}}}
\providecommand{\doarXiv}[1]{\href{https://arxiv.org/abs/#1}{\nolinkurl{https://arxiv.org/abs/#1}}}

\bibitem[{Abbott {et~al.}(2016{\natexlab{a}})}]{Abbott:2016blz}
Abbott, B.~P., {et~al.} 2016{\natexlab{a}}, Phys. Rev. Lett., 116, 061102,
  \dodoi{10.1103/PhysRevLett.116.061102}

\bibitem[{Abbott {et~al.}(2016{\natexlab{b}})}]{TheLIGOScientific:2016wfe}
---. 2016{\natexlab{b}}, Phys. Rev. Lett., 116, 241102,
  \dodoi{10.1103/PhysRevLett.116.241102}

\bibitem[{Abbott {et~al.}(2017{\natexlab{a}})}]{TheLIGOScientific:2017qsa}
---. 2017{\natexlab{a}}, Phys. Rev. Lett., 119, 161101,
  \dodoi{10.1103/PhysRevLett.119.161101}

\bibitem[{Abbott {et~al.}(2017{\natexlab{b}})}]{Evans:2016mbw}
---. 2017{\natexlab{b}}, Class. Quant. Grav., 34, 044001,
  \dodoi{10.1088/1361-6382/aa51f4}

\bibitem[{Abbott {et~al.}(2020)}]{Abbott_2020}
---. 2020, Living Reviews in Relativity, 23, \dodoi{10.1007/s41114-020-00026-9}

\bibitem[{Akutsu {et~al.}(2020)}]{akutsu2020overviewkagracalibration}
Akutsu, T., {et~al.} 2020, arXiv:2009.09305 [astro-ph, physics:gr-qc].
\newblock \doarXiv{2009.09305}

\bibitem[{Alcubierre {et~al.}(2003)Alcubierre, Br{\"u}gmann, Diener, Koppitz,
  Pollney, {et~al.}}]{Alcubierre:2002kk}
Alcubierre, M., Br{\"u}gmann, B., Diener, P., {et~al.} 2003, Phys.Rev., D67,
  084023, \dodoi{10.1103/PhysRevD.67.084023}

\bibitem[{Alcubierre {et~al.}(2004)}]{Alcubierre:2003pc}
Alcubierre, M., {et~al.} 2004, Class. Quant. Grav., 21, 589,
  \dodoi{10.1088/0264-9381/21/2/019}

\bibitem[{Alfieri {et~al.}(2018)Alfieri, Bernuzzi, Perego, \&
  Radice}]{Alfieri:2018a}
Alfieri, R., Bernuzzi, S., Perego, A., \& Radice, D. 2018, Journal of Low Power
  Electronics and Applications, 8, \dodoi{10.3390/jlpea8020015}

\bibitem[{{Amaro-Seoane}
  {et~al.}(2017)}]{amaroseoane2017laserinterferometerspace}
{Amaro-Seoane}, {et~al.} 2017, arXiv:1702.00786 [astro-ph].
\newblock \doarXiv{1702.00786}

\bibitem[{Ansorg {et~al.}(2004)Ansorg, Br{\"u}gmann, \& Tichy}]{Ansorg:2004ds}
Ansorg, M., Br{\"u}gmann, B., \& Tichy, W. 2004, Phys. Rev., D70, 064011,
  \dodoi{10.1103/PhysRevD.70.064011}

\bibitem[{Arnowitt {et~al.}(1959)Arnowitt, Deser, \& Misner}]{Arnowitt:1959ah}
Arnowitt, R.~L., Deser, S., \& Misner, C.~W. 1959, Phys. Rev., 116, 1322,
  \dodoi{10.1103/PhysRev.116.1322}

\bibitem[{Arnowitt {et~al.}(2008)Arnowitt, Deser, \& Misner}]{Arnowitt:1962hi}
---. 2008, Gen. Rel. Grav., 40, 1997, \dodoi{10.1007/s10714-008-0661-1}

\bibitem[{Babiuc {et~al.}(2008)}]{Babiuc:2007vr}
Babiuc, M., {et~al.} 2008, Class. Quant. Grav., 25, 125012,
  \dodoi{10.1088/0264-9381/25/12/125012}

\bibitem[{Baiotti {et~al.}(2009)Baiotti, Bernuzzi, Corvino, De~Pietri, \&
  Nagar}]{Baiotti:2008nf}
Baiotti, L., Bernuzzi, S., Corvino, G., De~Pietri, R., \& Nagar, A. 2009, Phys.
  Rev., D79, 024002, \dodoi{10.1103/PhysRevD.79.024002}

\bibitem[{Baker {et~al.}(2007)Baker, van Meter, McWilliams, Centrella, \&
  Kelly}]{Baker:2006ha}
Baker, J.~G., van Meter, J.~R., McWilliams, S.~T., Centrella, J., \& Kelly,
  B.~J. 2007, Phys.Rev.Lett., 99, 181101, \dodoi{10.1103/PhysRevLett.99.181101}

\bibitem[{Baumgarte \& Shapiro(2010)}]{Baumgarte:2010}
Baumgarte, T., \& Shapiro, S. 2010, Numerical Relativity (Cambridge: Cambridge
  University Press)

\bibitem[{Baumgarte \& Shapiro(1999)}]{Baumgarte:1998te}
Baumgarte, T.~W., \& Shapiro, S.~L. 1999, Phys. Rev., D59, 024007,
  \dodoi{10.1103/PhysRevD.59.024007}

\bibitem[{{Berger} \& {Colella}(1989)}]{Berger:1989a}
{Berger}, M.~J., \& {Colella}, P. 1989, Journal of Computational Physics, 82,
  64, \dodoi{10.1016/0021-9991(89)90035-1}

\bibitem[{Berger \& Oliger(1984)}]{Berger:1984zza}
Berger, M.~J., \& Oliger, J. 1984, J.Comput.Phys., 53, 484

\bibitem[{Bernuzzi(2020)}]{Bernuzzi:2020tgt}
Bernuzzi, S. 2020, Invited Review for GERG.
\newblock \doarXiv{2004.06419}

\bibitem[{Bernuzzi \& Dietrich(2016)}]{Bernuzzi:2016pie}
Bernuzzi, S., \& Dietrich, T. 2016, Phys. Rev., D94, 064062,
  \dodoi{10.1103/PhysRevD.94.064062}

\bibitem[{Bernuzzi \& Hilditch(2010)}]{Bernuzzi:2009ex}
Bernuzzi, S., \& Hilditch, D. 2010, Phys. Rev., D81, 084003,
  \dodoi{10.1103/PhysRevD.81.084003}

\bibitem[{Bernuzzi {et~al.}(2014)Bernuzzi, Nagar, Balmelli, Dietrich, \&
  Ujevic}]{Bernuzzi:2014kca}
Bernuzzi, S., Nagar, A., Balmelli, S., Dietrich, T., \& Ujevic, M. 2014,
  Phys.Rev.Lett., 112, 201101, \dodoi{10.1103/PhysRevLett.112.201101}

\bibitem[{Bernuzzi {et~al.}(2012)Bernuzzi, Thierfelder, \&
  Br{\"u}gmann}]{Bernuzzi:2011aq}
Bernuzzi, S., Thierfelder, M., \& Br{\"u}gmann, B. 2012, Phys.Rev., D85,
  104030, \dodoi{10.1103/PhysRevD.85.104030}

\bibitem[{Berrut \&
  Trefethen(2004)}]{berrut2004barycentriclagrangeinterpolation}
Berrut, J.-P., \& Trefethen, L.~N. 2004, SIAM Review, 46, 501,
  \dodoi{10.1137/S0036144502417715}

\bibitem[{Bona {et~al.}(2010)Bona, {Bona-Casas}, \&
  Palenzuela}]{bona2010actionprinciplenumericalrelativity}
Bona, C., {Bona-Casas}, C., \& Palenzuela, C. 2010, Physical Review D, 82,
  124010, \dodoi{10.1103/PhysRevD.82.124010}

\bibitem[{Bona {et~al.}(2003)Bona, Ledvinka, Palenzuela, \&
  Zacek}]{Bona:2003fj}
Bona, C., Ledvinka, T., Palenzuela, C., \& Zacek, M. 2003, Phys. Rev., D67,
  104005, \dodoi{10.1103/PhysRevD.67.104005}

\bibitem[{Bona {et~al.}(1995)Bona, Mass{\'o}, Seidel, \& Stela}]{Bona:1994b}
Bona, C., Mass{\'o}, J., Seidel, E., \& Stela, J. 1995, Phys. Rev. Lett., 75,
  600

\bibitem[{Bowen \& York(1980)}]{Bowen:1980yu}
Bowen, J.~M., \& York, Jr., J.~W. 1980, Phys. Rev., D21, 2047,
  \dodoi{10.1103/PhysRevD.21.2047}

\bibitem[{Boyle {et~al.}(2007)Boyle, Lindblom, Pfeiffer, Scheel, \&
  Kidder}]{boyle2007testingaccuracystability}
Boyle, M., Lindblom, L., Pfeiffer, H., Scheel, M., \& Kidder, L.~E. 2007,
  Physical Review D, 75, 024006, \dodoi{10.1103/PhysRevD.75.024006}

\bibitem[{Boyle {et~al.}(2019)}]{Boyle:2019kee}
Boyle, M., {et~al.} 2019, Class. Quant. Grav., 36, 195006,
  \dodoi{10.1088/1361-6382/ab34e2}

\bibitem[{Brandt \& Br{\"u}gmann(1997)}]{Brandt:1997tf}
Brandt, S., \& Br{\"u}gmann, B. 1997, Phys. Rev. Lett., 78, 3606,
  \dodoi{10.1103/PhysRevLett.78.3606}

\bibitem[{Brown {et~al.}(2009)Brown, Diener, Sarbach, Schnetter, \&
  Tiglio}]{brown2009turduckeningblackholes}
Brown, D., Diener, P., Sarbach, O., Schnetter, E., \& Tiglio, M. 2009, Physical
  Review D, 79, 044023, \dodoi{10.1103/PhysRevD.79.044023}

\bibitem[{Br{\"u}gmann {et~al.}(2008)Br{\"u}gmann, Gonzalez, Hannam, Husa,
  Sperhake, {et~al.}}]{Brugmann:2008zz}
Br{\"u}gmann, B., Gonzalez, J.~A., Hannam, M., {et~al.} 2008, Phys.Rev., D77,
  024027, \dodoi{10.1103/PhysRevD.77.024027}

\bibitem[{Bugner {et~al.}(2016)Bugner, Dietrich, Bernuzzi, Weyhausen, \&
  Br{\"u}gmann}]{Bugner:2015gqa}
Bugner, M., Dietrich, T., Bernuzzi, S., Weyhausen, A., \& Br{\"u}gmann, B.
  2016, Phys. Rev., D94, 084004, \dodoi{10.1103/PhysRevD.94.084004}

\bibitem[{Burstedde {et~al.}(2019)Burstedde, Holke, \&
  Isaac}]{burstedde2019numberfaceconnectedcomponents}
Burstedde, C., Holke, J., \& Isaac, T. 2019, Foundations of Computational
  Mathematics, 19, 843, \dodoi{10.1007/s10208-018-9400-5}

\bibitem[{Burstedde {et~al.}(2011)Burstedde, Wilcox, \&
  Ghattas}]{Burstedde:2011a}
Burstedde, C., Wilcox, L.~C., \& Ghattas, O. 2011, SIAM Journal on Scientific
  Computing, 33, 1103, \dodoi{10.1137/100791634}

\bibitem[{Campanelli {et~al.}(2006)Campanelli, Lousto, Marronetti, \&
  Zlochower}]{Campanelli:2005dd}
Campanelli, M., Lousto, C.~O., Marronetti, P., \& Zlochower, Y. 2006, Phys.
  Rev. Lett., 96, 111101, \dodoi{10.1103/PhysRevLett.96.111101}

\bibitem[{Cao \& Hilditch(2012)}]{Cao:2011fu}
Cao, Z., \& Hilditch, D. 2012, Phys.Rev., D85, 124032,
  \dodoi{10.1103/PhysRevD.85.124032}

\bibitem[{Cao {et~al.}(2008)Cao, Yo, \&
  Yu}]{cao2008reinvestigationmovingpunctured}
Cao, Z., Yo, H.-J., \& Yu, J.-P. 2008, Physical Review D, 78, 124011,
  \dodoi{10.1103/PhysRevD.78.124011}

\bibitem[{Carter~Edwards {et~al.}(2014)Carter~Edwards, Trott, \&
  Sunderland}]{carteredwards2014kokkosenablingmanycore}
Carter~Edwards, H., Trott, C.~R., \& Sunderland, D. 2014, Journal of Parallel
  and Distributed Computing, 74, 3202, \dodoi{10.1016/j.jpdc.2014.07.003}

\bibitem[{Chirvasa \& Husa(2010)}]{chirvasa2010finitedifferencemethods}
Chirvasa, M., \& Husa, S. 2010, Journal of Computational Physics, 229, 2675,
  \dodoi{10.1016/j.jcp.2009.12.016}

\bibitem[{Clough {et~al.}(2015)Clough, Figueras, Finkel, Kunesch, Lim, \&
  Tunyasuvunakool}]{Clough:2015sqa}
Clough, K., Figueras, P., Finkel, H., {et~al.} 2015.
\newblock \doarXiv{1503.03436}

\bibitem[{Damour {et~al.}(2008)Damour, Nagar, Hannam, Husa, \&
  Br{\"u}gmann}]{Damour:2008te}
Damour, T., Nagar, A., Hannam, M., Husa, S., \& Br{\"u}gmann, B. 2008, Phys.
  Rev., D78, 044039, \dodoi{10.1103/PhysRevD.78.044039}

\bibitem[{Daverio {et~al.}(2018)Daverio, Dirian, \& Mitsou}]{Daverio:2018tjf}
Daverio, D., Dirian, Y., \& Mitsou, E. 2018.
\newblock \doarXiv{1810.12346}

\bibitem[{Dietrich \& Bernuzzi(2015)}]{Dietrich:2014wja}
Dietrich, T., \& Bernuzzi, S. 2015, Phys.Rev., D91, 044039,
  \dodoi{10.1103/PhysRevD.91.044039}

\bibitem[{Dietrich {et~al.}(2018)Dietrich, Radice, Bernuzzi, Zappa, Perego,
  Brügmann, Chaurasia, Dudi, Tichy, \& Ujevic}]{Dietrich:2018phi}
Dietrich, T., Radice, D., Bernuzzi, S., {et~al.} 2018, Class. Quant. Grav., 35,
  24LT01, \dodoi{10.1088/1361-6382/aaebc0}

\bibitem[{Felker \& Stone(2018)}]{felker2018fourthorderaccuratefinite}
Felker, K.~G., \& Stone, J.~M. 2018, Journal of Computational Physics, 375,
  1365, \dodoi{10.1016/j.jcp.2018.08.025}

\bibitem[{Fernando {et~al.}(2018)Fernando, Neilsen, Lim, Hirschmann, \&
  Sundar}]{Fernando:2018mov}
Fernando, M., Neilsen, D., Lim, H., Hirschmann, E., \& Sundar, H. 2018,
  \dodoi{10.1137/18M1196972}

\bibitem[{Friedrich(1985)}]{Friedrich:1985}
Friedrich, H. 1985, Communications in Mathematical Physics, 100, 525,
  \dodoi{10.1007/BF01217728}

\bibitem[{Galaviz {et~al.}(2010)Galaviz, Bruegmann, \&
  Cao}]{galaviz2010numericalevolutionmultiple}
Galaviz, P., Bruegmann, B., \& Cao, Z. 2010, Physical Review D, 82, 024005,
  \dodoi{10.1103/PhysRevD.82.024005}

\bibitem[{Goldberg {et~al.}(1967)Goldberg, MacFarlane, Newman, Rohrlich, \&
  Sudarshan}]{Goldberg:1966uu}
Goldberg, J.~N., MacFarlane, A.~J., Newman, E.~T., Rohrlich, F., \& Sudarshan,
  E. C.~G. 1967, J. Math. Phys., 8, 2155

\bibitem[{Goodale {et~al.}(2003)Goodale, Allen, Lanfermann, Mass{\'o}, Radke,
  Seidel, \& Shalf}]{Goodale:2002a}
Goodale, T., Allen, G., Lanfermann, G., {et~al.} 2003, in Vector and Parallel
  Processing -- VECPAR'2002, 5th International Conference, Lecture Notes in
  Computer Science (Berlin: Springer)

\bibitem[{Grete {et~al.}(2019)Grete, Glines, \&
  O'Shea}]{grete2019kathenaperformanceportable}
Grete, P., Glines, F.~W., \& O'Shea, B.~W. 2019, arXiv:1905.04341 [astro-ph,
  physics:physics].
\newblock \doarXiv{1905.04341}

\bibitem[{Gundlach {et~al.}(2005)Gundlach, Martin-Garcia, Calabrese, \&
  Hinder}]{Gundlach:2005eh}
Gundlach, C., Martin-Garcia, J.~M., Calabrese, G., \& Hinder, I. 2005, Class.
  Quant. Grav., 22, 3767, \dodoi{10.1088/0264-9381/22/17/025}

\bibitem[{Gustafsson {et~al.}(2013)Gustafsson, Kreiss, \&
  Oliger}]{Gustafsson:2013td}
Gustafsson, B., Kreiss, H.-O., \& Oliger, J. 2013, {Time-dependent problems and
  difference methods; 2nd ed.}, Pure and applied mathematics a wiley series of
  texts, monographs and tracts (Somerset: Wiley).
\newblock \url{https://cds.cern.ch/record/2122877}

\bibitem[{Hannam {et~al.}(2010)Hannam, Husa, Ohme, M{\"u}ller, \&
  Br{\"u}gmann}]{Hannam:2010ec}
Hannam, M., Husa, S., Ohme, F., M{\"u}ller, D., \& Br{\"u}gmann, B. 2010, Phys.
  Rev., D82, 124008, \dodoi{10.1103/PhysRevD.82.124008}

\bibitem[{Healy {et~al.}(2019)Healy, Lousto, Lange, O'Shaughnessy, Zlochower,
  \& Campanelli}]{Healy:2019jyf}
Healy, J., Lousto, C.~O., Lange, J., {et~al.} 2019, Phys. Rev. D, 100, 024021,
  \dodoi{10.1103/PhysRevD.100.024021}

\bibitem[{Herrmann {et~al.}(2007)Herrmann, Hinder, Shoemaker, \&
  Laguna}]{herrmann2007unequalmassbinary}
Herrmann, F., Hinder, I., Shoemaker, D., \& Laguna, P. 2007, Classical and
  Quantum Gravity, 24, S33, \dodoi{10.1088/0264-9381/24/12/S04}

\bibitem[{Hilditch {et~al.}(2013)Hilditch, Bernuzzi, Thierfelder, Cao, Tichy,
  \& Bruegmann}]{Hilditch:2012fp}
Hilditch, D., Bernuzzi, S., Thierfelder, M., {et~al.} 2013, Phys. Rev., D88,
  084057, \dodoi{10.1103/PhysRevD.88.084057}

\bibitem[{Hilditch \& Ruiz(2018)}]{Hilditch:2016xos}
Hilditch, D., \& Ruiz, M. 2018, Class. Quant. Grav., 35, 015006,
  \dodoi{10.1088/1361-6382/aa96c6}

\bibitem[{Hilditch {et~al.}(2016)Hilditch, Weyhausen, \&
  Br{\"u}gmann}]{Hilditch:2015aba}
Hilditch, D., Weyhausen, A., \& Br{\"u}gmann, B. 2016, Phys. Rev., D93, 063006,
  \dodoi{10.1103/PhysRevD.93.063006}

\bibitem[{Holmstr{\"o}m(1999)}]{holmstrom1999solvinghyperbolicpdes}
Holmstr{\"o}m, M. 1999, SIAM Journal on Scientific Computing, 21, 405,
  \dodoi{10.1137/S1064827597316278}

\bibitem[{Huerta {et~al.}(2019)Huerta, Haas, Jha, Neubauer, \&
  Katz}]{huerta2019supportinghighperformancehighthroughput}
Huerta, E.~A., Haas, R., Jha, S., Neubauer, M., \& Katz, D.~S. 2019, Computing
  and Software for Big Science, 3, 5, \dodoi{10.1007/s41781-019-0022-7}

\bibitem[{Husa {et~al.}(2008)Husa, Gonz{\'a}lez, Hannam, Br{\"u}gmann, \&
  Sperhake}]{Husa:2007hp}
Husa, S., Gonz{\'a}lez, J.~A., Hannam, M., Br{\"u}gmann, B., \& Sperhake, U.
  2008, Class. Quant. Grav., 25, 105006, \dodoi{10.1088/0264-9381/25/10/105006}

\bibitem[{Jani {et~al.}(2016)Jani, Healy, Clark, London, Laguna, \&
  Shoemaker}]{Jani:2016wkt}
Jani, K., Healy, J., Clark, J.~A., {et~al.} 2016, Class. Quant. Grav., 33,
  204001, \dodoi{10.1088/0264-9381/33/20/204001}

\bibitem[{Ketcheson(2010)}]{ketcheson2010rungekuttamethods}
Ketcheson, D.~I. 2010, Journal of Computational Physics, 229, 1763,
  \dodoi{10.1016/j.jcp.2009.11.006}

\bibitem[{Kidder {et~al.}(2017)}]{Kidder:2016hev}
Kidder, L.~E., {et~al.} 2017, J. Comput. Phys., 335, 84,
  \dodoi{10.1016/j.jcp.2016.12.059}

\bibitem[{Kreiss \& Oliger(1973)}]{Kreiss:1973}
Kreiss, H.~O., \& Oliger, J. 1973, Methods for the approximate solution of time
  dependent problems (Geneva: International Council of Scientific Unions, World
  Meteorological Organization)

\bibitem[{{LIGO Scientific Collaboration}(2018)}]{lalsuite}
{LIGO Scientific Collaboration}. 2018, {LIGO} {A}lgorithm {L}ibrary -
  {LALS}uite, free software (GPL), \dodoi{10.7935/GT1W-FZ16}

\bibitem[{Lindblom {et~al.}(2006)Lindblom, Scheel, Kidder, Owen, \&
  Rinne}]{Lindblom:2005qh}
Lindblom, L., Scheel, M.~A., Kidder, L.~E., Owen, R., \& Rinne, O. 2006,
  Class.Quant.Grav., 23, S447, \dodoi{10.1088/0264-9381/23/16/S09}

\bibitem[{Loffler {et~al.}(2012)}]{Loffler:2011ay}
Loffler, F., {et~al.} 2012, Class. Quant. Grav., 29, 115001,
  \dodoi{10.1088/0264-9381/29/11/115001}

\bibitem[{Lousto {et~al.}(2010)Lousto, Nakano, Zlochower, \&
  Campanelli}]{Lousto:2010qx}
Lousto, C.~O., Nakano, H., Zlochower, Y., \& Campanelli, M. 2010, Phys.Rev.,
  D82, 104057, \dodoi{10.1103/PhysRevD.82.104057}

\bibitem[{Mewes {et~al.}(2020)Mewes, Zlochower, Campanelli, Baumgarte, Etienne,
  Lopez~Armengol, \& Cipolletta}]{Mewes:2020vic}
Mewes, V., Zlochower, Y., Campanelli, M., {et~al.} 2020, Phys. Rev. D, 101,
  104007, \dodoi{10.1103/PhysRevD.101.104007}

\bibitem[{Mewes {et~al.}(2018)Mewes, Zlochower, Campanelli, Ruchlin, Etienne,
  \& Baumgarte}]{mewes2018numericalrelativityspherical}
---. 2018, Physical Review D, 97, 084059, \dodoi{10.1103/PhysRevD.97.084059}

\bibitem[{Miller {et~al.}(2021)Miller, Dolence, Gaspar, Grete, Glines, \&
  et~al.}]{parthenon:web}
Miller, J., Dolence, J., Gaspar, A., {et~al.} 2021, Parthenon performance
  portable AMR framework, \url{https://github.com/lanl/parthenon}

\bibitem[{Morton(1966)}]{morton1966computer}
Morton, G.~M. 1966, {A computer oriented geodetic data base and a new technique
  in file sequencing}, Tech. rep.

\bibitem[{M{\"o}sta {et~al.}(2014)M{\"o}sta, Mundim, Faber, Haas, Noble,
  {et~al.}}]{Moesta:2013dna}
M{\"o}sta, P., Mundim, B.~C., Faber, J.~A., {et~al.} 2014, Class.Quant.Grav.,
  31, 015005, \dodoi{10.1088/0264-9381/31/1/015005}

\bibitem[{M{\"u}ller \& Br{\"u}gmann(2010)}]{Muller:2009jx}
M{\"u}ller, D., \& Br{\"u}gmann, B. 2010, Class. Quant. Grav., 27, 114008,
  \dodoi{10.1088/0264-9381/27/11/114008}

\bibitem[{Nagar {et~al.}(2018)}]{Nagar:2018zoe}
Nagar, A., {et~al.} 2018, Phys. Rev., D98, 104052,
  \dodoi{10.1103/PhysRevD.98.104052}

\bibitem[{Nakamura {et~al.}(1987)Nakamura, Oohara, \& Kojima}]{Nakamura:1987zz}
Nakamura, T., Oohara, K., \& Kojima, Y. 1987, Prog. Theor. Phys. Suppl., 90, 1

\bibitem[{Nakano(2015)}]{Nakano_2015}
Nakano, H. 2015, Classical and Quantum Gravity, 32, 177002,
  \dodoi{10.1088/0264-9381/32/17/177002}

\bibitem[{Nakano {et~al.}(2011)Nakano, Zlochower, Lousto, \&
  Campanelli}]{Nakano:2011pb}
Nakano, H., Zlochower, Y., Lousto, C.~O., \& Campanelli, M. 2011, Phys.Rev.,
  D84, 124006, \dodoi{10.1103/PhysRevD.84.124006}

\bibitem[{Peters(1964)}]{Peters:1964zz}
Peters, P.~C. 1964, Phys. Rev., 136, B1224, \dodoi{10.1103/PhysRev.136.B1224}

\bibitem[{Peters \& Mathews(1963)}]{Peters:1963ux}
Peters, P.~C., \& Mathews, J. 1963, Phys. Rev., 131, 435,
  \dodoi{10.1103/PhysRev.131.435}

\bibitem[{Pollney {et~al.}(2011)Pollney, Reisswig, Schnetter, Dorband, \&
  Diener}]{Pollney:2009yz}
Pollney, D., Reisswig, C., Schnetter, E., Dorband, N., \& Diener, P. 2011,
  Phys. Rev., D83, 044045, \dodoi{10.1103/PhysRevD.83.044045}

\bibitem[{Pretorius(2005)}]{Pretorius:2005gq}
Pretorius, F. 2005, Phys. Rev. Lett., 95, 121101,
  \dodoi{10.1103/PhysRevLett.95.121101}

\bibitem[{Punturo {et~al.}(2010)Punturo, Abernathy, Acernese, Allen, Andersson,
  {et~al.}}]{Punturo:2010zz}
Punturo, M., Abernathy, M., Acernese, F., {et~al.} 2010, Class.Quant.Grav., 27,
  194002, \dodoi{10.1088/0264-9381/27/19/194002}

\bibitem[{Purrer {et~al.}(2012)Purrer, Husa, \& Hannam}]{Purrer:2012wy}
Purrer, M., Husa, S., \& Hannam, M. 2012, Phys. Rev. D, 85, 124051,
  \dodoi{10.1103/PhysRevD.85.124051}

\bibitem[{Radice {et~al.}(2020)Radice, Bernuzzi, \& Perego}]{Radice:2020ddv}
Radice, D., Bernuzzi, S., \& Perego, A. 2020, Ann. Rev. Nucl. Part. Sci., 70,
  \dodoi{10.1146/annurev-nucl-013120-114541}

\bibitem[{Radice {et~al.}(2014)Radice, Rezzolla, \& Galeazzi}]{Radice:2013xpa}
Radice, D., Rezzolla, L., \& Galeazzi, F. 2014, Class.Quant.Grav., 31, 075012,
  \dodoi{10.1088/0264-9381/31/7/075012}

\bibitem[{Randall {et~al.}(2002)Randall, Ringler, Heikes, Jones, \&
  Baumgardner}]{Randall:2002}
Randall, D.~A., Ringler, T.~D., Heikes, R., Jones, P., \& Baumgardner, J. 2002,
  Comput. Sci. Eng., 4, 32

\bibitem[{Reisswig {et~al.}(2013)Reisswig, Haas, Ott, Abdikamalov, Mösta,
  Pollney, \& Schnetter}]{Reisswig:2012nc}
Reisswig, C., Haas, R., Ott, C.~D., {et~al.} 2013, Phys. Rev., D87, 064023,
  \dodoi{10.1103/PhysRevD.87.064023}

\bibitem[{Reisswig \& Pollney(2011)}]{Reisswig:2010di}
Reisswig, C., \& Pollney, D. 2011, Class.Quant.Grav., 28, 195015,
  \dodoi{10.1088/0264-9381/28/19/195015}

\bibitem[{Rinne {et~al.}(2009)Rinne, Buchman, Scheel, \&
  Pfeiffer}]{Rinne:2008vn}
Rinne, O., Buchman, L.~T., Scheel, M.~A., \& Pfeiffer, H.~P. 2009,
  Class.Quant.Grav., 26, 075009, \dodoi{10.1088/0264-9381/26/7/075009}

\bibitem[{Ruchlin {et~al.}(2018)Ruchlin, Etienne, \&
  Baumgarte}]{ruchlin2018senrnrpynumericala}
Ruchlin, I., Etienne, Z.~B., \& Baumgarte, T.~W. 2018, Physical Review D, 97,
  064036, \dodoi{10.1103/PhysRevD.97.064036}

\bibitem[{Ruiz {et~al.}(2011)Ruiz, Hilditch, \& Bernuzzi}]{Ruiz:2010qj}
Ruiz, M., Hilditch, D., \& Bernuzzi, S. 2011, Phys. Rev., D83, 024025,
  \dodoi{10.1103/PhysRevD.83.024025}

\bibitem[{Schnetter {et~al.}(2004)Schnetter, Hawley, \&
  Hawke}]{Schnetter:2003rb}
Schnetter, E., Hawley, S.~H., \& Hawke, I. 2004, Class.Quant.Grav., 21, 1465,
  \dodoi{10.1088/0264-9381/21/6/014}

\bibitem[{Shibata \& Nakamura(1995)}]{Shibata:1995we}
Shibata, M., \& Nakamura, T. 1995, Phys. Rev., D52, 5428,
  \dodoi{10.1103/PhysRevD.52.5428}

\bibitem[{Shibata \& Taniguchi(2011)}]{Shibata:2011jka}
Shibata, M., \& Taniguchi, K. 2011, Living Rev. Rel., 14, 6,
  \dodoi{10.12942/lrr-2011-6}

\bibitem[{Shibata \& Uryu(2000)}]{Shibata:1999wm}
Shibata, M., \& Uryu, K. 2000, Phys. Rev., D61, 064001,
  \dodoi{10.1103/PhysRevD.61.064001}

\bibitem[{Sperhake(2007)}]{sperhake2007binaryblackholeevolutions}
Sperhake, U. 2007, Physical Review D, 76, 104015,
  \dodoi{10.1103/PhysRevD.76.104015}

\bibitem[{Stone {et~al.}(2008)Stone, Gardiner, Teuben, Hawley, \&
  Simon}]{stone2008athenanewcode}
Stone, J.~M., Gardiner, T.~A., Teuben, P., Hawley, J.~F., \& Simon, J.~B. 2008,
  The Astrophysical Journal Supplement Series, 178, 137, \dodoi{10.1086/588755}

\bibitem[{Stone {et~al.}(2020)Stone, Tomida, White, \&
  Felker}]{stone2020athenamathplusmathplus}
Stone, J.~M., Tomida, K., White, C.~J., \& Felker, K.~G. 2020, The
  Astrophysical Journal Supplement Series, 249, 4,
  \dodoi{10.3847/1538-4365/ab929b}

\bibitem[{Stout {et~al.}(1997)Stout, De~Zeeuw, Gombosi, Groth, Marshall, \&
  Powell}]{stout1997adaptiveblockshigh}
Stout, Q.~F., De~Zeeuw, D.~L., Gombosi, T.~I., {et~al.} 1997, in Proceedings of
  the 1997 {{ACM}}/{{IEEE}} Conference on {{Supercomputing}}, {{SC}} '97 ({New
  York, NY, USA}: {Association for Computing Machinery}), 1--10,
  \dodoi{10.1145/509593.509650}

\bibitem[{Szilagyi {et~al.}(2009)Szilagyi, Lindblom, \&
  Scheel}]{Szilagyi:2009qz}
Szilagyi, B., Lindblom, L., \& Scheel, M.~A. 2009, Phys. Rev., D80, 124010,
  \dodoi{10.1103/PhysRevD.80.124010}

\bibitem[{Thierfelder {et~al.}(2011)Thierfelder, Bernuzzi, \&
  Br{\"u}gmann}]{Thierfelder:2011yi}
Thierfelder, M., Bernuzzi, S., \& Br{\"u}gmann, B. 2011, Phys.Rev., D84,
  044012, \dodoi{10.1103/PhysRevD.84.044012}

\bibitem[{Trefethen(2013)}]{trefethen2013approximation}
Trefethen, L.~N. 2013, {Approximation Theory and Approximation Practice}, Other
  Titles in Applied Mathematics (Society for Industrial and Applied
  Mathematics)

\bibitem[{{Wang} \& {Lee}(2011)}]{Wang:2011}
{Wang}, N., \& {Lee}, J.-L. 2011, SIAM Journal of Scientific Computing, 33,
  2536

\bibitem[{Weyhausen {et~al.}(2012)Weyhausen, Bernuzzi, \&
  Hilditch}]{Weyhausen:2011cg}
Weyhausen, A., Bernuzzi, S., \& Hilditch, D. 2012, Phys. Rev., D85, 024038,
  \dodoi{10.1103/PhysRevD.85.024038}

\bibitem[{White {et~al.}(2016)White, Stone, \&
  Gammie}]{white2016extensionathenacode}
White, C.~J., Stone, J.~M., \& Gammie, C.~F. 2016, The Astrophysical Journal
  Supplement Series, 225, 22, \dodoi{10.3847/0067-0049/225/2/22}

\bibitem[{York(1979)}]{York:1979}
York, J.~W. 1979, in Sources of gravitational radiation, ed. L.~L. Smarr
  (Cambridge, UK: Cambridge University Press), 83--126

\bibitem[{Zlochower {et~al.}(2005)Zlochower, Baker, Campanelli, \&
  Lousto}]{zlochower2005accurateblackhole}
Zlochower, Y., Baker, J.~G., Campanelli, M., \& Lousto, C.~O. 2005, Physical
  Review D, 72, 024021, \dodoi{10.1103/PhysRevD.72.024021}

\end{thebibliography}

\end{document}